\documentclass[a4paper,9pt]{article}
\usepackage{jheppub}
\usepackage[utf8x]{inputenc}
\usepackage[T1]{fontenc}
\usepackage{amsfonts}
\usepackage{amsmath,bm}
\usepackage{mathtools}
\usepackage{amssymb}
\usepackage[titletoc,toc,title]{appendix}
\usepackage{braket}
\usepackage{cancel}
\usepackage[normalem]{ulem}
\usepackage[font={footnotesize}]{caption}
\usepackage{color}
\usepackage{comment}
\usepackage{enumitem}
\usepackage{epsfig}
\usepackage{epstopdf}
\usepackage{float}
\usepackage{graphicx}
\usepackage{hyperref}
\usepackage{cleveref}
\usepackage{mathtools}
\usepackage{ragged2e}
\usepackage{subcaption}

\def\Tr{{{\rm Tr~ }}}

\usepackage[a4paper,
bindingoffset=0.2in,
left=1.6in,
right=-.2in,
top=1.5in,
bottom=-.4in,
footskip=.4in]{geometry}

\hypersetup{
	colorlinks=true,
	citecolor=red,
	filecolor=red,
	linkcolor=blue,
	linktocpage=true,
	urlcolor=blue
}

\begin{document}
	
\title{ Islands for Entanglement Negativity in Communicating Black Holes }

\author{Mir Afrasiar,}
\author{Jaydeep Kumar Basak,}
\author{Ashish Chandra}
\author{and Gautam Sengupta}

\affiliation{
Department of Physics,\\
Indian Institute of Technology,\\ 
Kanpur 208 016, India
}
	
\emailAdd{afrasiar@iitk.ac.in}
\emailAdd{jaydeep@iitk.ac.in}
\emailAdd{achandra@iitk.ac.in}
\emailAdd{sengupta@iitk.ac.in}

\abstract{We obtain the holographic entanglement negativity for bipartite mixed states at a finite temperature in baths described by conformal field theories dual to configurations involving two communicating black holes in a braneworld geometry. In this context, we analyze the mixed state entanglement structure characterized by the information transfer between the black holes. The model corresponds to a configuration of two dimensional eternal JT black holes in a braneworld geometry involving two Planck branes coupled through shared bath systems described by $CFT_2$s. Our results reproduce analogue of the Page curves for the entanglement negativity obtained earlier in the context of random matrix theory and from geometric evaporation in JT black hole configurations.}

	\maketitle
	
	\flushbottom

\section{Introduction} \label{sec:intro}
Over the last few decades, the black hole information loss paradox \cite{Hawking:1975vcx,Hawking:1976ra} has been one of the most engaging and fascinating issues in the quest for a quantum theory of gravity. The central element of this puzzle
involves the monotonic increase in the entanglement entropy of the Hawking radiation from an evaporating black hole resulting in the fine grained entropy to dominate the coarse grained entropy at late times which leads to a violation of unitarity. It could be shown that the unitarity of the black hole evaporation process required the corresponding entanglement entropy of the Hawking radiation to follow a Page curve \cite{Page:1993wv}. Very recently this issue has been addressed through the fascinating development of the island proposal which has led to the exciting possibility of a resolution of this long standing paradox \cite{Engelhardt:2014gca,Almheiri:2019psf,Penington:2019npb,Almheiri:2019hni,Almheiri:2020cfm,Almheiri:2019yqk,Penington:2019kki,Almheiri:2019qdq}. The island or the quantum extremal surface (QES) formula for entanglement entropy was motivated from the quantum corrected Ryu-Takayanagi (RT) proposal \cite{Ryu:2006bv,Ryu:2006ef,Hubeny:2007xt,Faulkner:2013ana} and this may be obtained by extremizing the generalized fine-grained entanglement entropy of the Hawking radiation. In this context it could be shown that at late times the entanglement entropy of
a bath subsystem in the radiation flux of the black hole receives contributions from a region in the black hole geometry termed the entanglement island\footnote{Recently, there has been a rich development in these directions which can be found in \cite{Almheiri:2019psy,Anderson:2020vwi,Chen:2019iro,Balasubramanian:2020hfs,Chen:2020wiq,Gautason:2020tmk,Bhattacharya:2020ymw,Anegawa:2020ezn,Hashimoto:2020cas,Hartman:2020swn,Krishnan:2020oun,Alishahiha:2020qza,Geng:2020qvw,Li:2020ceg,Chandrasekaran:2020qtn,Bak:2020enw,Krishnan:2020fer,Karlsson:2020uga,Hartman:2020khs,Balasubramanian:2020coy,Balasubramanian:2020xqf,Sybesma:2020fxg,Chen:2020hmv,Ling:2020laa,Hernandez:2020nem,Marolf:2020rpm,Matsuo:2020ypv,Akal:2020twv,Caceres:2020jcn,Raju:2020smc,Deng:2020ent,Anous:2022wqh,Bousso:2022gth,Hu:2022ymx,Grimaldi:2022suv,Akers:2022max,Yu:2021rfg,Geng:2021mic,Chou:2021boq,Hollowood:2021lsw,He:2021mst,Arefeva:2021kfx,Ling:2021vxe,Bhattacharya:2021dnd,Azarnia:2021uch,Saha:2021ohr,Hollowood:2021wkw,Sun:2021dfl,Li:2021dmf,Aguilar-Gutierrez:2021bns,Ahn:2021chg,Yu:2021cgi,Lu:2021gmv,Caceres:2021fuw,Akal:2021foz,Arefeva:2022cam,Arefeva:2022guf,Bousso:2022ntt,Krishnan:2021ffb,Zeng:2021kyb,Teresi:2021qff,Okuyama:2021bqg,Chen:2021jzx,Pedraza:2021ssc,Guo:2021blh,Kibe:2021gtw,Renner:2021qbe,Dong:2021oad,Raju:2021lwh,Nam:2021bml,Kames-King:2021etp,Chen:2021lnq,Sato:2021ftf,Kudler-Flam:2021alo,Wang:2021afl,Ageev:2021ipd,Buoninfante:2021ijy,Cadoni:2021ypx,Marolf:2021ghr,Chu:2021gdb,Urbach:2021zil,Li:2021lfo,Neuenfeld:2021bsb,Aalsma:2021bit,Ghosh:2021axl,Bhattacharya:2021jrn,Geng:2021wcq,Krishnan:2021faa,Verheijden:2021yrb,Bousso:2021sji,Karananas:2020fwx,Goto:2020wnk,Bhattacharya:2020uun,Chen:2020jvn,Agon:2020fqs,Laddha:2020kvp,Akers:2019nfi,Chen:2019uhq,Basu:2022reu,Uhlemann:2021nhu,Uhlemann:2021itz,Germani:2022rac,Yadav:2022fmo,Omidi:2021opl} and the references therein.}. Specifically the bath subsystem and the island regions were shown to be a part of the same entanglement wedge in higher dimension as described in \cite{Almheiri:2019hni}. The corresponding island formula for the generalized fine-grained entropy of a subsystem 
$\mathcal{R}$ in the radiation bath is given as,
\begin{align}\label{IsformEE}
	S[\mathcal{R}]=\min \left\{\operatorname{ext}_{Is(\mathcal{R})}\left[\frac{\operatorname{Area}[\partial Is(\mathcal{R})]}{4 G_{N}}+S_{eff}[\operatorname{\mathcal{R}} \cup Is(\mathcal{R})]\right]\right\},
\end{align} 
where $Is(\mathcal{R})$ is the island region in the black hole geometry corresponding to the subsystem $\mathcal{R}$ in the radiation bath \footnote{Higher dimensional generalization of the island construction for the entanglement entropy has been studied in some recent papers \cite{Almheiri:2019psy,Chen:2020uac,Chen:2020hmv,Ling:2020laa,Krishnan:2020fer}.}.

From the perspective described above, an extremely interesting and elegant model has been proposed in \cite{Balasubramanian:2021xcm} similar to the one described in \cite{Almheiri:2020cfm}. However the authors of \cite{Balasubramanian:2021xcm} considered two copies of finite sized reservoirs described by $CFT_2$s each with two quantum dots located at their boundaries at a finite temperature. The holographic dual of these quantum dots are described by Planck branes in the bulk $AdS_3$ space time which supports $AdS_2$ geometries with Jackiw-Teitelboim (JT) gravity \cite{Teitelboim:1983ux,Jackiw:1984je}. Hence the two Planck branes may involve eternal JT black holes which communicate with each other through the common radiation reservoirs. Note that from the perspective of each brane, the bath together with the other Planck brane appears as a gravitating configuration \cite{Balasubramanian:2021xcm}. For this configuration, the authors of \cite{Balasubramanian:2021xcm} have computed the generalized entanglement entropy for a finite subsystem in the radiation reservoirs which once again characterizes the communication between the two eternal JT black holes on the Planck branes. 

In a separate context, it is well known in quantum information theory that the entanglement entropy is an appropriate measure for the characterization of pure state entanglement, however for mixed states
it receives contributions from irrelevant classical and quantum correlations. A consistent and computable measure for the characterization of mixed state entanglement which serves as an upper bound to the distillable entanglement is described by the entanglement negativity introduced in \cite{Vidal:2002zz,PhysRevLett.95.090503}. In \cite{Calabrese:2014yza,Calabrese:2012nk,Calabrese:2012ew} the authors established a replica technique to compute the entanglement negativity for bipartite pure and mixed states in $CFT_2$. The first holographic computation for the entanglement negativity for the pure vacuum state in $CFT$s was described in \cite{Rangamani:2014ywa}.
Subsequently general holographic proposals\footnote{Motivated by the developments in \cite{Dong:2021clv}, a heuristic proof of these proposals has been presented in \cite{KumarBasak:2020ams}.} for the entanglement negativity of bipartite pure and mixed states in $CFT_2$s involving specific algebraic sums of the lengths of geodesics homologous to various combinations of subsystems were introduced in \cite{Chaturvedi:2016rcn,Jain:2017aqk,Malvimat:2018txq}\footnote{For further developments see also \cite{Chaturvedi:2017znc,Jain:2017uhe,Malvimat:2018ood,Chaturvedi:2016rft,Jain:2017xsu,KumarBasak:2020viv,Afrasiar:2021hld,Basu:2021awn,Basu:2021axf,Basu:2022nds}.}. It is interesting to note that the authors of \cite{Kudler-Flam:2018qjo} had proposed an alternative prescription for the holographic entanglement negativity of bipartite mixed states in $CFT$s in terms of the backreacted minimal entanglement wedge cross section (EWCS) in the context of the $AdS/CFT$ scenario, further refined in \cite{KumarBasak:2020eia}. For spherically entangling surfaces of the subsystems considered in the dual $CFT$s, the backreaction parameter was described by an overall numerical factor which is dependent on the dimension of the $CFT$s. A proof for this duality between the holographic entanglement negativity and the bulk EWCS was further established in \cite{Kusuki:2019zsp} involving the idea of reflected entropy described in \cite{Dutta:2019gen}.

A generalization of these proposals for the entanglement negativity in $CFT_2$s coupled to semiclassical gravity was advanced in \cite{KumarBasak:2020ams} with a possible derivation using the replica wormhole contributions to the gravitational path integral involving replica symmetry breaking saddles as discussed in \cite{Dong:2021clv}. In the present article, we focus on the computation of the holographic entanglement negativity for various bipartite mixed states in a braneworld model \cite{Balasubramanian:2021xcm} mentioned earlier. We consider different scenarios involving the subsystem sizes and the time for two adjacent and disjoint subsystems located in the bath regions at a finite temperature. Furthermore, we discuss the behaviours of the entanglement negativity profiles obtained in these scenarios in terms of the Hawking radiation. We observe interesting similarities between our results with those described in \cite{KumarBasak:2021rrx,Shapourian:2020mkc}.

In the appendix \ref{model1}, we explore a similar model of communicating black holes described extensively in \cite{Geng:2021iyq} where the authors have considered a $BCFT_2$ on a manifold with two boundaries in the context of the $AdS_3/BCFT_2$ scenario. The holographic dual of this configuration is described by a wedge enclosed within the two KR branes in the bulk $AdS_3$ braneworld geometry. The KR branes involve $CFT_2$ matter fields with a constant Lagrangian which is connected to the $CFT_2$ on the asymptotic boundary of the dual $AdS_3$ geometry through transparent boundary conditions \cite{Almheiri:2019yqk,Almheiri:2019hni}. At a finite temperature, black holes may be induced on these two KR branes from the higher dimensional eternal $AdS_3$ BTZ black hole. In this model, we compute the holographic entanglement negativity and obtain the corresponding Page curves for various bipartite mixed states of two adjacent and disjoint subsystems in the bath $BCFT_2$s at a finite temperature.

This article is organized as follows. In \cref{review1}, we review the relevant works discussed in the braneworld model \cite{Balasubramanian:2021xcm} and holographic entanglement negativity described in \cite{Jain:2017aqk,Malvimat:2018txq}. Next in \cref{model2}, we apply the results for the generalized entanglement entropy discussed in \cref{review1} for different cases of adjacent and disjoint subsystems and obtain the corresponding holographic entanglement negativity. In appendix \ref{model1}, we first compute the entanglement entropy for a generic subsystem in the context of another model of braneworld geometry \cite{Geng:2021iyq}. Subsequently, we apply these results to further compute the holographic entanglement negativity for various bipartite mixed states using the holographic proposals described in \cref{review1}. In appendix \ref{formula_model2}, we list the results for the holographic entanglement negativity for all the different scenarios discussed in \cref{model2}. Finally in the \cref{discussion}, we summarize and discuss our results and future open issues.


\section{Review of earlier results}\label{review1}

\subsection{Braneworld model}
In this subsection we will be reviewing an intriguing model of finite sized non-gravitating reservoirs, coupled with two quantum dots at its two boundaries \cite{Balasubramanian:2021xcm}. The holographic dual of these quantum dots are Plank branes described by $AdS_2$ geometries. The maximal extension of the Penrose diagram in this context consists of two eternal JT black holes in the $AdS_2$ space time located on the Planck branes. In this case the usual reflecting boundaries are coupled to two non-gravitating reservoirs through transparent boundary conditions \cite{Almheiri:2019yqk,Almheiri:2019qdq} such that the two black holes are connected to each other through the shared reservoirs. This construction involves  identical matter $CFT_2$s on both the reservoirs and the gravity regions. Interestingly, a single brane together with a reservoir appears to be a gravitating one from the perspective of each brane. In this framework, the authors of \cite{Balasubramanian:2021xcm} have considered the two black holes at different temperatures and obtained the generalized entanglement entropy of a finite region in the two reservoirs by utilizing \cref{IsformEE}. The first term corresponds to the value of the dilaton field in the JT gravity with a constant $\phi_0$ added to it whereas the second term follows from the usual computation of the entanglement entropy \cite{Ryu:2006bv} of a segment $\mathcal{R} \cup Is(\mathcal{R})$. The generalized entanglement entropy of the finite region in the reservoirs characterizes the communication between the two JT black holes on the Planck branes. In what follows we review the computation of the generalized entanglement entropy for a subsystem in the radiation reservoirs while considering the JT black holes on the Planck branes at a same temperature.

\subsubsection{Entanglement entropy in the two black hole configuration} 
In this subsection, we consider two eternal JT black holes at the same temperature and describe the explicit computation for the generalized entanglement entropy for a subsystem $A$ consisting of the union of two identical line segments in the two copies of the reservoirs. 

For this configuration, the metrics referring to the exterior regions of the two eternal JT black holes may be written as
\begin{eqnarray}
	ds_1^2 &=& \frac{4\pi^2}{\beta^2} \frac{-dt^2+d\xi^2}{\sinh^2 \frac{2\pi \xi}{\beta}} , \quad \xi \in \left(-\infty, - \epsilon\right]\,, \label{MetricPhysL1b1}\\
	ds_2^2 &=& \frac{4\pi^2}{\beta^2} \frac{-dt^2+d\xi^2}{\sinh^2 \frac{2\pi }{\beta}(\xi - L)},\quad \xi \in \left[L + \epsilon, +\infty\right) \label{MetricPhysL2b2}\,,
\end{eqnarray} 
where $L$ is the length of each radiation reservoir with a metric defined by
\begin{equation}\label{MetricBath}
	ds_{R}^2 = \frac{-d t^2 + d\xi^2}{\epsilon^2}\,, \quad \xi \in \left[- \epsilon, L+\epsilon\right]\, 
\end{equation}
where the reservoir is glued continuously to the surfaces $\xi=-\epsilon$ and $L+\epsilon$. The dilaton profiles for the two eternal JT black holes on the Planck branes are then given as follows
\begin{eqnarray}\label{dilaton2BH}
	\phi_{a}(\xi) &=& \frac{2\pi\phi_r}{\beta}  \coth \frac{2\pi \xi}{\beta}\,, \label{dilaton-BH-1-b1r}\\
	\phi_{b}(\xi) &=& \frac{2\pi\phi_r}{\beta}  \coth \frac{2\pi}{\beta}\left(\xi-L\right). \end{eqnarray}

We now compute the generalized entanglement entropy for the subsystem described by the union of two identical segments $A=[p_1,p_2]\cup[p_3,p_4]$ in the radiation regions of the two TFD copies. The end points of the two corresponding segments in the ($\xi, t$) coordinates are specified as follows \cite{Balasubramanian:2021xcm}
\begin{equation}
	p_1 = (v, -t+i\frac{\beta}{2}), \quad p_2 = (u, -t+i\frac{\beta}{2}), \quad p_3 = (u, t), \quad p_4 = (v, t)\,.
\end{equation}
In this case there are seven possible contributions to the corresponding entanglement entropy due to the different structures of the RT surfaces supported by the subsystems mentioned above. In what follows, we describe these distinct contributions in detail.

\begin{figure}[H]
	\centering
	\begin{subfigure}[b]{0.45\textwidth}
		\centering
		\includegraphics[width=\textwidth]{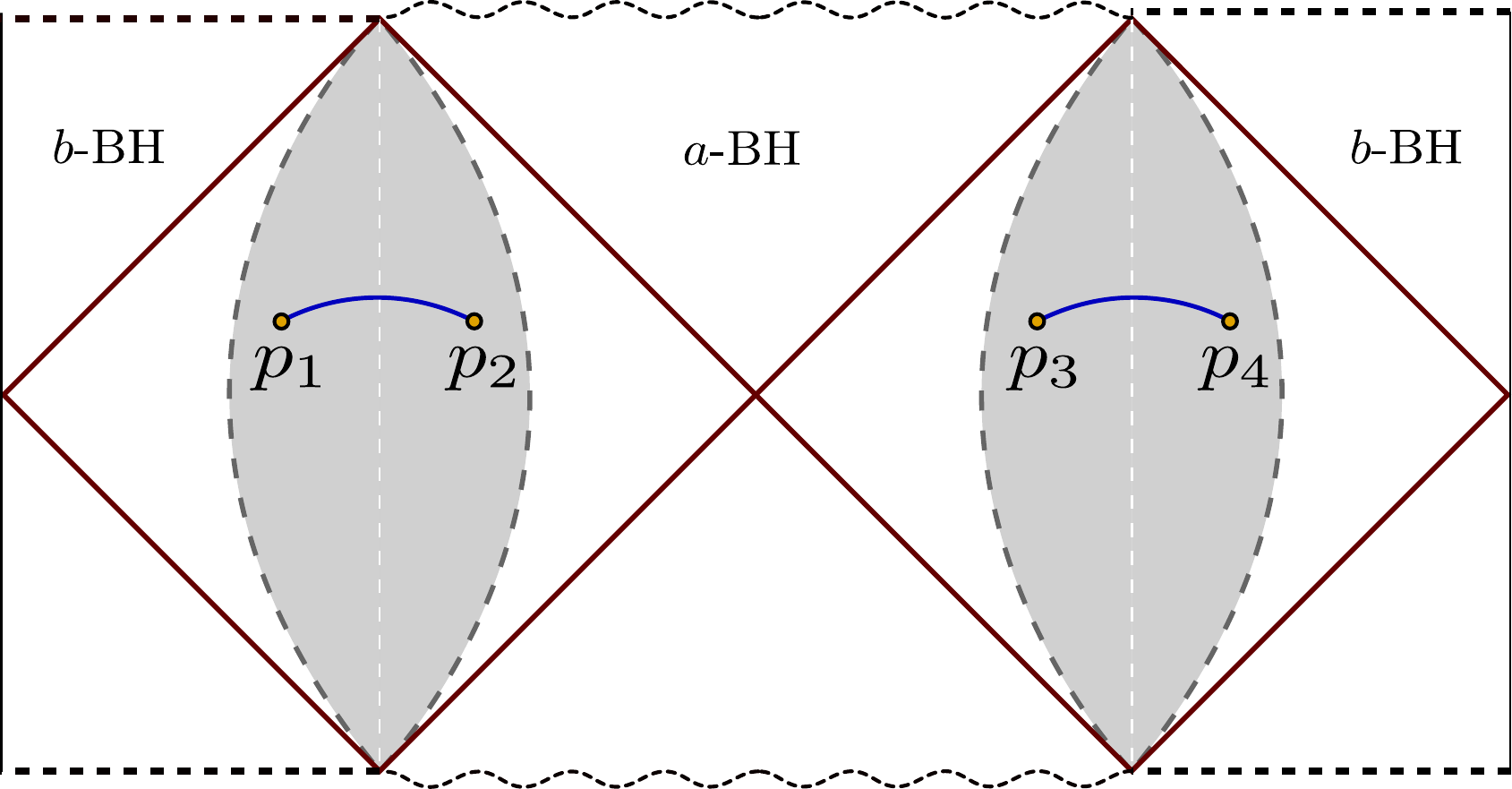}
		\caption{}
		\label{a}
	\end{subfigure}
	\hspace{.5cm}
	\begin{subfigure}[b]{0.45\textwidth}
		\centering
		\includegraphics[width=\textwidth]{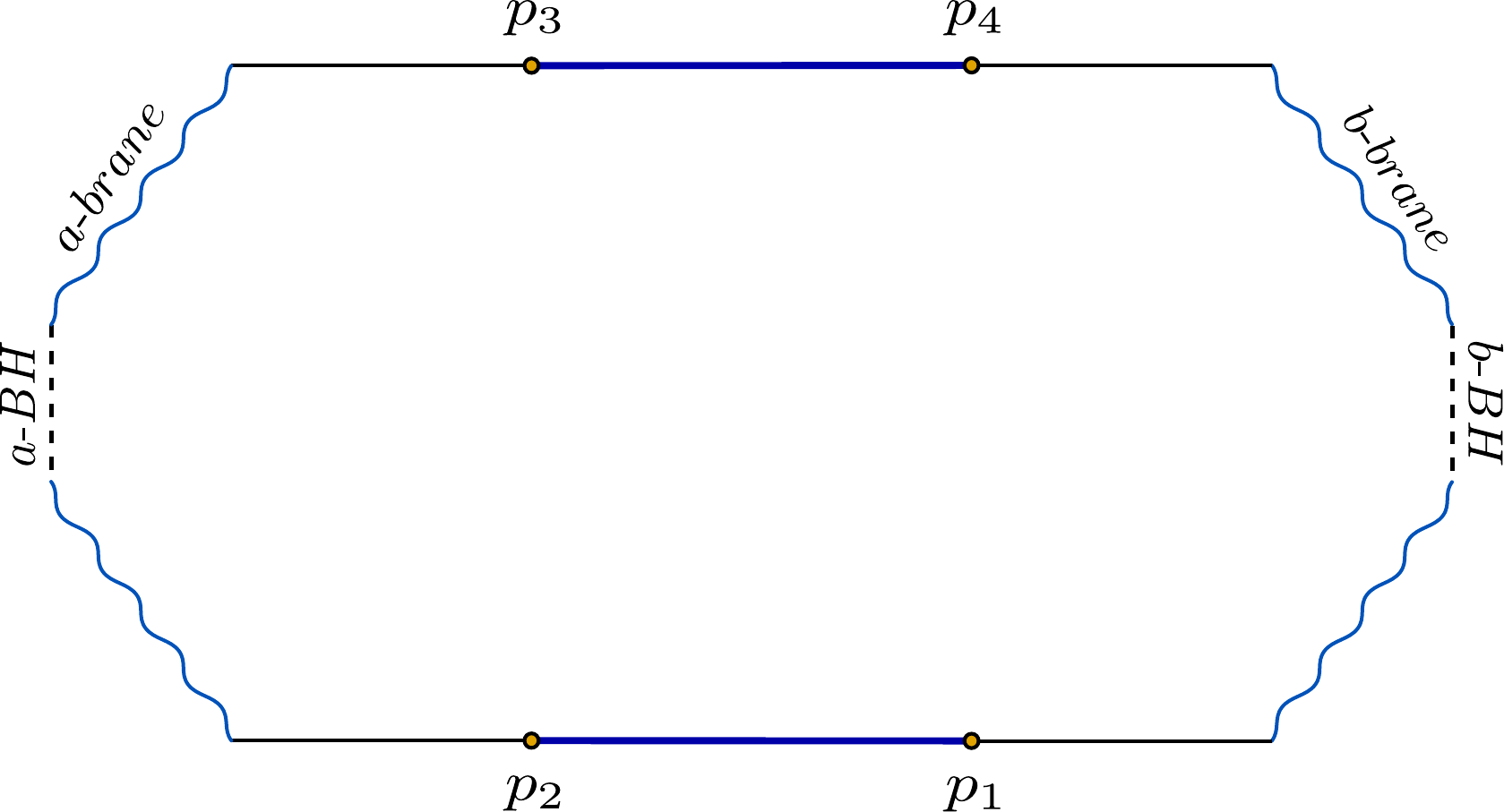}
		\caption{}
		\label{b}
	\end{subfigure}
	\caption{(a) The maximal extension of the Penrose diagram for two $AdS_2$ eternal black holes. A subsystem given by the union of two segments is considered in the radiation reservoirs (shaded regions) with end points $p_1$, $p_2$, $p_3$, $p_4$. Note that the left and the right most black lines are identified in this diagram. (b) A constant time slice of the $AdS_2$ eternal black holes where the Planck branes are denoted by the wigly lines and the black hole intereiors are shown by the dotted lines. (Figures modified from \cite{Balasubramanian:2021xcm,Geng:2021hlu})}\label{penrose2}
\end{figure}

$\bm{(a)}$ We first discuss the configuration which is completely connected and does not include any island region in the gravity sector. The end points of the two segments of the subsystem $A$ are connected to each other $p_1 \leftrightarrow p_4$ and $p_2 \leftrightarrow p_3$ by two geodesics in the 3-dimensional bulk as depicted in \cref{RTa}. We term these geodesics as bulk-type RT surfaces. The expression for the corresponding generalized entanglement entropy may be obtained utilizing \cref{IsformEE} as follows
\begin{eqnarray}\label{Sbulk}
	\mathcal{S}_{A}^{\text{bulk}} = 2\frac{c}{3} \log \left[\frac{\beta }{\pi } \cosh\frac{2\pi t}{\beta }\right].
\end{eqnarray}

$\bm{(b)}$ The second configuration also corresponds to a fully connected one which includes island regions from both the JT black holes. We may obtain the generalized entanglement entropy for the subsystem $A$ following a procedure analogous to the single black hole case as discussed in \cite{Balasubramanian:2021xcm}. However for this configuration, the computation involves an extremization of \cref{IsformEE} over two island regions located on the $a$ and the $b$-black holes. The subsystem $A$ in this context admits RT surfaces which start from $\partial A$ and intersect the exterior regions of both the black holes. We call these RT surfaces $ab$-type which are depicted in \cref{RTb}. The corresponding generalized entanglement entropy for the subsystem $A$ may then be expressed as
\begin{eqnarray}\label{S-ab}
	\mathcal{S}_{A}^{ab} &=& 4 \phi_0 + \frac{4 \pi  \phi_r }{\beta } \coth \left(\frac{2 \pi }{\beta } u + \log  \frac{24 \pi  \phi }{c \beta } \right) + \frac{c}{3} \log \left[ \frac{\beta}{\pi} \; \frac{\cosh \left(\frac{4 \pi }{\beta } u + \log  \frac{24 \pi  \phi }{c \beta } \right)-1}{ \sinh \left(\frac{2 \pi }{\beta } u + \log  \frac{24 \pi  \phi }{c \beta } \right)} \right]\nonumber\\
	&+& \frac{4 \pi  \phi_r }{\beta } \coth \left(\frac{2 \pi}{\beta } (L-v) + \log \frac{24 \pi  \phi }{c \beta }\right) + \frac{c}{3} \log \left[ \frac{\beta}{\pi}\; \frac{\cosh \left(\frac{4 \pi}{\beta } (L-v)+ \log \frac{24 \pi  \phi }{c \beta }\right)-1}{  \sinh \left( \frac{2 \pi}{\beta } (L-v)+ \log \frac{24 \pi  \phi }{c \beta }\right)} \right].\nonumber\\
	&&
\end{eqnarray}

$\bm{(c)}$ We now discuss a disconnected configuration which does not include any island region as shown in \cref{RTc}. Here the entanglement entropy for the subsystem $A$ may be obtained from the geodesics which are homologous to each of the segments in the two TFD copies separately. We term these geodesics as dome-type RT surfaces. The corresponding generalized entanglement entropy for this configuration is given by the following expression
\begin{eqnarray}\label{S-dome}
	\mathcal{S}_{A}^{\text{dome}} = 2 \frac{c}{3} \log \left( \frac{\beta }{\pi } \sinh \frac{\pi }{\beta } |u-v| \right).
\end{eqnarray}

$\bm{(d)}$ This configuration includes an island region in the $a$-black hole corresponding to two geodesics which start from the end points $p_2$, $p_3$ of each of the segments and intersect the two exterior regions of the $a$-black hole. However, the other end points of the subsystem $A$ are connected to each other $ p_4 \leftrightarrow p_1 $ by another geodesic. In this connected configuration, we term the corresponding geodesics as $a$-bulk type RT surfaces which are depicted in \cref{RTd}. The expression for the generalized entanglement entropy for the subsystem $A$ may be computed using \cref{IsformEE} as follows
\begin{eqnarray}\label{S-abulk}
	\mathcal{S}_{A}^{a\text{-bulk}} &=&  4 \phi_0 + \frac{4 \pi  \phi_r }{\beta } \coth \left(\frac{2 \pi }{\beta } u + \log  \frac{24 \pi  \phi }{c \beta } \right)\nonumber\\
	&+& \frac{c}{3} \log \left[ \frac{\beta}{\pi}\;\frac{\cosh \left(\frac{4 \pi }{\beta } u + \log  \frac{24 \pi  \phi }{c \beta }\right) -1}{  \sinh \left(\frac{2 \pi }{\beta } u + \log  \frac{24 \pi  \phi }{c \beta } \right)} \right] + \frac{c}{3} \log \left[\frac{\beta }{\pi } \cosh\frac{2\pi t}{\beta }\right].
\end{eqnarray}

$\bm{(e)}$  Similar to the previous case, we now discuss another connected configuration which admits an island region in the $b$-black hole geometry only. The geodesics in this case start from the end points $p_1$, $p_4$ of the two segments and intersect the two exterior regions of the $b$-black hole. However another geodesic connects the other end points $p_2$, $p_3$ of the subsystem $A$. In contrast to the previous case, these geodesics are termed as $b$-bulk type RT surfaces (\cref{RTe}). The corresponding generalized entanglement entropy for the subsystem $A$ in this case may be obtained utilizing \cref{IsformEE} as follows

\begin{eqnarray}\label{S-bbulk}
	\mathcal{S}_{A}^{b\text{-bulk}} &=& 4 \phi_0 + \frac{4 \pi  \phi_r }{\beta } \coth \left(\frac{2 \pi}{\beta } (L-v) + \log \frac{24 \pi  \phi }{c \beta }\right)\nonumber\\
	&+& \frac{c}{3} \log \left[ \frac{\beta}{\pi}\; \frac{\cosh \left(\frac{4 \pi}{\beta } (L-v)+ \log \frac{24 \pi  \phi }{c \beta }\right)-1}{  \sinh \left( \frac{2 \pi}{\beta } (L-v)+ \log \frac{24 \pi  \phi }{c \beta }\right)} \right] + \frac{c}{3} \log\left[ \frac{\beta }{\pi } \cosh\frac{2\pi t}{\beta }\right].
\end{eqnarray}

$\bm{(f)}$ Another disconnected configuration includes two island regions in the gravity sector where the corresponding geodesics, which are termed as $aa$-type RT surfaces, start from $\partial A$ and intersect the two exterior regions of the $a$-black hole (\cref{RTf}). The generalized entanglement entropy for this disconnected configuration is computed by extremizing \cref{IsformEE} over the two island regions and is given as
\begin{eqnarray}\label{S-aa}
	\mathcal{S}_{A}^{aa} &=& 4 \phi_0 + \frac{4 \pi  \phi_r }{\beta } \coth \left(\frac{2 \pi }{\beta } u + \log  \frac{24 \pi  \phi }{c \beta } \right) + \frac{c}{3} \log \left[\frac{\beta}{\pi}\; \frac{\cosh \left(\frac{4 \pi }{\beta } u + \log  \frac{24 \pi  \phi }{c \beta }-1\right)}{  \sinh \left(\frac{2 \pi }{\beta } u + \log  \frac{24 \pi  \phi }{c \beta } \right)} \right]\nonumber\\
	&+& \frac{4 \pi  \phi_r }{\beta } \coth \left(\frac{2 \pi }{\beta } v + \log  \frac{24 \pi  \phi }{c \beta } \right) +\frac{c}{3} \log \left[ \frac{\beta}{\pi}\;\frac{\cosh \left(\frac{4 \pi }{\beta } v + \log  \frac{24 \pi  \phi }{c \beta }-1\right)}{  \sinh \left(\frac{2 \pi }{\beta } v + \log  \frac{24 \pi  \phi }{c \beta } \right)} \right].
\end{eqnarray}

$\bm{(g)}$ Finally we consider the last configuration which is similar to the preceding one but with the island regions in the 
$b$-black hole geometry as shown in \cref{RTg}. In this disconnected configuration, the RT surfaces supported by the subsystem $A$ are termed $bb$-type. Once again, the corresponding generalized entanglement entropy may be obtained utilizing \cref{IsformEE} as follows

\begin{align}\label{S-bb}
	\mathcal{S}_{A}^{bb} =& \,4 \phi_0 + \frac{4 \pi  \phi_r }{\beta } \coth \left(\frac{2 \pi}{\beta } (L-v) + \log \frac{24 \pi  \phi }{c \beta }\right)\nonumber\\
	+&\frac{c}{3} \log \left[\frac{\beta}{\pi}\; \frac{\cosh \left(\frac{4 \pi}{\beta } (L-v)+ \log \frac{24 \pi  \phi }{c \beta }\right)-1}{  \sinh \left( \frac{2 \pi}{\beta } (L-v)+ \log \frac{24 \pi  \phi }{c \beta }\right)} \right]\nonumber\\
	+& \frac{4 \pi  \phi_r }{\beta } \coth \left(\frac{2 \pi}{\beta } (L-u) + \log \frac{24 \pi  \phi }{c \beta }\right) + \frac{c}{3} \log \left[ \frac{\beta}{\pi}\;\frac{\cosh \left(\frac{4 \pi}{\beta } (L-u)+ \log \frac{24 \pi  \phi }{c \beta }\right)-1}{  \sinh \left( \frac{2 \pi}{\beta } (L-v)+ \log \frac{24 \pi  \phi }{c \beta }\right)} \right].\nonumber\\
	&
\end{align}

The generalized entanglement entropy for the subsystem $A$ in the radiation reservoirs may now be determined 
from the minimum of all the above possible contributions as follows
\begin{equation}\label{HEE-Shaghoulian}
	S_A=\text{min}\left(\mathcal{S}_{A}^{\text{bulk}},\mathcal{S}_{A}^{ab},\mathcal{S}_{A}^{\text{dome}},\mathcal{S}_{A}^{a\text{-bulk}},\mathcal{S}_{A}^{b\text{-bulk}},\mathcal{S}_{A}^{aa},\mathcal{S}_{A}^{bb}\right)\,.
\end{equation}

In what follows we plot the generalized entanglement entropies for the subsystem $A$ with respect to the time and its size for all the above possible configurations obtained from the respective structures of the corresponding RT surfaces.
\begin{figure}[H]
	\centering
	\begin{subfigure}[t]{.45\textwidth}
		\includegraphics[width=1\linewidth]{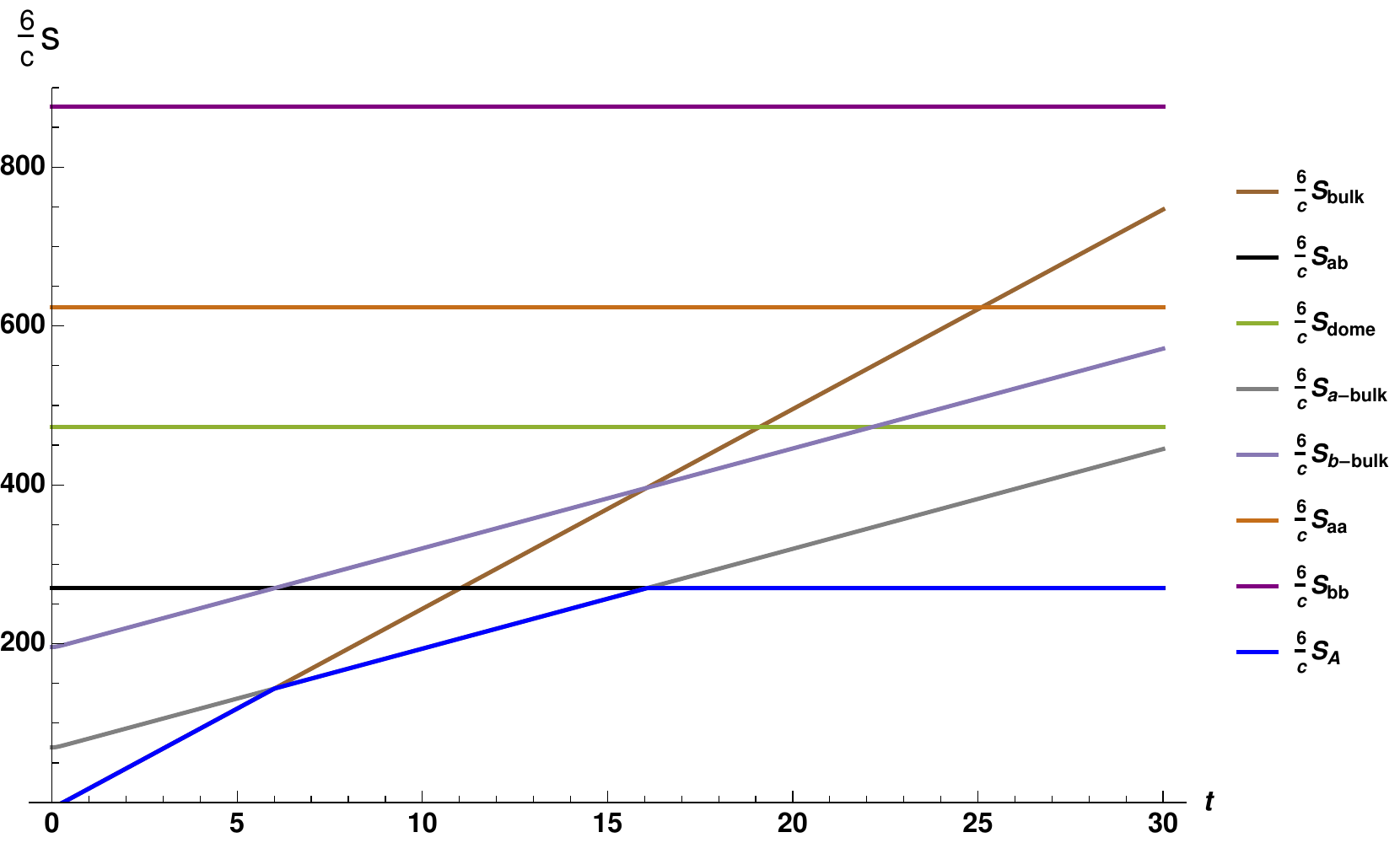}
		\caption{Entanglement entropies corresponding to the different RT surfaces w.r.t time. Here $A=[.02L,.78L]$ and time $t$ is varied from [0,\,30].}
		\label{Entropy_t}
	\end{subfigure}
	\hspace{.1cm}
	\begin{subfigure}[t]{.45\textwidth}
		\includegraphics[width=1\linewidth]{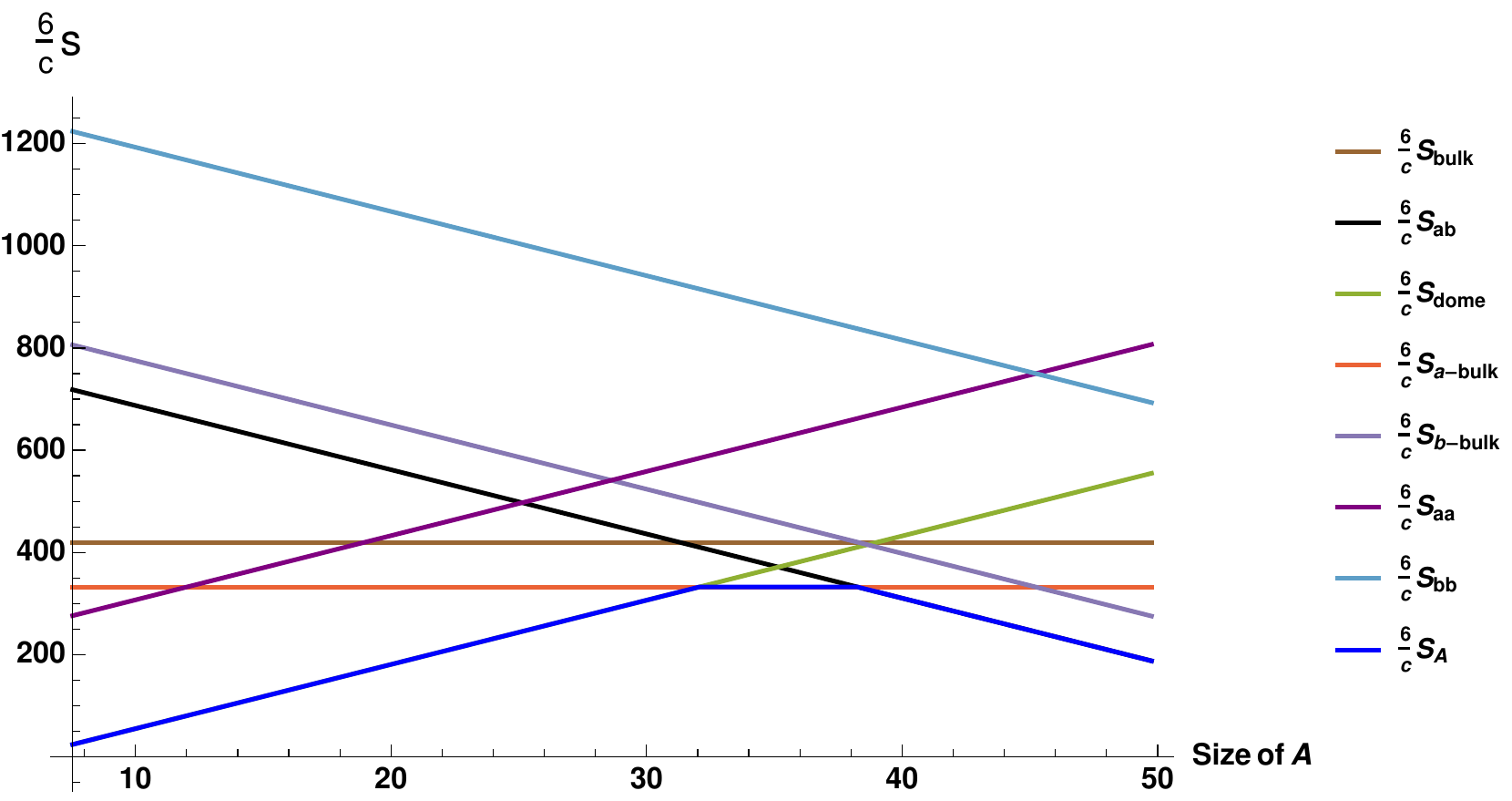}
		\caption{Entanglement entropies corresponding to the different RT surfaces w.r.t the size of the subsystem $A$. Here $t=17$ and the size of $A$ is varied from $[.1L,.99L]$.}
		\label{Entropy_r}
	\end{subfigure}
	\caption{In the above figures, we have chosen $\beta=1$, $c=500$, $\phi_0= \frac{30c}{6}$, $\phi_r= \frac{30}{\pi}$, $L=\frac{16\pi}{\beta}$, $\epsilon=.001$.}
	\label{holographicEEShag}
\end{figure}
\begin{figure}[H]
	\centering
	\vspace{1.2cm}
	\begin{subfigure}[b]{0.45\textwidth}
		\centering
		\includegraphics[width=\textwidth]{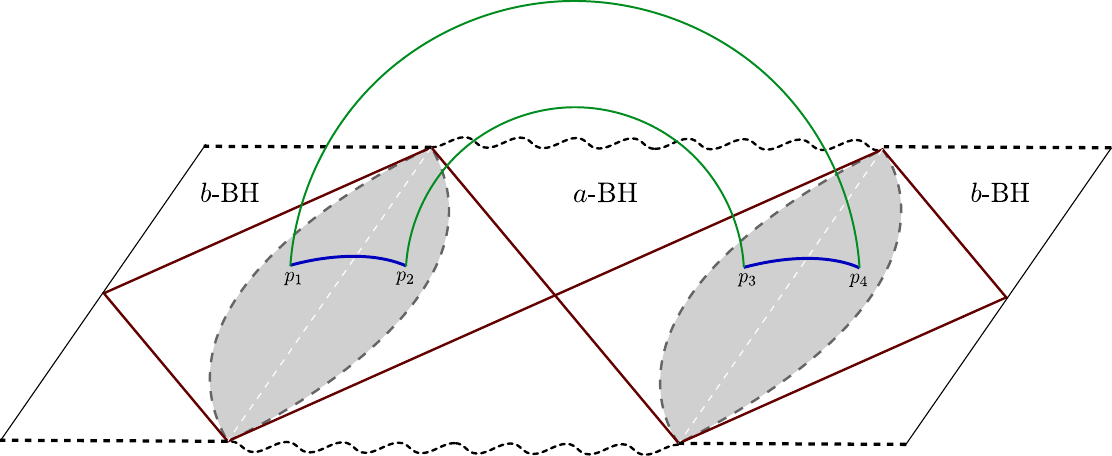}
		\caption{}
		\label{RTa}
	\end{subfigure}
	\hspace{.1cm}
	\vspace{.4cm}
	\begin{subfigure}[b]{0.45\textwidth}
		\centering
		\includegraphics[width=\textwidth]{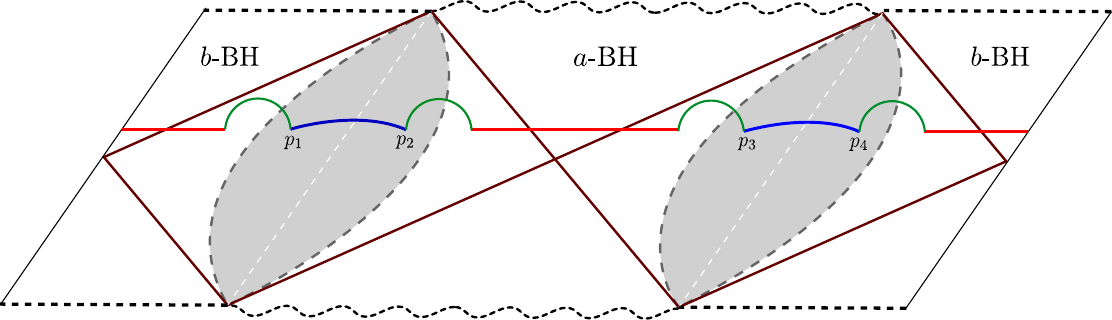}
		\caption{}
		\label{RTb}
	\end{subfigure}
	\hspace{.1cm}
	\vspace{1.3cm}
	\begin{subfigure}[b]{0.45\textwidth}
		\centering
		\includegraphics[width=\textwidth]{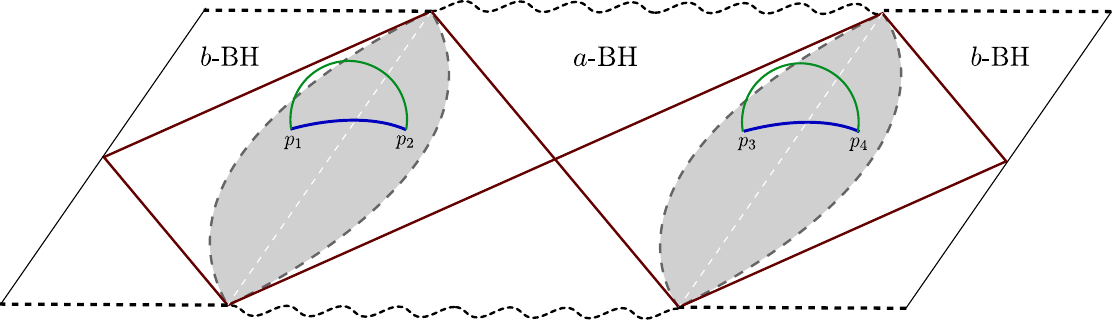}
		\caption{}
		\label{RTc}
	\end{subfigure}
	\hspace{.1cm}
	\begin{subfigure}[b]{0.45\textwidth}
		\centering
		\includegraphics[width=\textwidth]{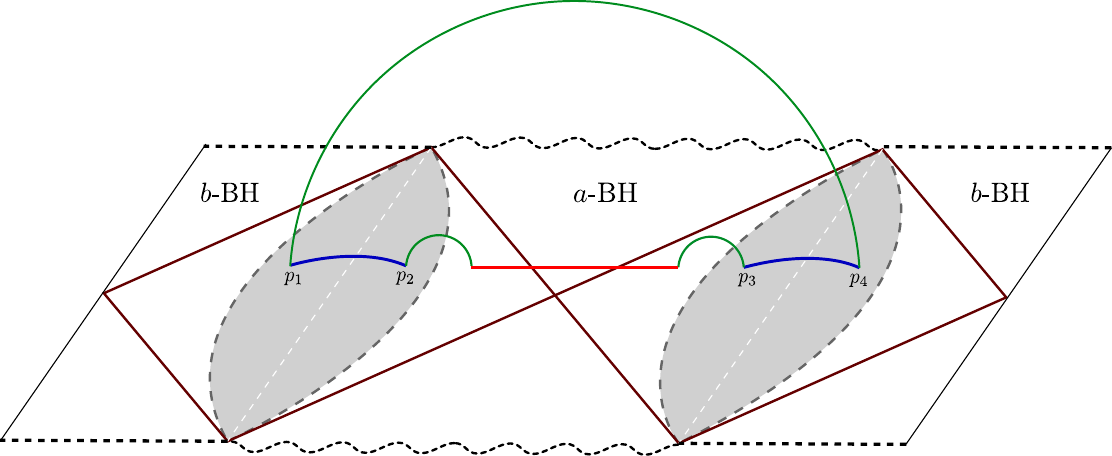}
		\caption{}
		\label{RTd}
	\end{subfigure}
	\hspace{.1cm}
	\vspace{1.5cm}
	\begin{subfigure}[b]{0.45\textwidth}
		\centering
		\includegraphics[width=\textwidth]{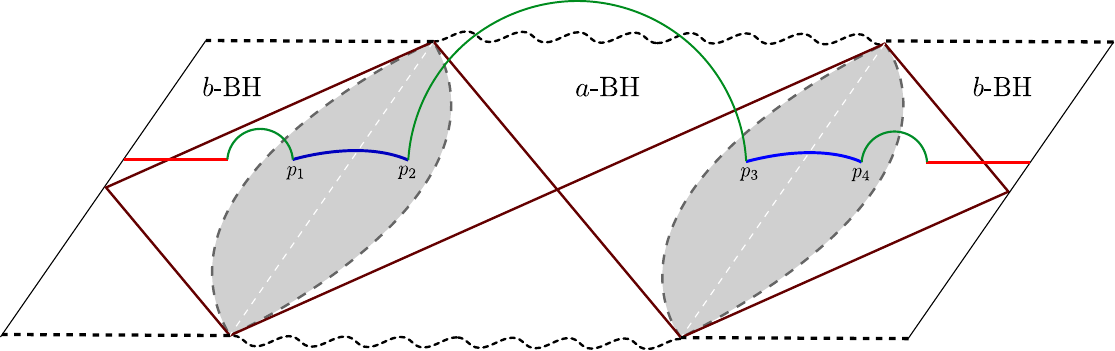}
		\caption{}
		\label{RTe}
	\end{subfigure}
	\hspace{.1cm}
	\begin{subfigure}[b]{0.45\textwidth}
		\centering
		\includegraphics[width=\textwidth]{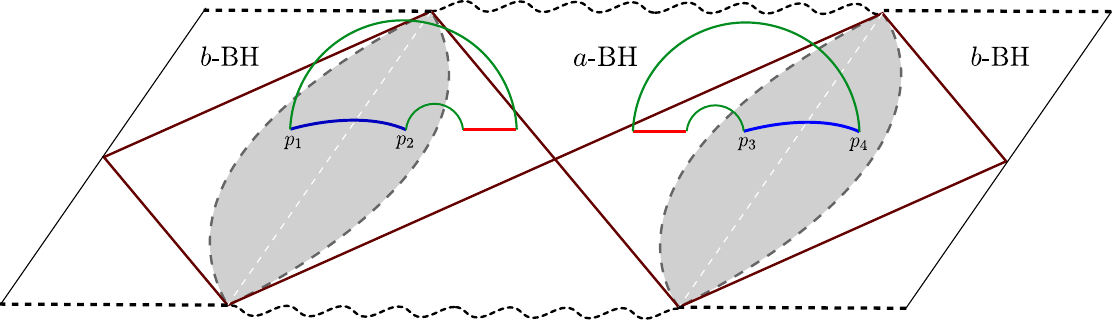}
		\caption{}
		\label{RTf}
	\end{subfigure}
	\hspace{.1cm}
	\begin{subfigure}[b]{0.45\textwidth}
		\centering
		\includegraphics[width=\textwidth]{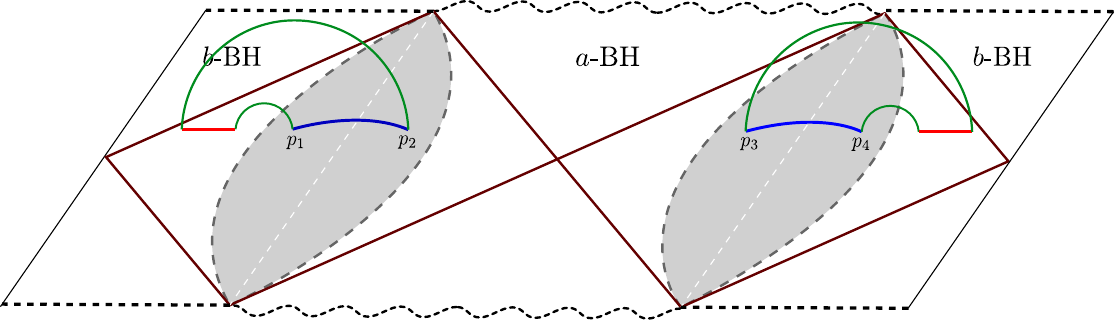}
		\caption{}
		\label{RTg}
	\end{subfigure}
	\caption{The schematic depicts the possibile contributions to the entanglement entropy arising from different RT surfaces (green curves) supported by the subsystem $A$ in the radiation reservoirs. The subsystem $A$ is shown as a union of two segments (blue lines) in the two radiation reservoirs (shaded grey regions) and island regions are indicated by the red segments. (Figure modified from \cite{Balasubramanian:2021xcm})}
	\label{2BHshag}
\end{figure}


\subsection{Holographic entanglement negativity} 
In this subsection, we first provide the definition of the entanglement negativity in the context of quantum information theory\cite{Vidal:2002zz}. For this case, we consider a tripartite system in a pure state consisting of the subsystems $A_1$, $A_2$ and $B$ where $A=A_1\cup A_2$ and $B=A^c$ represents rest of the system. Consequently, the reduced density matrix for the bipartite mixed state configuration described by the subsystem $A$ may be obtained by tracing over the subsystem $B$ as $\rho_A=\Tr_B \rho$. Thus the entanglement negativity of the corresponding bipartite mixed state is defined as
\begin{equation}\label{EN_def}
	\mathcal{E}=\ln \Tr |\rho_A ^{T_2}|~,
\end{equation} 
where the trace norm $\Tr |\rho_A ^{T_2}|$ is described as sum of the absolute eigenvalues of $\rho_A ^{T_2}$. 
In \cref{EN_def}, the partial transpose of the reduced density matrix $\rho_A$ is defined as 
\begin{equation}
	\bra{e_i ^{(1)} e_j ^{(2)}}\rho_A ^{T_2}\ket{e_k ^{(1)} e_l ^{(2)}}=\bra{e_i ^{(1)} e_l ^{(2)}}\rho_A \ket{e_k ^{(1)} e_j ^{(2)}}\,.
\end{equation}
Here $\ket{e_i ^{(1)}}$ and $\ket{e_j ^{(1)}}$ are expressed as the basis vectors of the Hilbert spaces $\mathcal{H}_1$ and $\mathcal{H}_2$ respectively. 

For the case of $CFT_2$s, the authors of \cite{Calabrese:2012nk,Calabrese:2014yza,Calabrese:2012ew} described a suitable replica technique to compute the entanglement negativity for bipartite states which involves the quantity $\Tr (\rho_A ^{T_2})^n$ for the replica index $n$ being restricted to the even sequences $ n_e$. Finally in the replica limit $n_e \rightarrow 1$, we may obtain the expression of the entanglement negativity as follow
\begin{equation}\label{EN_cft}
	\mathcal{E}=\lim_{n_e \rightarrow1}\ln \Tr (\rho_A ^{T_2})^{n_e},
\end{equation} 
where the quantity $\Tr (\rho_A ^{T_2})^{n_e}$ can also be expressed as the twist field correlator for the appropriate bipartite states.

We now briefly recapitulate the holographic entanglement negativity proposals described in \cite{Chaturvedi:2016rcn,Jain:2017aqk,Malvimat:2018txq} for bipartite mixed states in $CFT_{2}$s in the $AdS_3/CFT_2$ scenario. The holographic proposal for the entanglement negativity involved an algebraic sum of  bulk geodesic lengths homologous to the subsystems for the mixed state configurations in question. For example in the case of two adjacent subsystems $A$ and $B$ the corresponding holographic entanglement negativity (HEN) is given as \cite{Jain:2017aqk}
\begin{equation}\label{NegAdj0}
	\mathcal{E}(A:B)=\frac{3}{16\pi G_N^{(3)}}\left(\mathcal{L}_A+\mathcal{L}_B-\mathcal{L}_{A\cup B}\right),
\end{equation}
where $\mathcal{L}_X$ is the bulk static minimal surface homologous to the subsystem $X$. The above equation may be expressed in terms of the entanglement entropies of the subsystems utilizing the RT formula \cite{Ryu:2006bv,Ryu:2006ef} as follows,
\begin{equation}\label{NegAdj1}
	\mathcal{E}(A:B)=\frac{3}{4}\left[S(A)+S(B)-S(A \cup B)\right]\,,
\end{equation}
where $S(X)$ is the corresponding EE for the subsystem $X$. In this connection the holographic R\'enyi entropy for the same in the dual $CFT_d$ is defined by \cite{Dong:2016fnf}
\begin{equation}\label{Renyi}
	S^{(n)}(X)=\frac{\mathcal{A}_X^{(n)}}{4 G_{N}^{(d+1)}}\,.
\end{equation}
In the above equation, $\mathcal{A}_X^{(n)}$ is related to the area of a back-reacting cosmic brane homologous to the subsystem $X$ in the bulk dual $AdS_{d+1}$ \cite{Dong:2016fnf}. In two dimensions, R\'enyi entropy of order half reads as
\begin{equation}\label{Renyi-half}
	S^{(1/2)}(X)=\frac{\mathcal{L}_X^{(1/2)}}{4 G_{N}^{(3)}}\,,
\end{equation}
where $\mathcal{L}_X^{(1/2)}$ corresponds to the length of a back-reacting cosmic brane in $AdS_3$ geometry. For the case of spherically entangling surfaces, the effect of the back-reaction in $AdS_3/CFT_2$ scenario may be characterized as \cite{Rangamani:2014ywa,Kudler-Flam:2018qjo,Hung:2011nu,Kusuki:2019zsp}
\begin{equation}\label{Backrreaction}
	\mathcal{L}^{(1/2)}_X=\frac{3}{2}	\mathcal{L}_X. 
\end{equation}
Subsequently, we may reframe the HEN proposal in \cref{NegAdj1} as a specific algebraic sum of R\'enyi entropies of order half \cite{KumarBasak:2020ams} for the corresponding subsystems utilizing the \cref{Renyi-half,Backrreaction}
\begin{equation}\label{NegAdj2}
	\mathcal{E}(A:B)=\frac{1}{2}\left[S^{(1 / 2)}(A)+S^{(1 / 2)}(B)-S^{(1 / 2)}(A \cup B)\right]\,.
\end{equation}

Following the same approach the above HEN proposal may be extended to the disjoint subsystems as \cite{Malvimat:2018txq,KumarBasak:2020ams}
\begin{equation}\label{NegDis0}
	\begin{aligned}
		\mathcal{E}(A:B)=&\frac{3}{4}\left[S(A \cup C) + S(B\cup C) - S(C) - S(A \cup B \cup C)\right]\\
		=&\frac{1}{2}\left[S^{(1 / 2)}(A\cup C)+S^{(1 / 2)}(B\cup C)-S^{(1 / 2)}(C)-S^{(1 / 2)}(A\cup B\cup C)\right]\,.
	\end{aligned}
\end{equation}
where the subsystem $C$ is sandwiched between the subsystems $A$ and $B$. In the following sections, we will use these proposals to compute the holographic entanglement negativity for various bipartite mixed state configurations.


\section{Entanglement negativity in braneworld model}\label{model2}
In this section, we consider the braneworld model of two finite sized non-gravitating reservoirs where each of the reservoirs are coupled to two quantum dots at its boundaries \cite{Balasubramanian:2021xcm}. These quantum dots constitute two copies of thermofield double states which are interacting through the common reservoirs. The holographic dual of these quantum dots are Planck branes described by $AdS_2$ geometries. The maximal extension of the Penrose diagram describes two eternal JT black holes\footnote{In this article, we label the two eternal JT black holes together with the two Planck branes as $a$ and $b$.} located on the Planck branes which are coupled to each other through the two copies of the shared reservoirs as depicted in \cref{penrose2}. This configuration also involves identical matter $CFT_2$s with
transparent boundary conditions in both the black hole and the reservoir regions \cite{Almheiri:2019yqk,Almheiri:2019hni}. For the above configuration, we investigate the entanglement entropy and the entanglement negativity of various bipartite mixed states in the radiation reservoirs which characterize information transfer between the two eternal JT black holes on the Planck branes.

\subsection{Holographic entanglement negativity and Page curve}
In the following subsections, we compute the holographic entanglement negativity for various bipartite mixed states in the non-gravitating radiation reservoirs in the context of the braneworld model utilizing the equations described in the \cref{NegAdj1,NegDis0,HEE-Shaghoulian}. Furthermore we analyze the behaviour of the entanglement negativity profiles with respect to the subsystem sizes and the time. 

\begin{figure}[H]
	\centering
	\includegraphics[scale=.7]{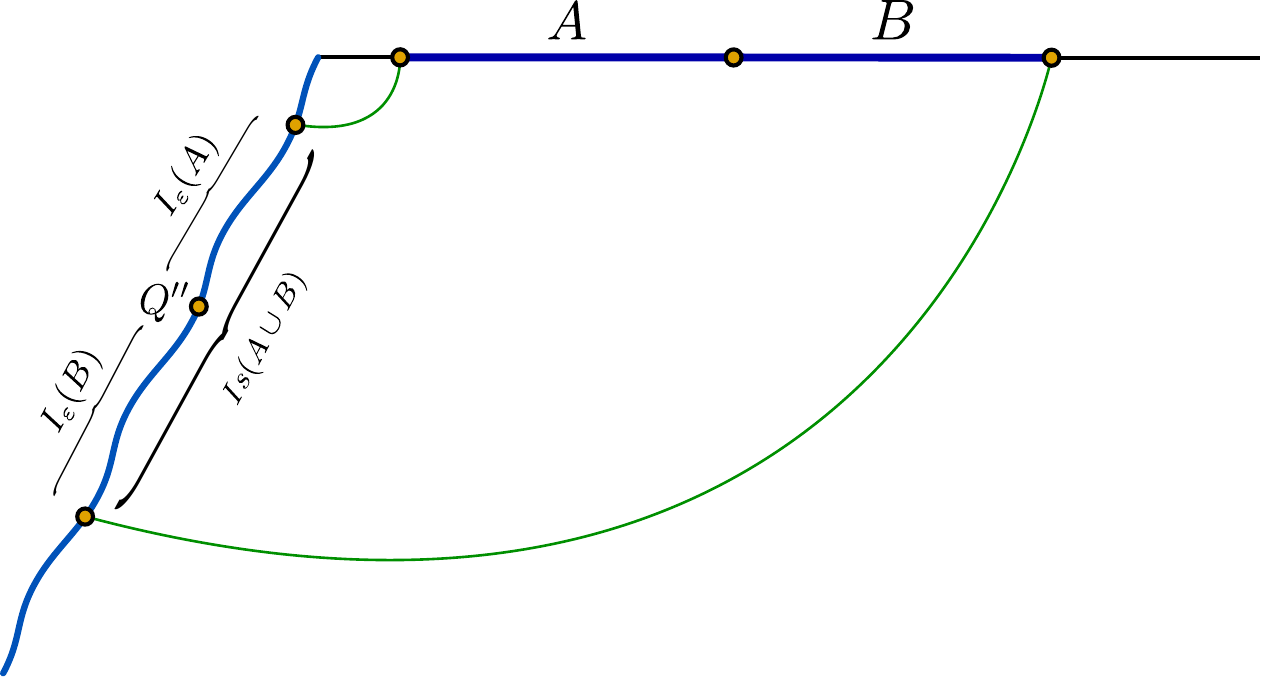}
	\caption{Schematic describes the appearance of the entanglement negativity island on the Planck brane for the case of two adjacent subsystems.}
	\label{QES}
\end{figure}

Note that in these subsections the behaviour of the various entanglement negativity profiles may be interpreted in terms of the entanglement negativity islands for the subsystems under consideration as described in \cite{KumarBasak:2020ams,KumarBasak:2021rrx}. The corresponding entanglement negativity including the island contribution for the mixed state configuration of two generic adjacent subsystems $A$ and $B$  is obtained as follows\footnote{In the present model, we do not utilize \cref{Ryu-FlamIslandgen} to compute the entanglement negativity between the subsystem $A$ and $B$.}
\begin{align}
	&{\cal E}^{gen}(A:B)=\frac{{\cal A}^{(1/2)}\left(Q^{\prime \prime}=\partial I_{\varepsilon}(A) \cap \partial I_{\varepsilon}(B)\right)}{4 G_{N}}+{\cal E}^{\mathrm{ eff}}\left(A\cup I_{\varepsilon}(A)  :B\cup I_{\varepsilon}(B)   \right)\notag\\
	&{\cal E}(A:B)	=\mathrm{min}(\mathrm{ext}_{Q^{\prime \prime}}\{ 	{\cal E}^{gen}(A:B)\}), \label{Ryu-FlamIslandgen}
\end{align}
\noindent
where, $Q^{\prime \prime}$ is the quantum extremal surface (QES) which is given by the intersection of the individual negativity islands $I_{\varepsilon}(A)$ and $I_{\varepsilon}(B)$ for the subsystems $A$ and $B$ as described in \cref{QES},
\begin{equation}
	Q^{\prime \prime}=\partial I_{\varepsilon}(A) \cap \partial I_{\varepsilon}(B).
\end{equation}
The second term ${\cal E}^{\mathrm{ eff}}$ in the above \cref{Ryu-FlamIslandgen} corresponds to the effective entanglement negativity between the quantum matter fields located in the regions $A\cup I_{\varepsilon}(A)$ and $B\cup I_{\varepsilon}(B)$. The entanglement negativity islands $I_{\varepsilon}(A)$ and $I_{\varepsilon}(B)$ obey the condition $I_{\varepsilon}(A)\cup I_{\varepsilon}(B)= Is(A\cup B)$, where $Is(A\cup B)$ is the entanglement entropy island for the subsystem $A\cup B$. In general, the islands for the entanglement negativity $I_{\varepsilon}(A)$ and $I_{\varepsilon}(B)$ do not correspond to the entanglement entropy islands $Is(A)$ and $Is(B)$ for the subsystems $A$ and $B$ respectively. 

\subsubsection{Adjacent subsystems}\label{Adj}
We start with two adjacent subsystems $A$ and $B$ of finite lengths $l_1$ and $l_2$ respectively in the radiation reservoirs and compute the holographic entanglement negativity between them using the \cref{NegAdj1,HEE-Shaghoulian}. In particular, we investigate the qualitative nature of the entanglement negativity profiles for three distinct scenarios involving the subsystem sizes and the time. In this context we utilize the structures of the various RT surfaces supported by the subsystems in question described earlier in the diagrams \cref{2BHshag}.

\subsubsection*{$\bm{(i)}$ Full system ($A\cup B$) fixed, common point varied}\label{adjcase1}
\begin{figure}[h!]
	\centering
	\begin{subfigure}[t]{.45\textwidth}
		\includegraphics[width=1\linewidth]{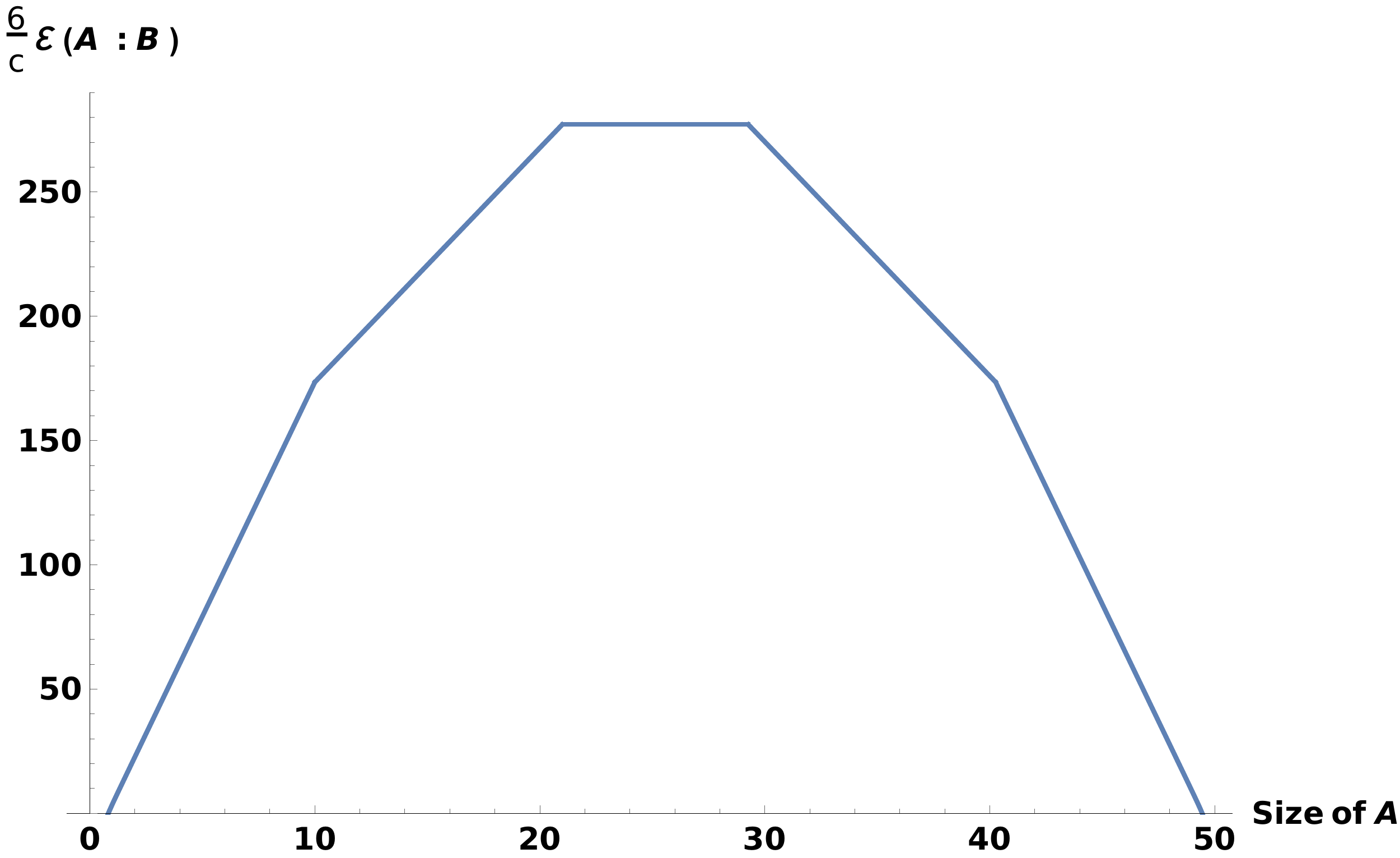}
		\caption{Page curve for the entanglement negativity where we have considered $A\cup B=[.01L,.99L]$ at $t=15$.}
		\label{fig:Adjcase1}
	\end{subfigure}
	\hspace{.1cm}
	\begin{subfigure}[t]{.45\textwidth}
		\includegraphics[width=1\linewidth]{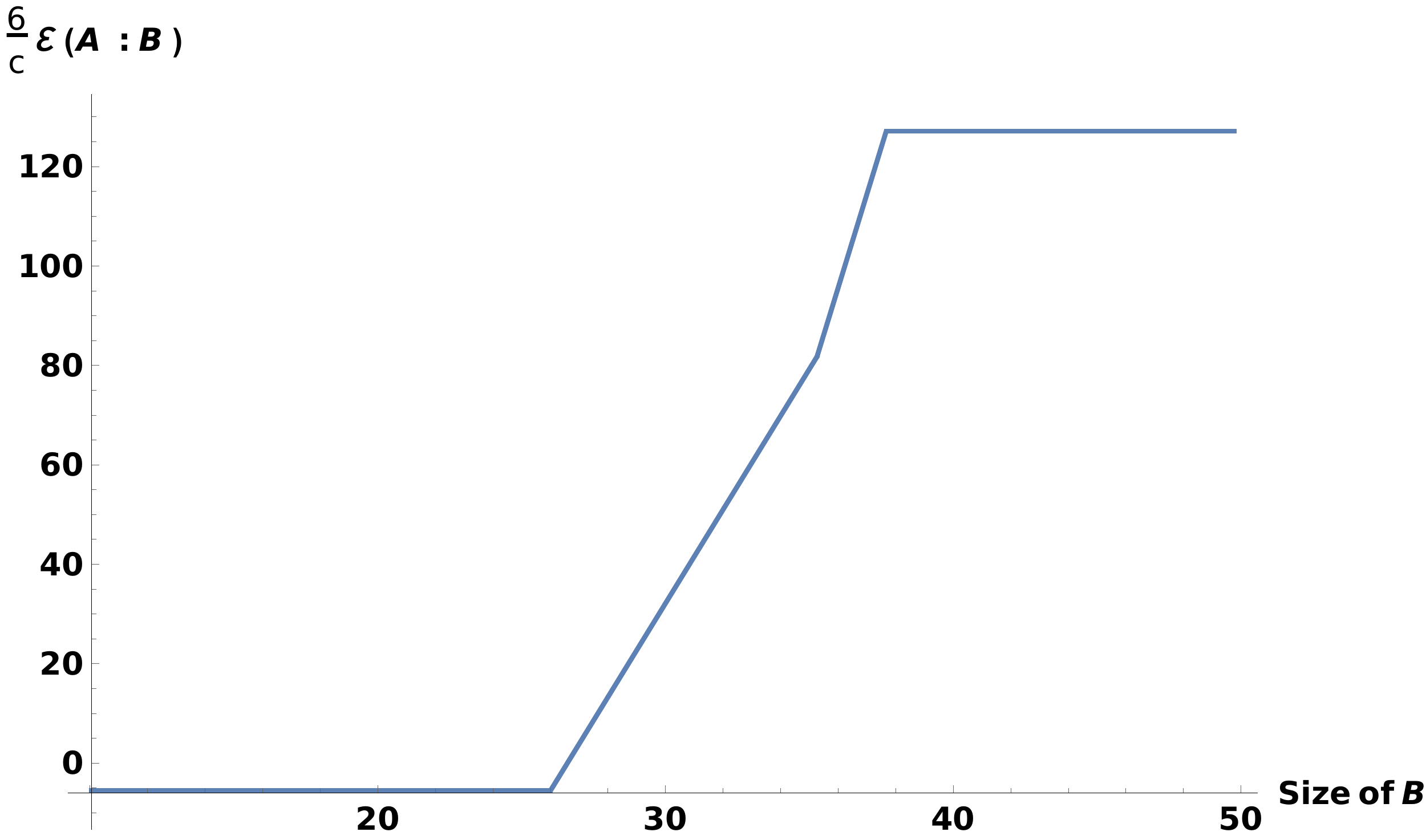}
		\caption{Profile of the entanglement negativity with respect to the size of $B$ where we have considered $A=[.01L,.15L]$ at $t=20$.}
		\label{fig:Adjcase2}
	\end{subfigure}
	\hspace{.1cm}
	\begin{subfigure}[t]{.6\textwidth}
		\includegraphics[width=1\linewidth]{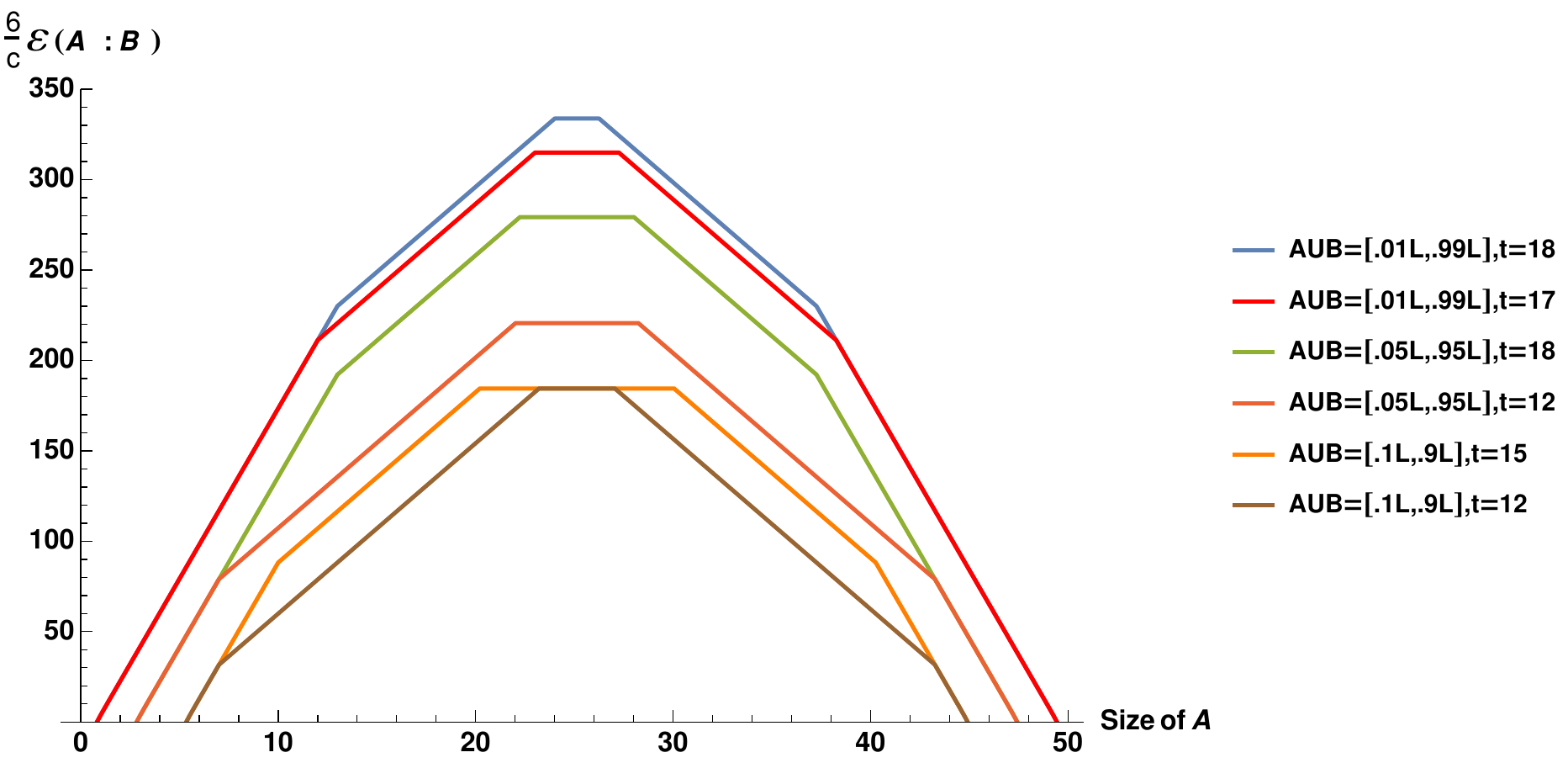}
		\caption{Page curves for entanglement negativity for different sizes of $(A\cup B)^c$ and different times.}
		\label{fig:Adjcase1plateau}
	\end{subfigure}
	\caption{Here $\beta=1$, $c=500$, $\phi_0= \frac{30c}{6}$, $\phi_r= \frac{30}{\pi}$, $L=\frac{16\pi}{\beta}$, $\epsilon=.001$.}
	\label{EN_Shag_adj}
\end{figure}

We first consider the case where the common point between the adjacent subsystems $A$ and $B$ is varied at a constant time slice while keeping the subsystem $A\cup B$ fixed which covers the entire reservoirs. In this scenario, we compute the holographic entanglement negativity between the subsystems $A$ and $B$ utilizing the \cref{NegAdj1,HEE-Shaghoulian}. We observe that our results in this context reproduces the analogue of the Page curve for the entanglement negativity as depicted in \cref{fig:Adjcase1}.

\begin{figure}[h!]
	\centering
	\begin{subfigure}[b]{0.45\textwidth}
		\centering
		\includegraphics[width=\textwidth]{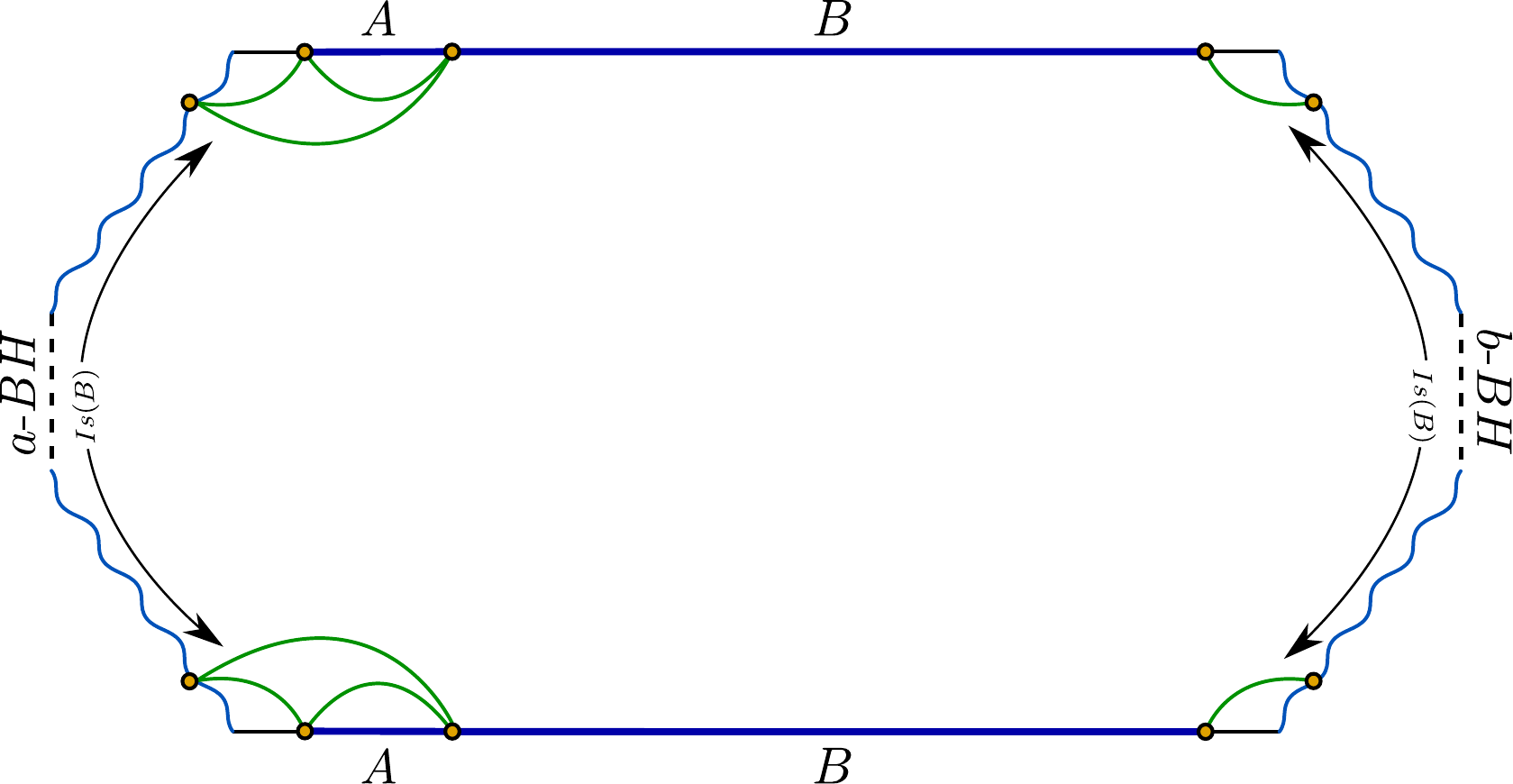}
		\caption{Phase-1}
		\label{}
	\end{subfigure}
	\vspace{.4cm}
	\hspace{.12cm}
	\begin{subfigure}[b]{0.45\textwidth}
		\centering
		\includegraphics[width=\textwidth]{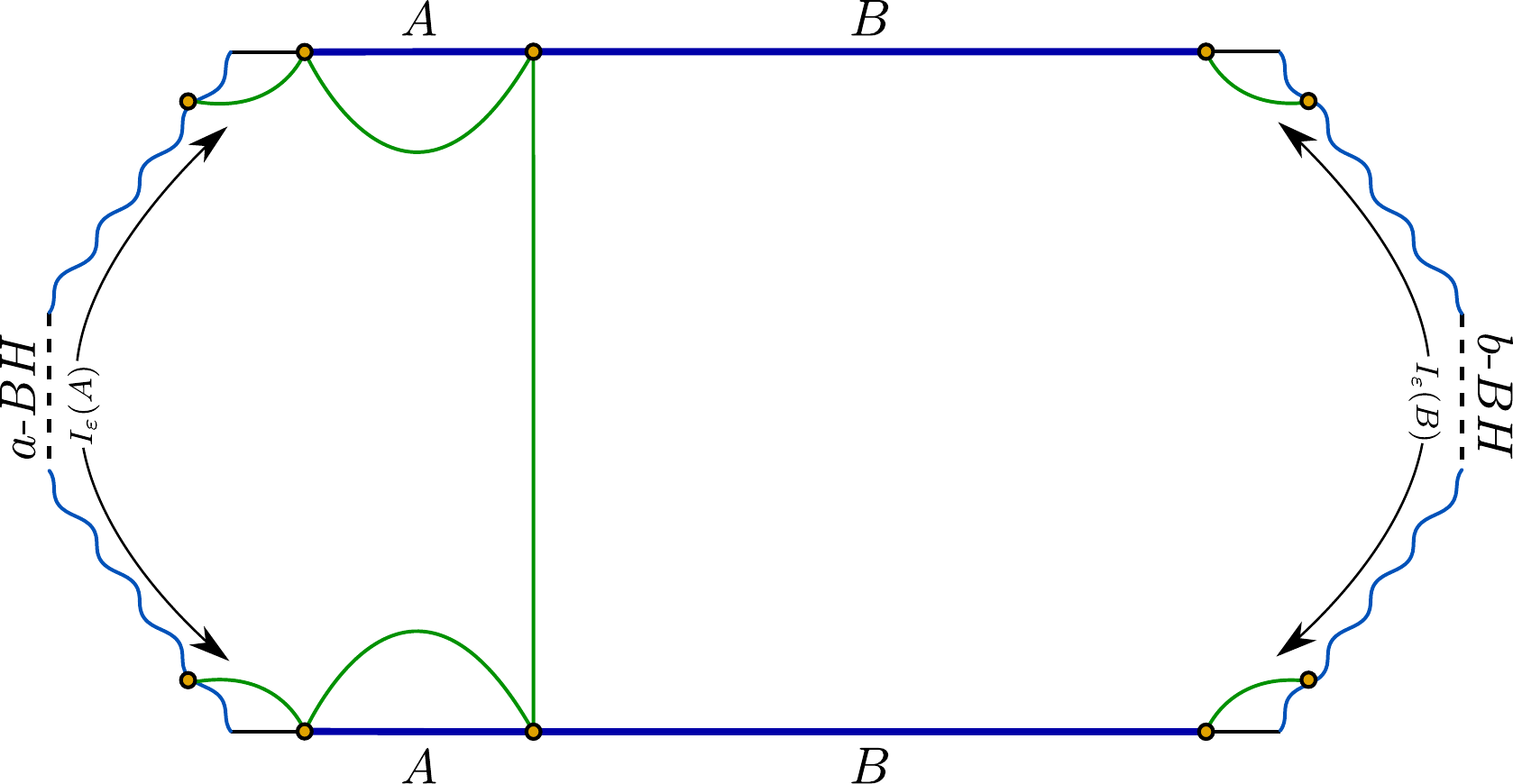}
		\caption{Phase-2}
		\label{}
	\end{subfigure}
	\vspace{.4cm}
	\begin{subfigure}[b]{0.45\textwidth}
		\centering
		\includegraphics[width=\textwidth]{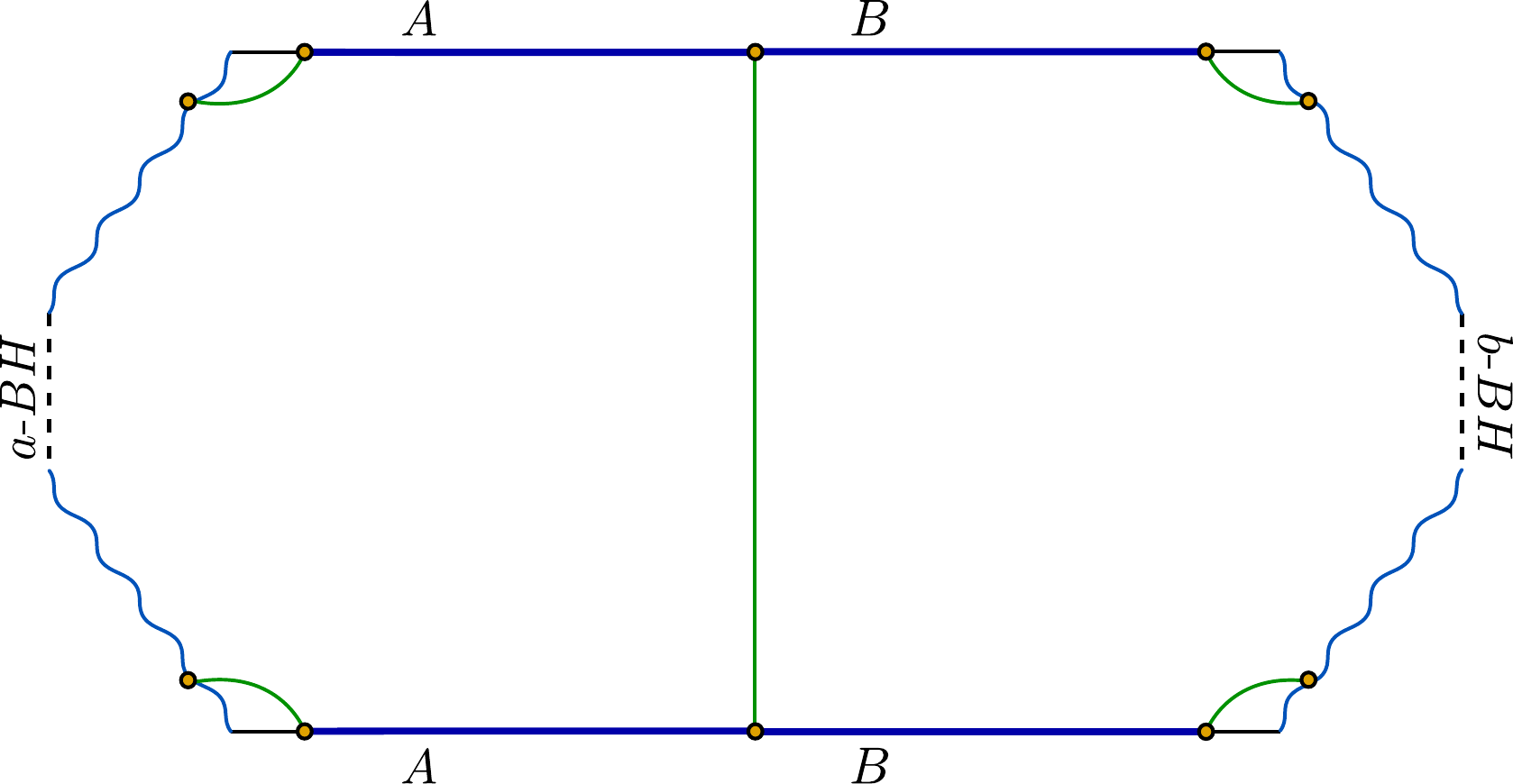}
		\caption{Phase-3}
		\label{}
	\end{subfigure}
	\hspace{.12cm}
	\begin{subfigure}[b]{0.45\textwidth}
		\centering
		\includegraphics[width=\textwidth]{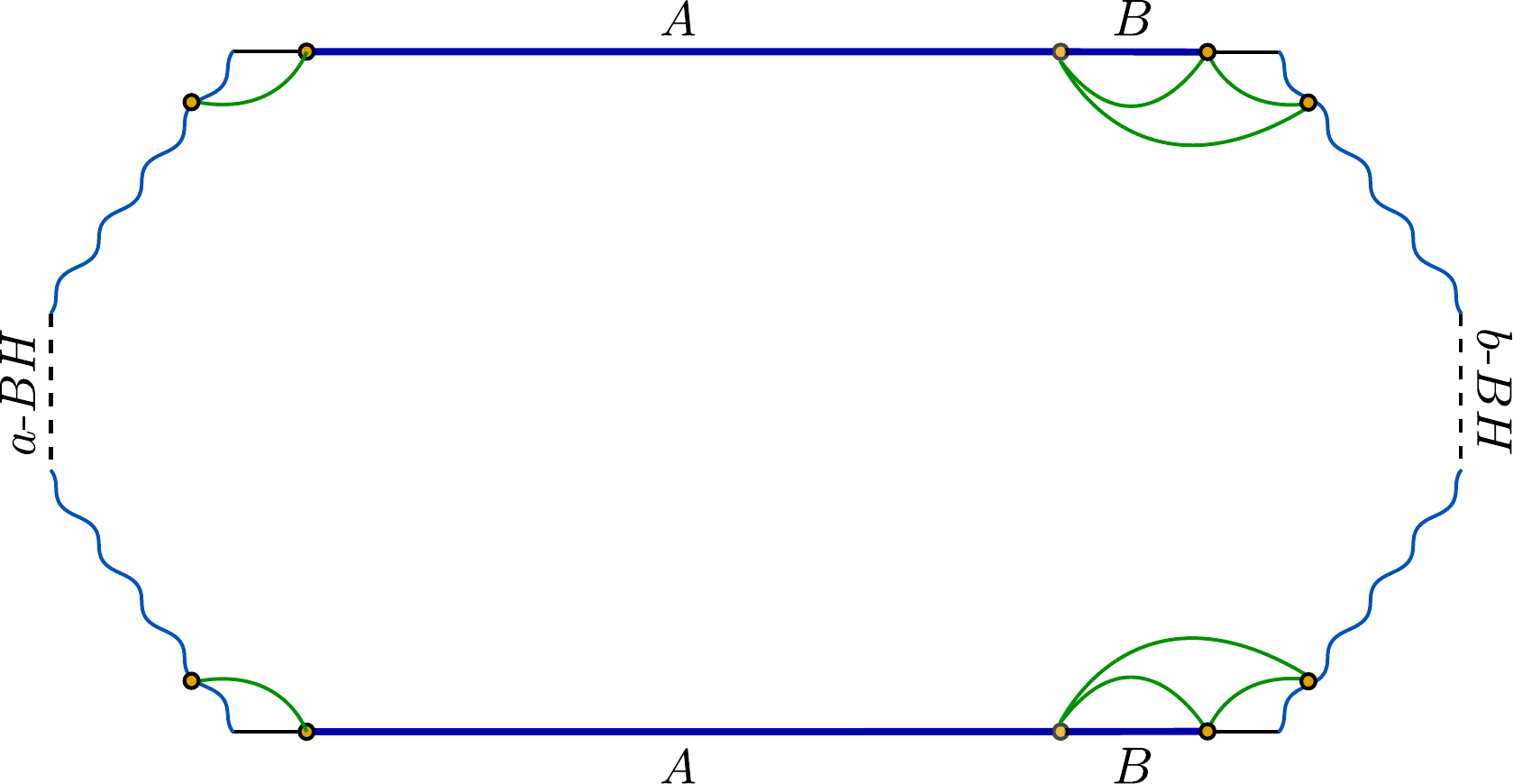}
		\caption{Phase-4}
		\label{}
	\end{subfigure}
	\begin{subfigure}[b]{0.45\textwidth}
		\centering
		\includegraphics[width=\textwidth]{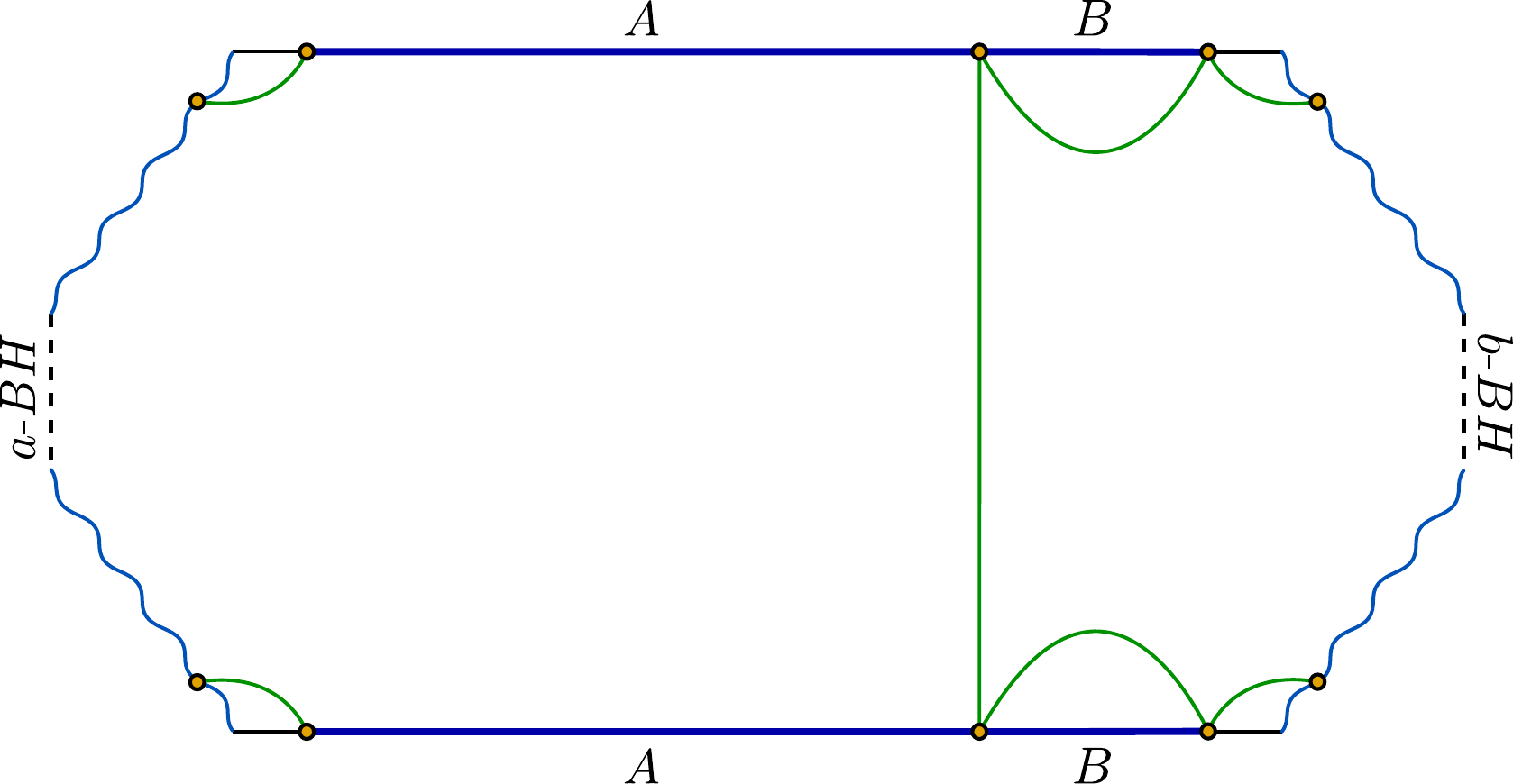}
		\caption{Phase-5}
		\label{}
	\end{subfigure}
	\caption{Schematic depicts all the phases of the entanglement negativity for two adjacent subsystems where the common point between them is varied. The RT surfaces supported by the adjacent subsystems are denoted by the green geodesics. Note that these geodesics corresponding to the subsystems indicate the locations of the island regions on the Planck branes.}
	\label{Adjcase1}
\end{figure}

In this scenario, the entanglement negativity profile consists of five phases due to the various structures of the RT surfaces supported by the subsystems under consideration. The expressions for the corresponding entanglement negativity in these phases are listed in the appendix \ref{Appendix1}. In what follows, we discuss these distinct phases of the entanglement negativity profile in details.

\noindent\\
\textbf{Phase 1:} In the first phase, the dominating contribution to the entanglement entropy for the subsystem $A$ arises from dome-type RT surfaces, whereas the subsystems $B$ and $A\cup B$ support $ab$-type RT surfaces each as depicted in \cref{Adjcase1}. Interestingly, here the size of the subsystem $A$ is very small compared to the size of $B$ such that this phase does not admit any island region corresponding to the subsystem $A$. However, this phase involves entanglement entropy islands corresponding to the subsystem $B$ and $A\cup B$ which are located on both the Planck branes. In this case the entanglement negativity between the subsystems $A$ and $B$ is governed by the degrees of freedom of the subsystem $A$. Note that as we shift the common point, the number of Hawking modes captured by the subsystem $A$ and its $CFT_2$-degrees of freedom increase accordingly. Hence we observe a linearly rising behaviour in the corresponding entanglement negativity profile with the increasing size of the subsystem $A$ as exhibited in \cref{fig:Adjcase1}.\\

\noindent\textbf{Phase 2:}  Next we proceed to the second phase where the subsystems $A$ and $A\cup B$ still support dome and $ab$-type RT surfaces respectively, whereas the subsystem $B$ now admits $b$-bulk type RT surfaces as shown in \cref{Adjcase1}. Consequently, this phase includes an entanglement negativity island $I_{\varepsilon}(A)$ corresponding to the subsystem $A$ which is located on the $a$-brane. This negativity island contains the entire interior region of the $a$-black hole. Note that in this phase, the entanglement negativity between the subsystems $A$ and $B$ is governed by the degrees of freedom present in the region $A\cup I_{\varepsilon}(A)$. This is determined by the number of interior Hawking modes captured by the negativity island $I_{\varepsilon}(A)$ whose corresponding pairs are located in $(A\cup I_{\varepsilon}(A))^c$ and by the $CFT_2$-degrees of freedom present in $A\cup I_{\varepsilon}(A)$. In this case, as we increase the size of the subsystem $A$, the number of interior Hawking modes of $A\cup I_{\varepsilon}(A)$ decreases due to a purification by their corresponding pairs from the exterior region which are now transferred to the region $A\cup I_{\varepsilon}(A)$. However, the number of Hawking modes coming from the $b$-black hole simultaneously increases in the region $A\cup I_{\varepsilon}(A)$ with its increasing size which tends to cancel the preceding decreasing effect.
Consequently, the degrees of freedom of the region $A\cup I_{\varepsilon}(A)$ are now determined only by the $CFT_2$-degrees of freedom which increases linearly with the shift of the common point between the adjacent subsystems. As a result, we observe a linearly rising behaviour in the corresponding entanglement negativity profile with a growth rate smaller than the previous phase.
\newline\\
\noindent
\textbf{Phase 3:} In the third phase (\cref{Adjcase1}), the RT surfaces supported by the subsystems $A$, $B$ and $A\cup B$ are $a$-$bulk$, $b$-$bulk$ and $ab$-type respectively. Here the size of the subsystem $A$ is comparable to the size of $B$ such that they are maximally entangled. This is a characteristic of  tripartite entanglement where the tripartition is defined by the subsystems $A$, $B$ and $(A\cup B)^c$. Hence in this phase, the degrees of freedom present in the subsystems $A$ and $B$ are entangled with the degrees of freedom of the subsystem $(A\cup B)^c$ as the common point is shifted. Therefore this corresponds to a constant behaviour of the entanglement negativity profile as depicted in \cref{fig:Adjcase1}.\\

\noindent\textbf{Phase 4 and Phase 5:} As depicted in \cref{Adjcase1}, the RT surfaces for the subsystems $A$ and $B$ in these two phases are interchanged with each other as compared to the first two phases. Hence, the corresponding entanglement negativity profile may be interpreted in a similar fashion as for the first two phases with the roles of the subsystems $A$ and $B$ exchanged.

We may further extend our analysis of the tripartite entanglement by considering different times and various sizes of the subsystem $(A\cup B)^c$. At a fixed time, we observe that the height of the plateau region of the corresponding entanglement negativity profile decreases with the increase in the size of the subsystem $(A\cup B)^c$ as exhibited in \cref{fig:Adjcase1plateau}. This behavior is consistent since the available degrees of freedom in the subsystems $A$ and $B$ entangled between themselves, decreases with the increasing size of $(A\cup B)^c$. On the other hand for a fixed size
of the subsystem $(A\cup B)^c$ the height of the plateau region rises with increasing time. Once again this is consistent since the number of Hawking modes in all the subsystems rise with increasing time which corresponds to a larger entanglement between the subsystems.

In addition, with smaller size of the subsystem $(A\cup B)^c$, the height of the plateau region changes rapidly with a small change in time. However for larger size of $(A\cup B)^c$, we need to increase the time sufficiently such that it changes the height of the plateau region. This character of the entanglement negativity profile is again consistent since the subsystem $(A\cup B)^c$ can accommodate a fewer number of Hawking modes when its size is very small whereas a large number of Hawking modes can be accumulated in the subsystems $A$ and $B$ due to their larger sizes. Consequently as time increases, most of the newly created Hawking modes are collected by the subsystems $A$ and $B$ and thus increase the height of the Plateau region rapidly. However, for a larger size of the subsystem $(A\cup B)^c$, the accommodation for the Hawking modes is larger such that it may now capture more newly created Hawking modes with increasing time and as a result the height of the plateau region changes very slowly as time increases (\cref{fig:Adjcase1plateau}).

\subsubsection*{$\bm{(ii)}$ Subsystem $\bm{A}$ fixed, $\bm{B}$ varied}\label{adjcase2}
In this scenario, we consider the length $l_1$ of the subsystem $A$ to be fixed at a constant time slice and investigate the behaviour of the holographic entanglement negativity while increasing the length $l_2$ of the subsystem $B$. The corresponding entanglement negativity profile is depicted in \cref{fig:Adjcase2} which consists of four distinct phases due to the various structures of the RT surfaces supported by the subsystems in question. The expressions for the holographic entanglement negativity between $A$ and $B$ in these phases, obtained through the \cref{NegAdj1,HEE-Shaghoulian}, are listed in the appendix \ref{Appendix2}. In what follows we now analyze these phases in detail.\\

\begin{figure}[h!]
	\centering
	\begin{subfigure}[b]{0.45\textwidth}
		\centering
		\includegraphics[width=\textwidth]{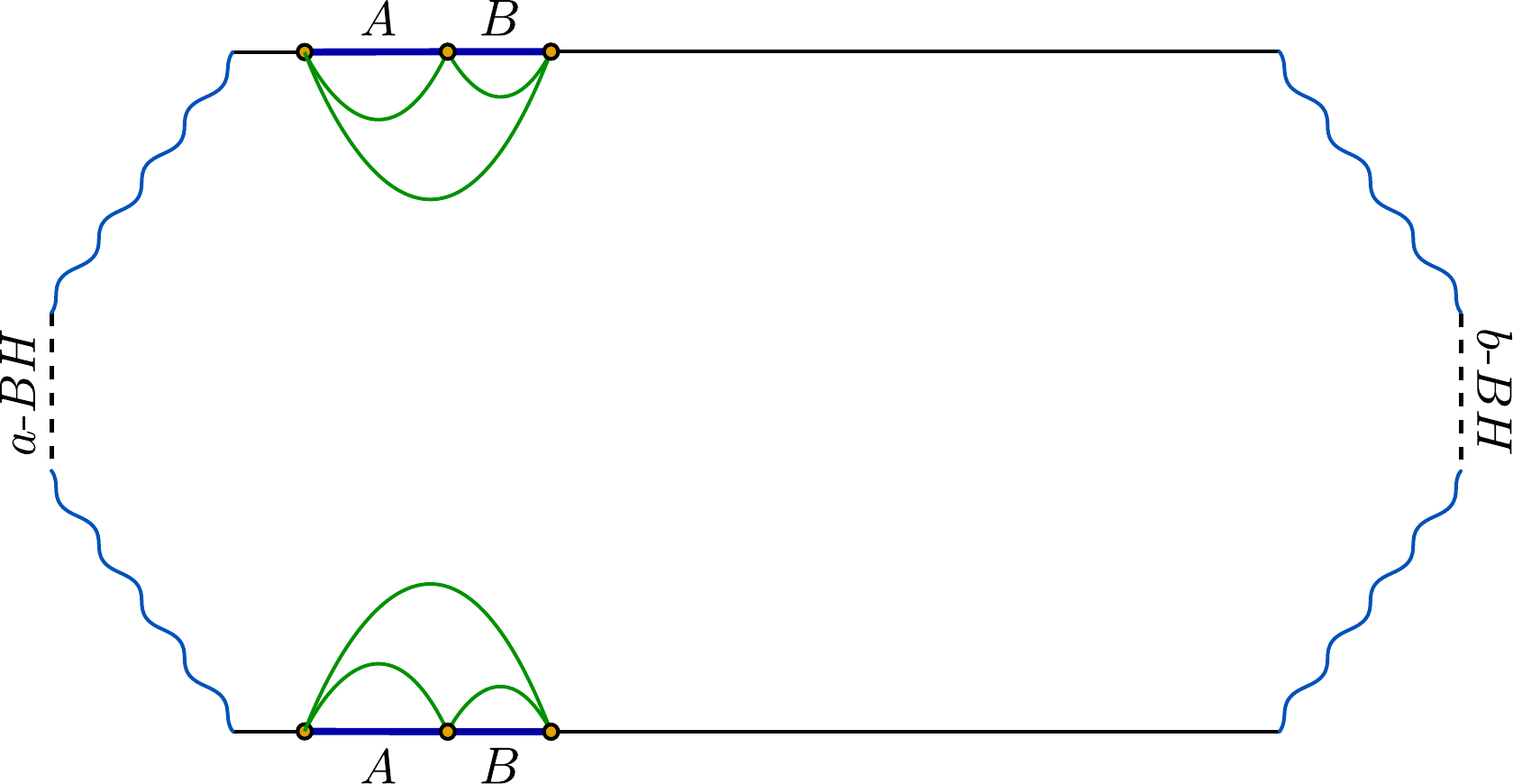}
		\caption{Phase-1}
		\label{}
	\end{subfigure}
	\vspace{.4cm}
	\hspace{.12cm}
	\begin{subfigure}[b]{0.45\textwidth}
		\centering
		\includegraphics[width=\textwidth]{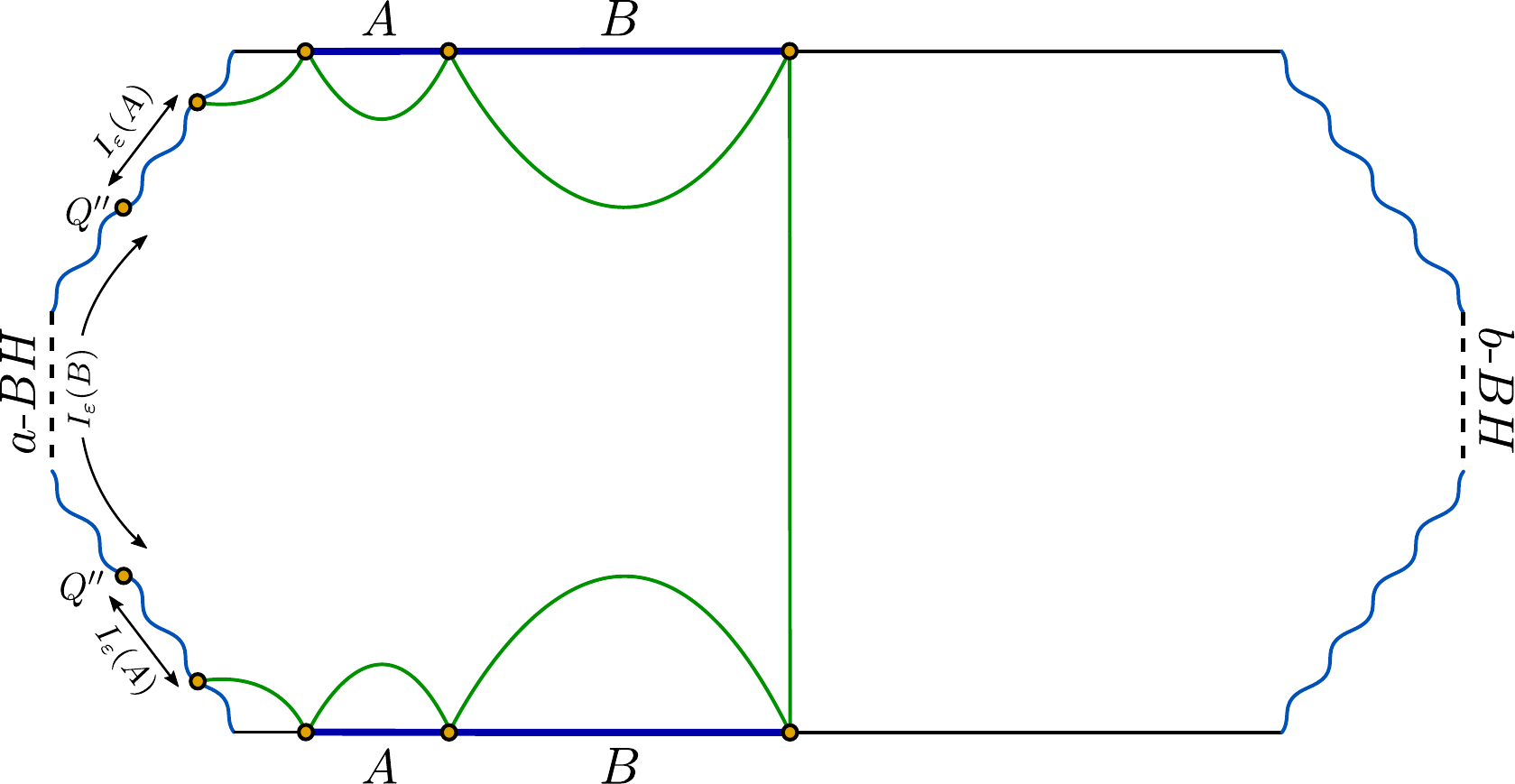}
		\caption{Phase-2}
		\label{}
	\end{subfigure}
	\begin{subfigure}[b]{0.45\textwidth}
		\centering
		\includegraphics[width=\textwidth]{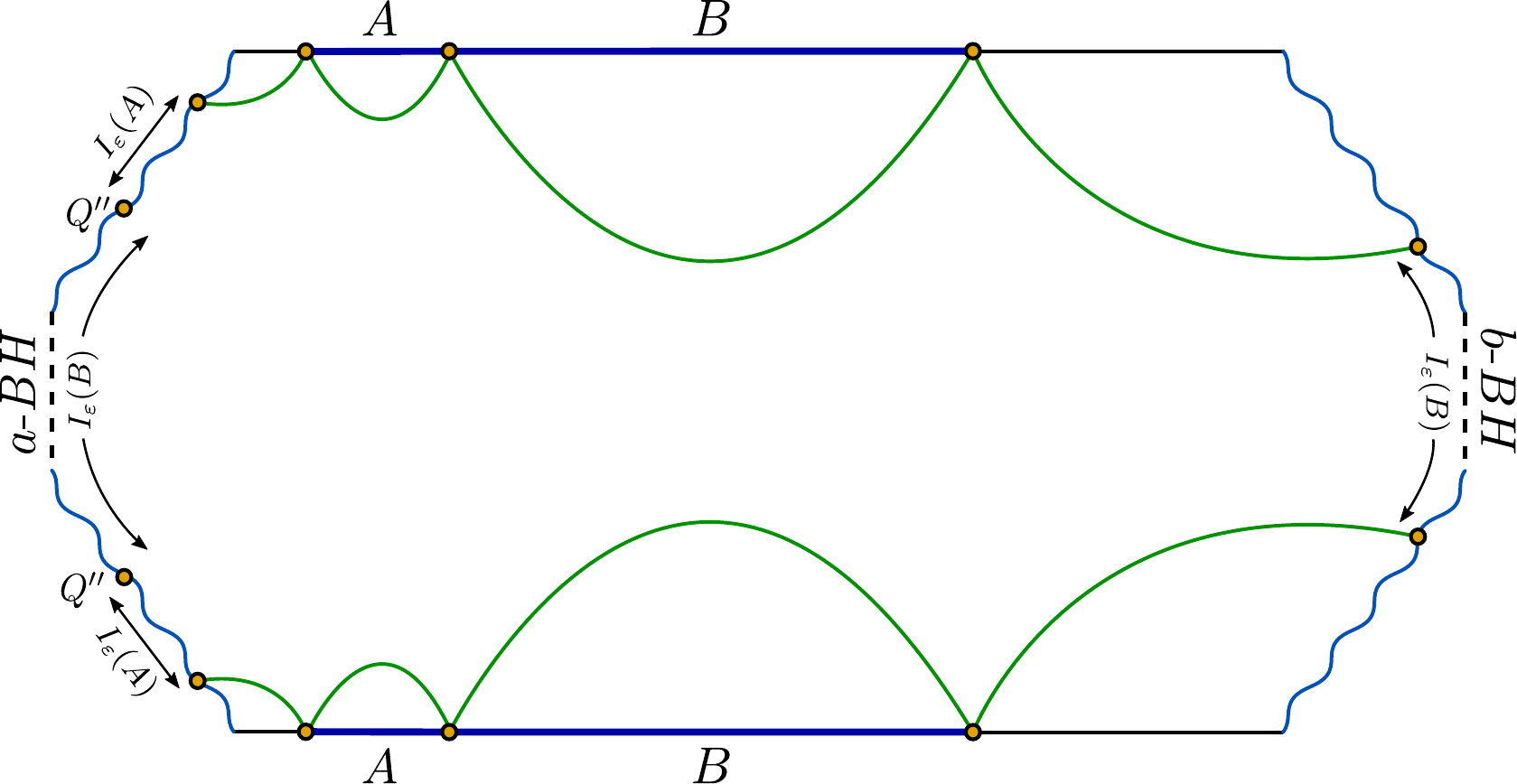}
		\caption{Phase-3}
		\label{}
	\end{subfigure}
	\hspace{.12cm}
	\begin{subfigure}[b]{0.45\textwidth}
		\centering
		\includegraphics[width=\textwidth]{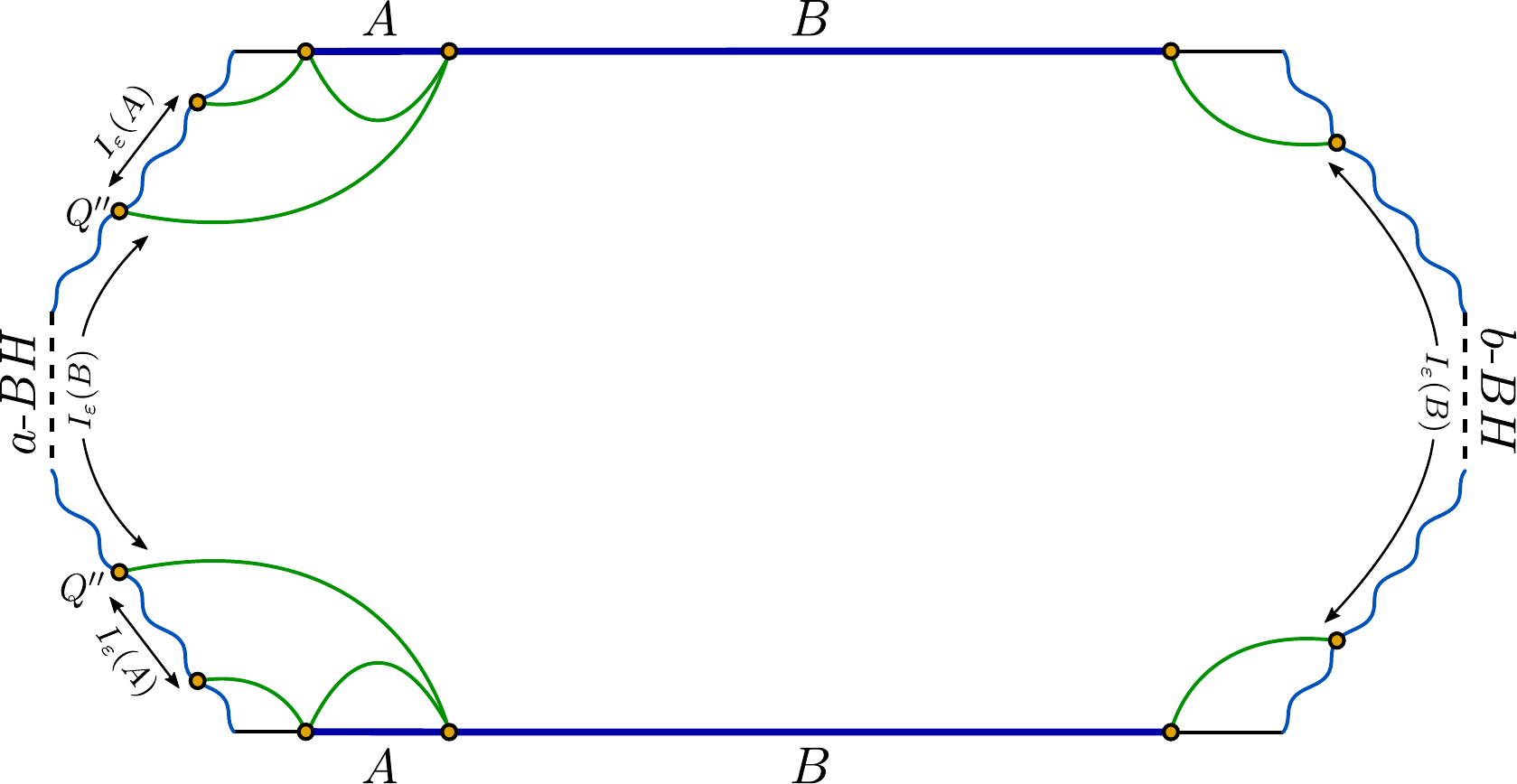}
		\caption{Phase-4}
		\label{}
	\end{subfigure}
	\caption{The diagram shows the possible phases of the entanglement negativity profile for the case of two adjacent subsystems where the size of the subsystem $B$ is varied.}
	\label{Adjcase2}
\end{figure}

\noindent
\textbf{Phase 1:}  We begin with the first phase where the RT surfaces for the subsystems $A$, $B$ and $A\cup B$ are $dome$-type each (\cref{Adjcase2}). Here the size of the subsystems $A$ and $B$ are very small and hence $A\cup B$ is far smaller than its complement $(A \cup B)^c$ which implies a vanishingly small entanglement negativity between the adjacent subsystems in question.\\

\noindent\textbf{Phase 2:}  In the second phase, the subsystems $A$ and $B$ still support dome-type RT surfaces whereas the subsystem $A\cup B$ admits $a$-bulk type RT surfaces. Consequently, this phase includes entanglement negativity islands $I_{\varepsilon}(A)$ and $I_{\varepsilon}(B)$ on the $a$-brane corresponding to the subsystems $A$ and $B$ respectively. The exterior regions of the $a$-black hole constitutes the island $I_{\varepsilon}(A)$ on the $a$-brane, whereas $I_{\varepsilon}(B)$ involves the entire interior region of the $a$-black hole. In this context, the ratio of $l_1$ to $l_2$ decreases with increasing length $l_2$ of the subsystem $B$ and consequently the ratio of the size of $I_{\varepsilon}(A)$ to the size of $I_{\varepsilon}(B)$ decreases accordingly. Note that in this phase the entanglement negativity between the subsystems $A$ and $B$ is governed by the degrees of freedom of $B \cup I_{\varepsilon}(B)$. Now as we increase $l_2$, the number of interior Hawking modes captured by the negativity island $I_{\varepsilon}(B)$ decreases due to a purification by their exterior partners which are located in the region $(B\cup I_{\varepsilon}(B))^c$. This purification arises since the exterior partners are transferred to the region $B\cup I_{\varepsilon}(B)$ as the
length $l_2$ of the subsystem $B$ increases. However, an equal number of Hawking modes from the $b$-black hole are transferred to the region $B\cup I_{\varepsilon}(B)$ simultaneously. Consequently, the degrees of freedom in the region $B\cup I_{\varepsilon}(B)$ are determined only by the $CFT_2$-degrees of freedom which increase with the increasing size of the subsystem $B$. Accordingly, the corresponding entanglement negativity profile rises linearly as depicted in \cref{fig:Adjcase2}.\\

\noindent\textbf{Phase 3:} In this phase, the dominant contributions to the entanglement entropies of the subsystems $A$ and $B$ still arise from $dome$-type RT surfaces, whereas the subsystem $A\cup B$ now supports $ab$-type RT surfaces. Hence this phase admits an entanglement negativity island $I_{\varepsilon}(B)$ located on both the branes corresponding to the subsystem $B$ as shown in \cref{Adjcase2}. This negativity island $I_{\varepsilon}(B)$ contains the entire interior regions of both the black holes. Furthermore this phase also includes an entanglement negativity island $I_{\varepsilon}(A)$ corresponding to the subsystem $A$ (\cref{Adjcase2}), which involves only the exterior regions of the $a$-black hole. In this scenario, the entanglement negativity between the subsystems $A$ and $B$ is governed by the degrees of freedom of the region $A\cup I_{\varepsilon}(A)$. Interestingly, with increasing length $l_2$ of the subsystem $B$, the island region $I_{\varepsilon}(A)$ increases in size. Consequently, the number of Hawking modes and the $CFT_2$-degrees of freedom present in $A\cup I_{\varepsilon}(A)$ increase with the increasing length $l_2$ of the subsystem $B$. Hence we observe the corresponding entanglement negativity profile to rise linearly with a growth rate higher than the previous phase.

\noindent\\
\textbf{Phase 4:} Finally in the last phase, the RT surfaces for the subsystems $B$ and $A\cup B$ are identified as $ab$-type each, however the subsystem $A$ still supports $dome$-type RT surfaces as depicted in \cref{Adjcase2}. Hence, this phase includes entanglement entropy islands for the subsystems $B$ and $A\cup B$, located on both of the Planck branes. Consequently, we have an entanglement negativity island $I_{\varepsilon}(A)$ corresponding to the subsystem $A$, which involves the exterior regions of the $a$-black hole. Once again the entanglement negativity in this phase is governed by the degrees of freedom of the region $A\cup I_{\varepsilon}(A)$. As we increase the length $l_2$, the size of the negativity island $I_{\varepsilon}(A)$ remains fixed which indicates that the number of the degrees of freedom in the region $A\cup I_{\varepsilon}(A)$is constant. Consequently, the corresponding entanglement negativity between the subsystems $A$ and $B$ exhibits a constant behaviour as shown in \cref{fig:Adjcase2}.

\subsubsection*{$\bm{(iii)}$ Subsystems $\bm{A}$ and $\bm{B}$ fixed, time varied}\label{adjcase3}

\begin{figure}[h!]
	\centering
	\includegraphics[width=10cm]{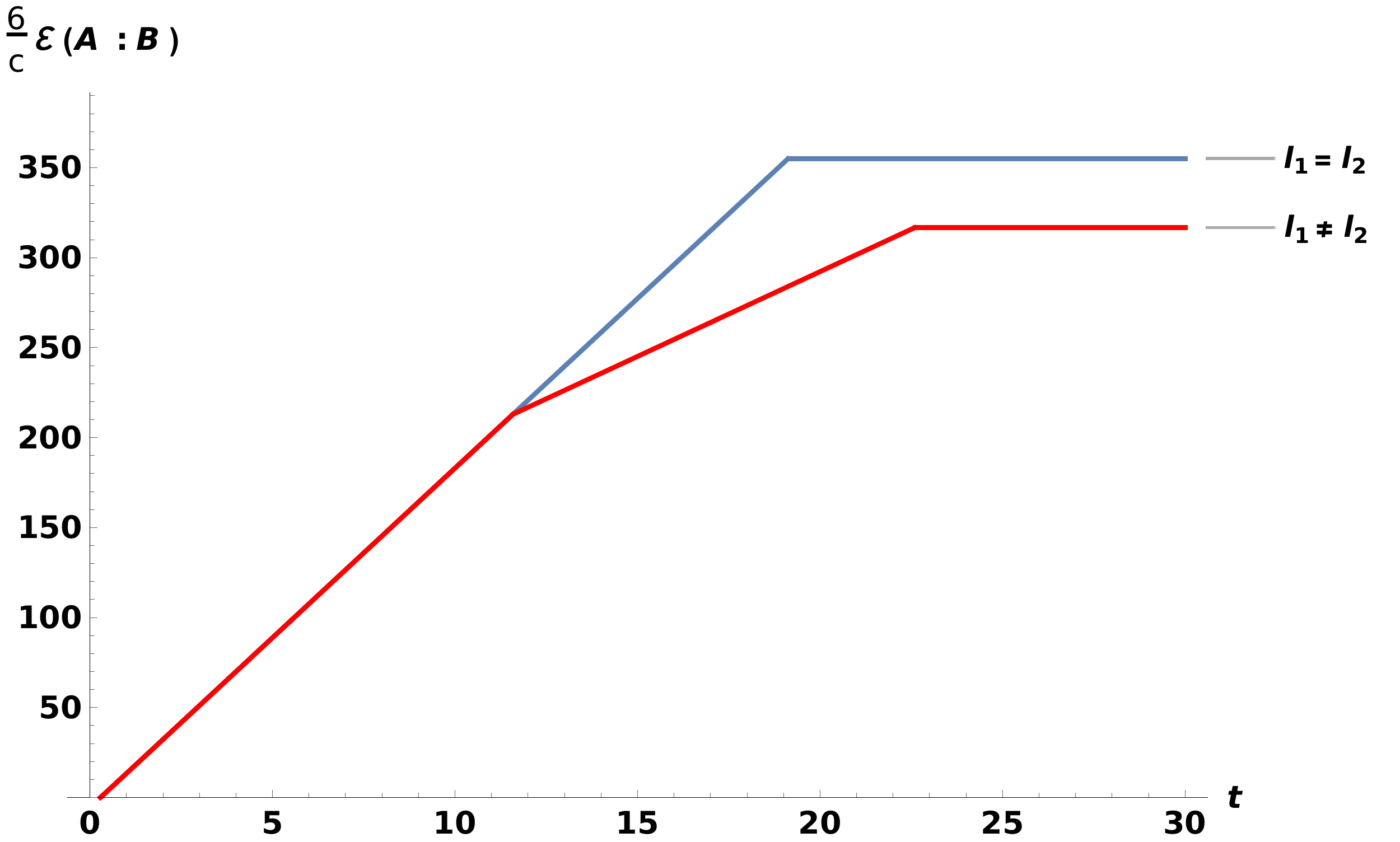}
	\caption{Page curves for entanglement negativity with respect to time $t$. Here $\beta=1$, $c=500$, $\phi_0= \frac{30c}{6}$, $\phi_r= \frac{30}{\pi}$, $L=\frac{16\pi}{\beta}$, $\epsilon=.001$, $A=[.01L,.5L]$ and $B=[.5L,.99L]$ (for $l_1=l_2$), $A=[.01L,.35L]$ and $B=[.35L,.99L]$ (for $l_1\neq l_2$).}
	\label{fig:Adjcase3}
\end{figure}

We conclude our analysis with the case where the lengths $l_1$ and $l_2$ of the two adjacent subsystems $A$ and $B$ respectively are fixed and the time $t$ is varied. In this context, we consider two sub cases with equal and unequal lengths of the subsystems $A$ and $B$ and obtain the corresponding holographic entanglement negativity between them utilizing the \cref{NegAdj1,HEE-Shaghoulian}. Since the subsystem sizes are fixed in this scenario, the $CFT_2$-degrees of freedom for the subsystems are also fixed and hence irrelevant for the description of the various phases of the entanglement negativity profiles. The only effect on the profiles arise from the Hawking modes arriving from both the black holes to the subsystems and this is utilized to analyze the corresponding entanglement negativity profiles depicted in \cref{fig:Adjcase3}.

\subsubsection*{(a) For $\bm{l_1=l_2}$}\label{p=1}
For the case of two equal lengths subsystems $A$ and $B$, we observe that the Page curve for the entanglement negativity consists of two phases as depicted in \cref{fig:Adjcase3}. The expressions for the holographic entanglement negativity in the corresponding phases are listed in the appendix \ref{Appendix3a}. In what follows, we comprehensively analyze these phases.
\begin{figure}[h!]
	\centering
	\begin{subfigure}[b]{0.45\textwidth}
		\centering
		\includegraphics[width=\textwidth]{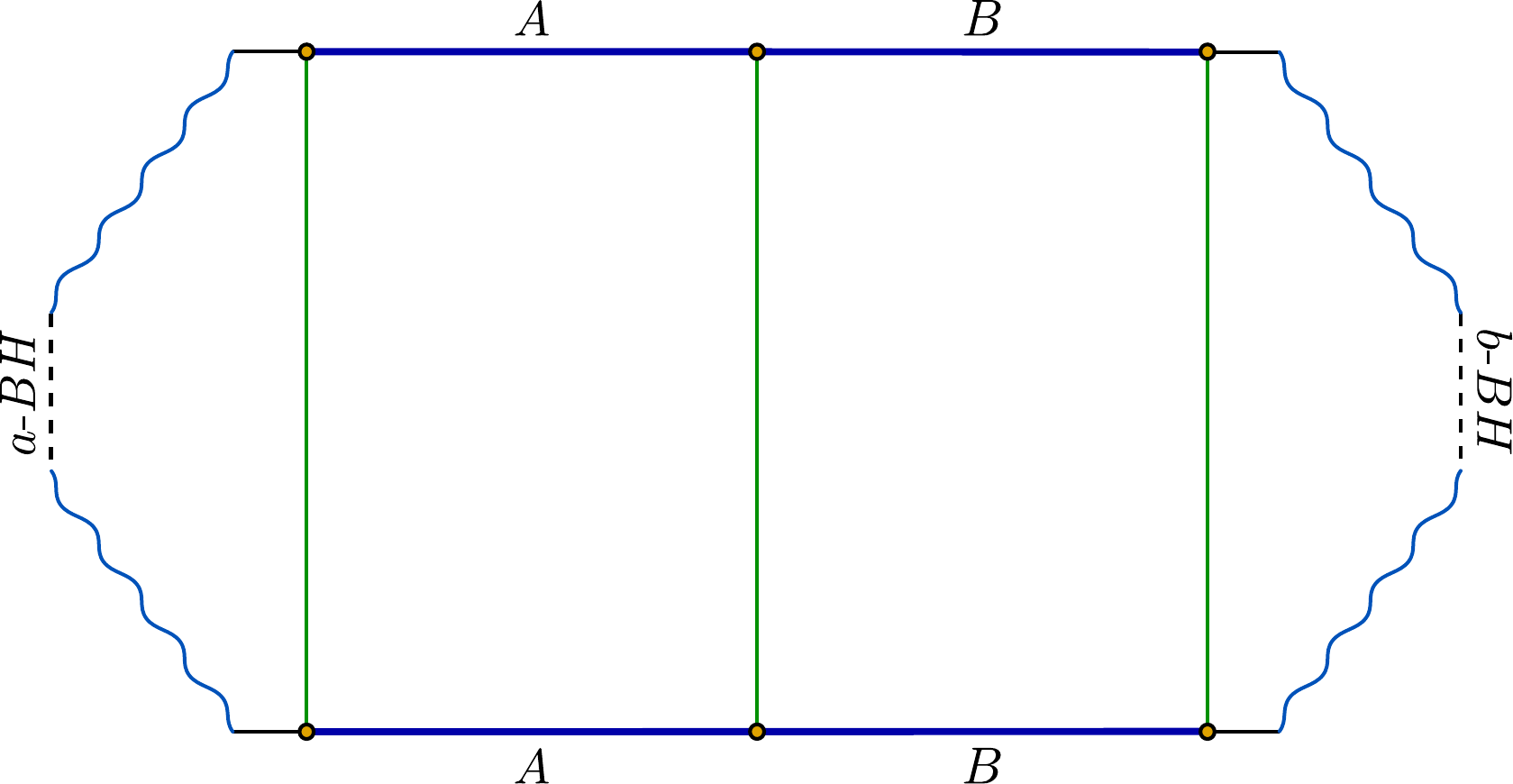}
		\caption{Phase-1(a)}
		\label{}
	\end{subfigure}
	\vspace{.4cm}
	\hspace{.12cm}
	\begin{subfigure}[b]{0.45\textwidth}
		\centering
		\includegraphics[width=\textwidth]{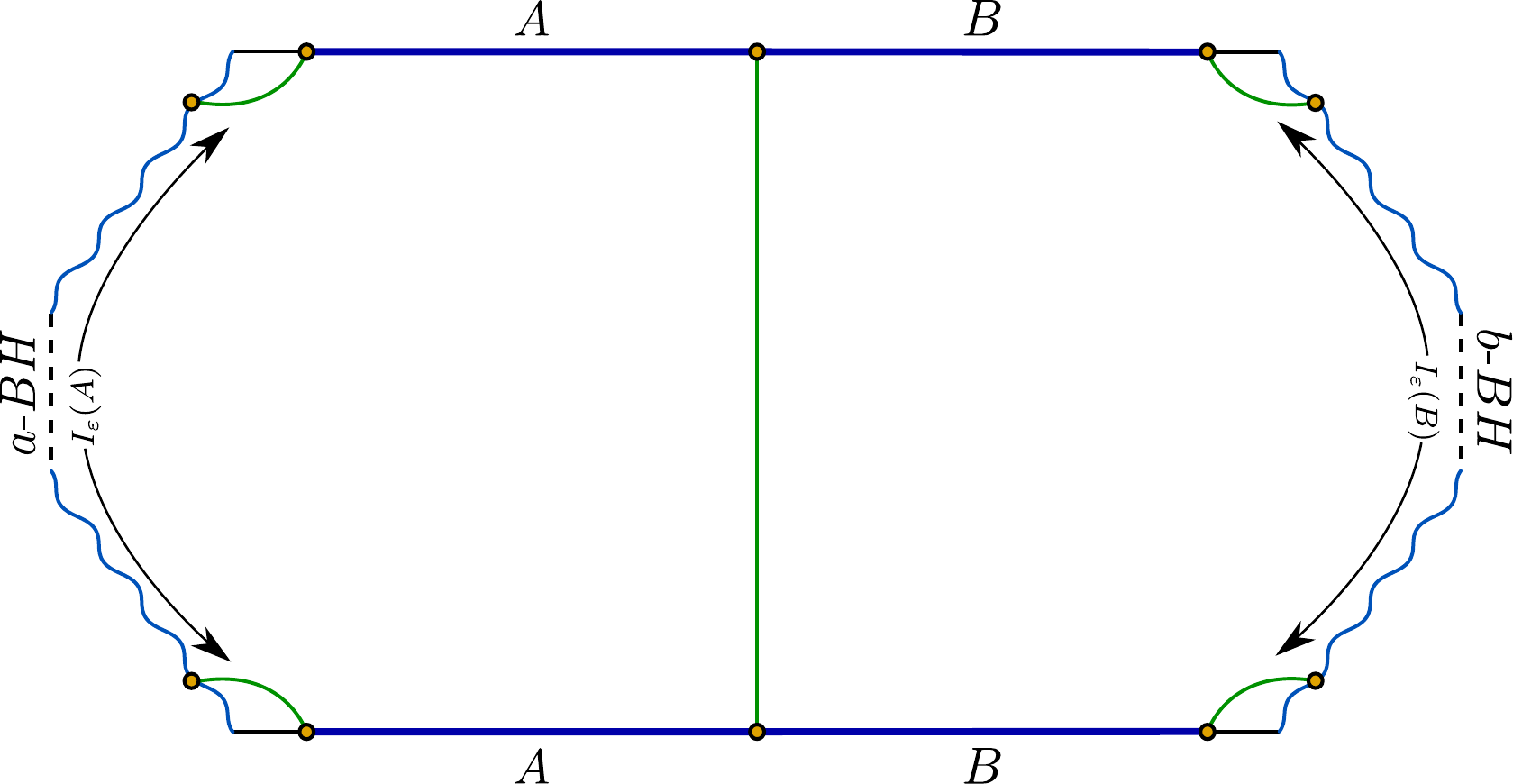}
		\caption{Phase-1(b)}
		\label{}
	\end{subfigure}
	\begin{subfigure}[b]{0.45\textwidth}
		\centering
		\includegraphics[width=\textwidth]{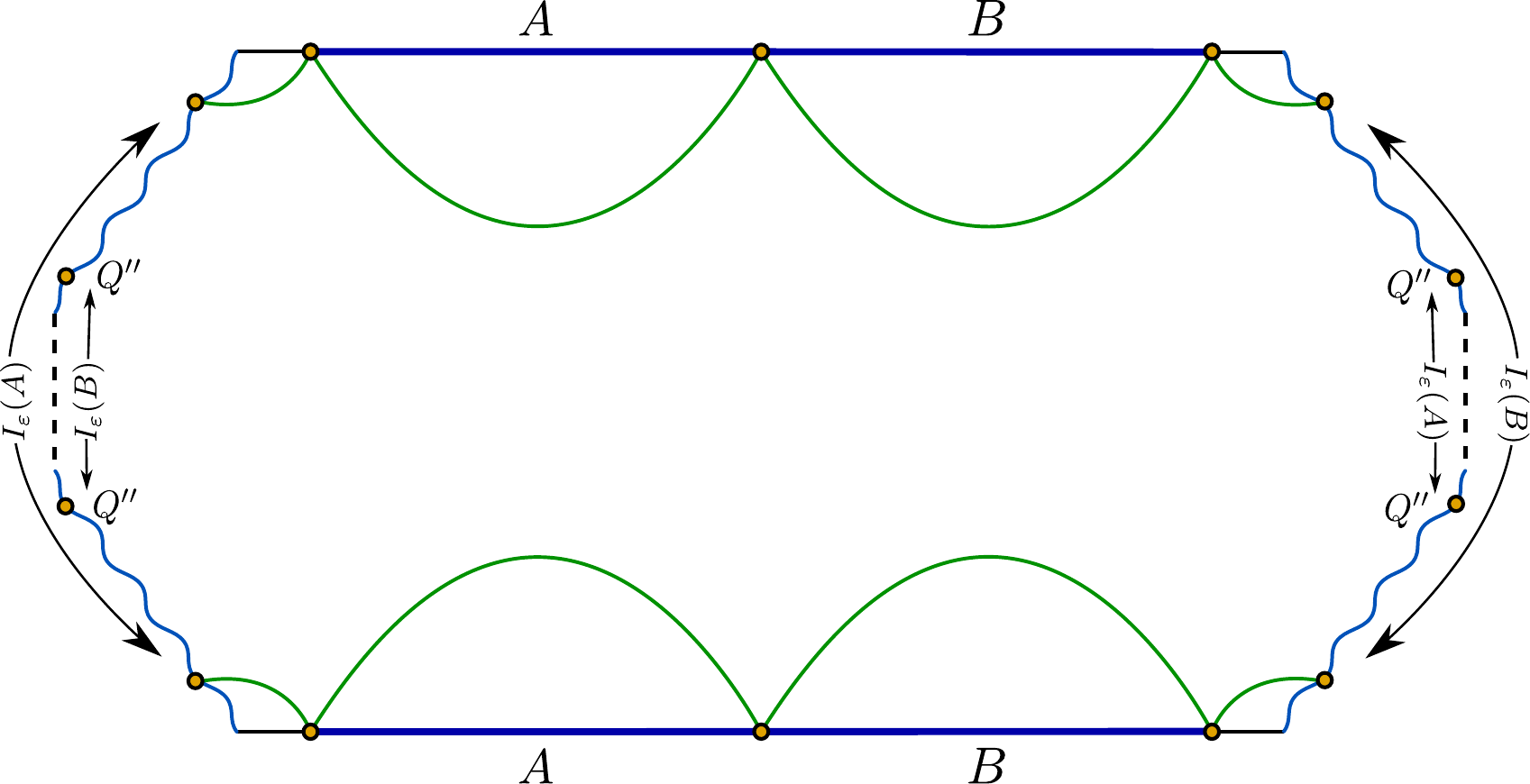}
		\caption{Phase-2}
		\label{}
	\end{subfigure}
	\caption{Schematic depicts all the possible phases of the entanglement negativity between two adjacent subsystems with equal sizes as time increases.}
	\label{Adjcase3}
\end{figure}
\noindent\\\\
\textbf{Phase 1:} This phase contains two intermediate sub phases due to the various structures of the RT surfaces supported by the subsystems in question. At early times, the subsystems $A$, $B$ and $A\cup B$ support bulk-type RT surfaces each (\cref{Adjcase3}). This sub phase does not include any island regions corresponding to the subsystems mentioned above. The number of Hawking modes from both the black holes accumulating in the two subsystems increases linearly with increasing time. Hence, the corresponding entanglement negativity profile rises linearly as time increases. We proceed to the second sub phase where the dominant contributions to the entanglement entropies of the subsystems $A$, $B$ and $A\cup B$ arise from $a$-$bulk$, $b$-$bulk$ and $ab$-type RT surfaces respectively. Hence this sub phase includes entanglement negativity islands $I_{\varepsilon}(A)$ and $I_{\varepsilon}(B)$ corresponding to the subsystems $A$ and $B$ respectively as shown in \cref{Adjcase3}. Here the island $I_{\varepsilon}(A)$ contains the entire interior region of the $a$-black hole whereas the other island $I_{\varepsilon}(B)$ includes the entire interior region of the $b$-black hole. 
In this case, the entanglement negativity between the subsystems is governed by the degrees of freedom in the region $A\cup I_{\varepsilon}(A)$ or $B\cup I_{\varepsilon}(B)$. Note that the degrees of freedom in the region $A\cup I_{\varepsilon}(A)$ are determined by the interior Hawking modes captured by the negativity island $I_{\varepsilon}(A)$ whose exterior partners are located in $(A\cup I_{\varepsilon}(A))^c$ and by the Hawking modes arriving from the $b$-black hole. These degrees of freedom increase with time which confirms the linear rise of the corresponding Page curve as depicted in \cref{fig:Adjcase3}. The description for the second sub phase from the perspective of the degrees of freedom of the region $B\cup I_{\varepsilon}(B)$ is analogous to the arguments mentioned above which again justifies the linear rising behaviour of the corresponding Page curve.\\

\noindent\textbf{Phase 2:} In the second phase, the subsystems $A$ and $B$ support $dome$-type RT surfaces each whereas the subsystem $A\cup B$ still admits $ab$-type RT surfaces (\cref{Adjcase3}). This phase includes an entanglement entropy island for the subsystem $A\cup B$ and hence we have entanglement negativity islands $I_{\varepsilon}(A)$ and $I_{\varepsilon}(B)$ corresponding to the subsystems $A$ and $B$ respectively. Here each of the negativity islands $I_{\varepsilon}(A)$ and $I_{\varepsilon}(B)$ contains the entire interior regions of both the black holes. Hence this phase corresponds to an overlap between the negativity island regions $I_{\varepsilon}(A)$ and $I_{\varepsilon}(B)$ as shown in \cref{Adjcase3}. This is an extremely interesting and novel scenario which has not been reported in earlier literatures on the island constructions\footnote{Although the island regions are overlapping, it does not correspond to common degrees of freedom of the regions $A\cup I_{\varepsilon}(A)$ and $B\cup I_{\varepsilon}(B)$ thus implying the consistency of the monogamy property of quantum entanglement. This can be understood since the exterior partners of the interior Hawking modes of both the regions $A\cup I_{\varepsilon}(A)$ and $B\cup I_{\varepsilon}(B)$ are distinct and located in the regions $(A\cup I_{\varepsilon}(A))^c$ and $(B\cup I_{\varepsilon}(B))^c$ respectively.}. Once again in this phase, the entanglement negativity between the subsystems is governed by the degrees of freedom of the region $A\cup I_{\varepsilon}(A)$ or $B\cup I_{\varepsilon}(B)$. Note that the degrees of freedom of the region $A\cup I_{\varepsilon}(A)$ are determined by the interior Hawking modes captured by the negativity island $I_{\varepsilon}(A)$ whose exterior partners are located in $(A\cup I_{\varepsilon}(A))^c$. At late times, the number of ingoing and outgoing Hawking modes in the region $A\cup I_{\varepsilon}(A)$ become equal such that its degrees of freedom remains constant throughout this phase. Consequently, the Page curve for the entanglement negativity exhibits a constant behaviour as depicted in \cref{fig:Adjcase3}. Once again, the description for this phase from the perspective of the degrees of freedom of the region $B\cup I_{\varepsilon}(B)$ is analogous to the arguments mentioned above which again justifies the constant behaviour of the corresponding Page curve. Note that in this phase, other candidates for the entanglement negativity islands corresponding to the subsystems $A$ and $B$ may be considered analogous to those observed in the phase-1(b). However, with those negativity islands, the corresponding entanglement negativity profile continues to rise linearly and thus do not provide a consistent interpretation for the phase-2.

\subsubsection*{(b) For $\bm{l_1\neq l_2}$}\label{p!=1} 
Next we consider two unequal lengths of the subsystems $A$ and $B$ and observe that the corresponding Page curve for the entanglement negativity consists of three consecutive phases. Again we refer to the appendix \ref{Appendix3b} for the entanglement negativity expressions obtained in these distinct phases. In the following, we now explore these phases in detail.\\

\begin{figure}[h!]
	\centering
	\begin{subfigure}[b]{0.45\textwidth}
		\centering
		\includegraphics[width=\textwidth]{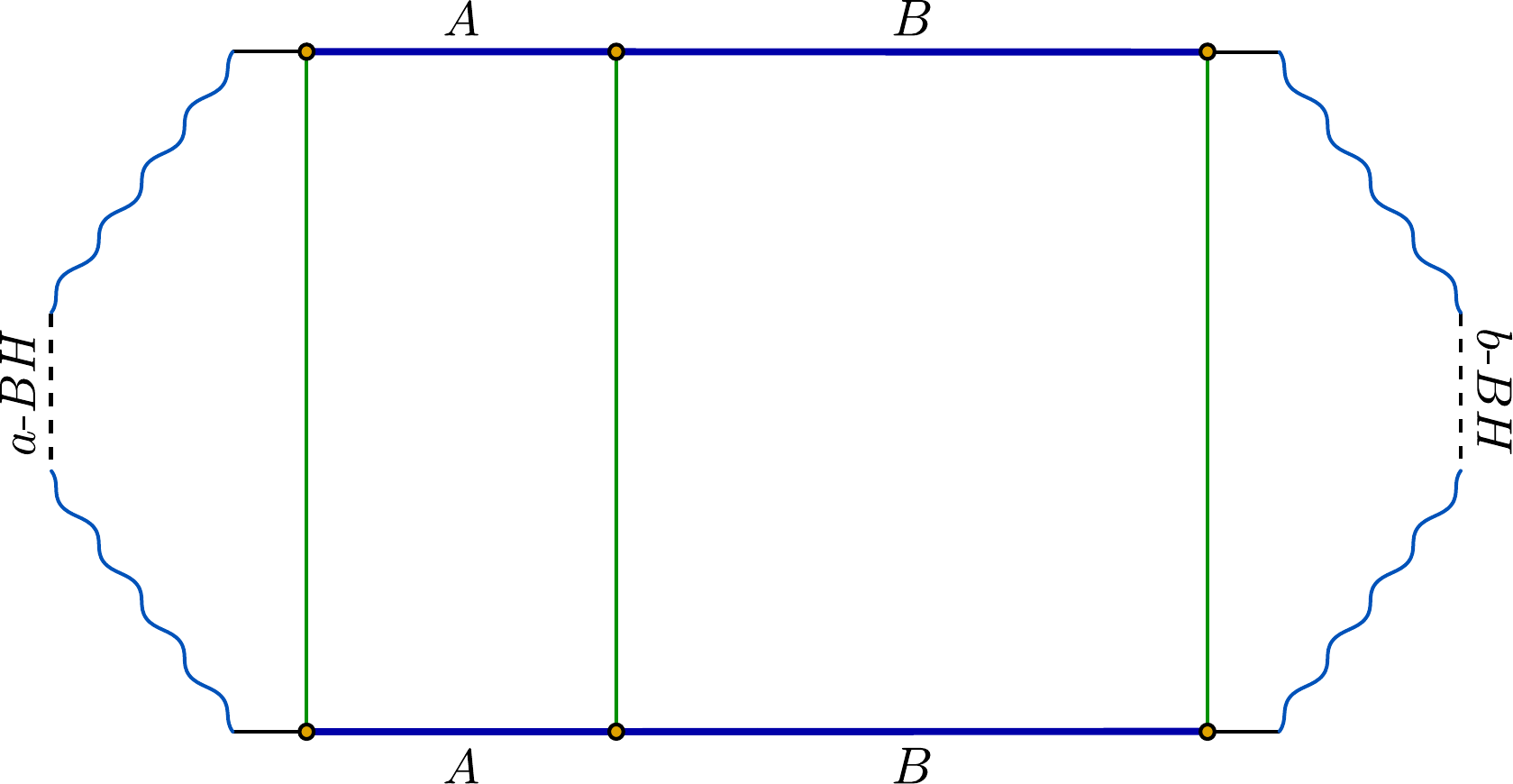}
		\caption{Phase-1(a)}
		\label{}
	\end{subfigure}
	\vspace{.4cm}
	\hspace{.12cm}
	\begin{subfigure}[b]{0.45\textwidth}
		\centering
		\includegraphics[width=\textwidth]{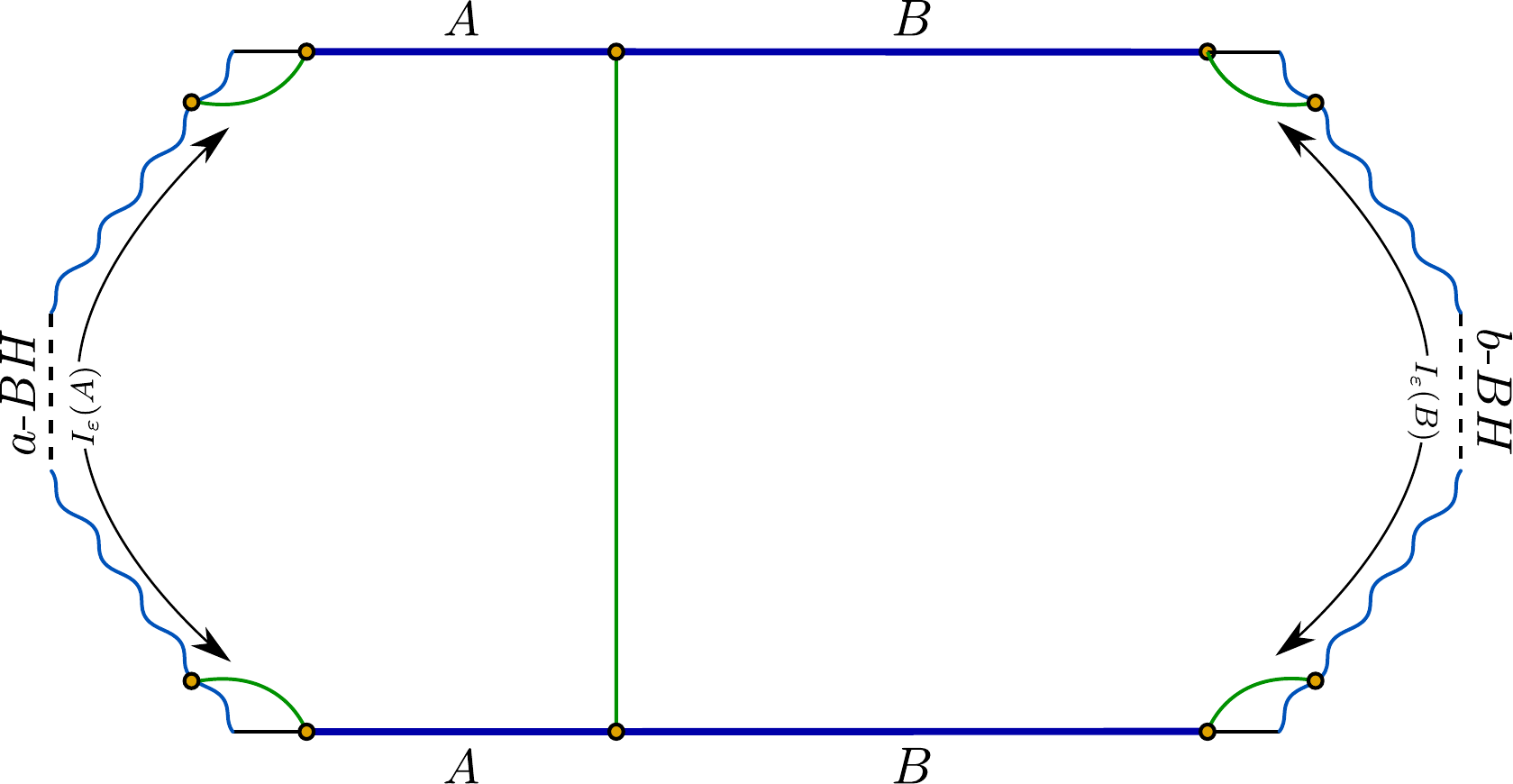}
		\caption{Phase-1(b)}
		\label{}
	\end{subfigure}
	\begin{subfigure}[b]{0.45\textwidth}
		\centering
		\includegraphics[width=\textwidth]{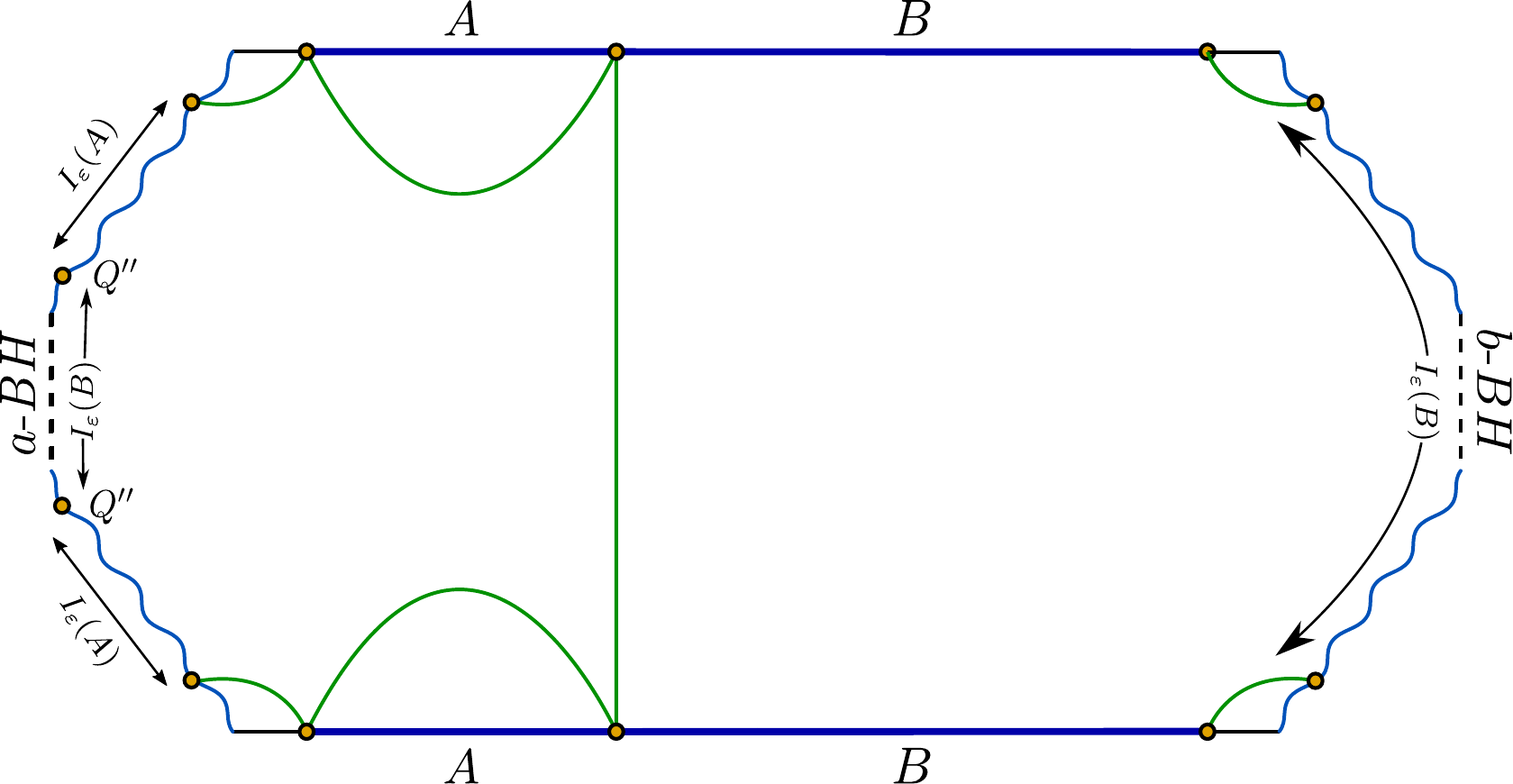}
		\caption{Phase-2}
		\label{}
	\end{subfigure}
	\hspace{.12cm}
	\begin{subfigure}[b]{0.45\textwidth}
		\centering
		\includegraphics[width=\textwidth]{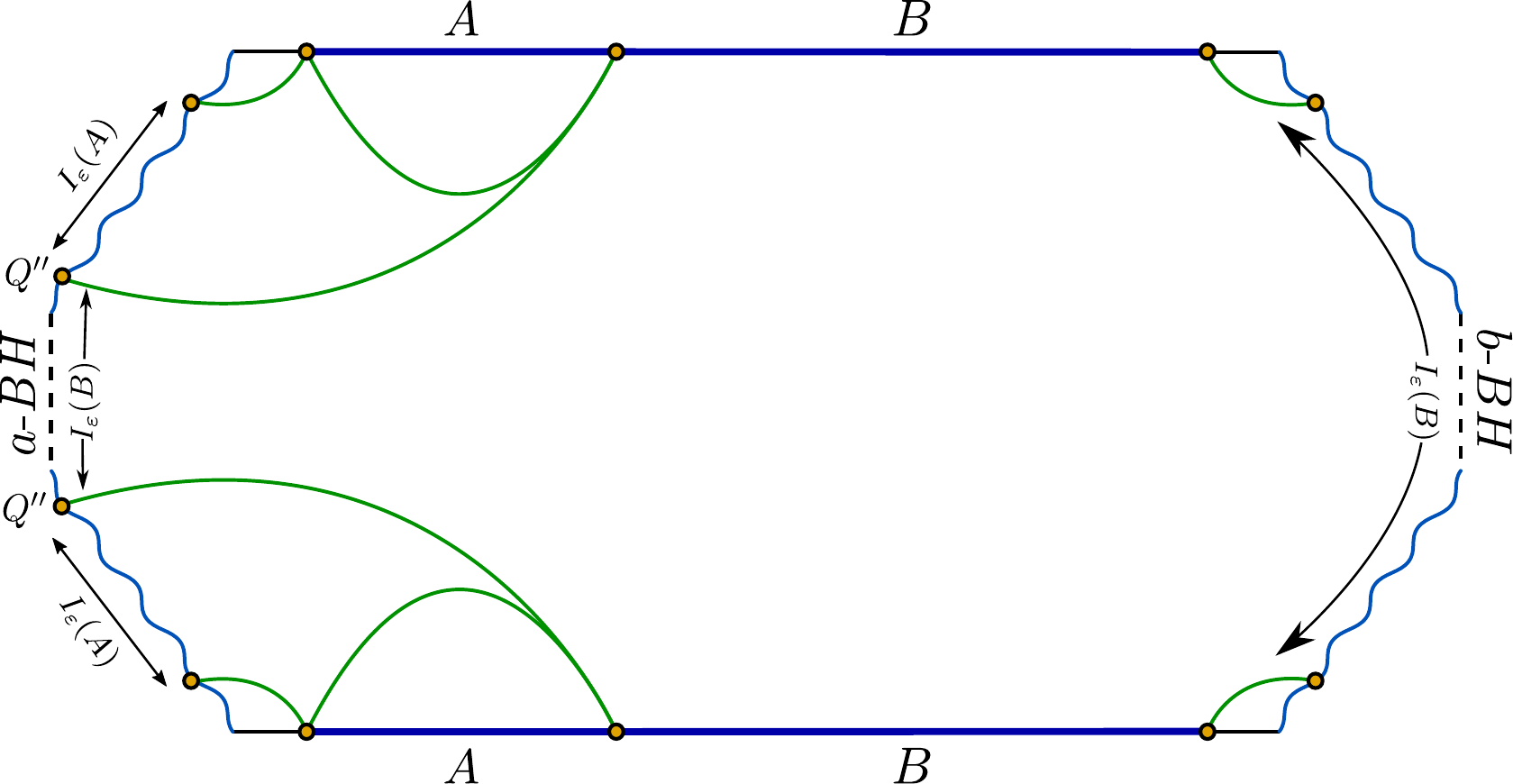}
		\caption{Phase-3}
		\label{}
	\end{subfigure}
	\caption{Different phases of the entanglement negativity between two adjacent subsystems with unequal sizes as time increases.}
	\label{Adjcase4}
\end{figure}

\noindent\textbf{Phase 1:} In this phase (\cref{Adjcase4}), the various structures of the RT surfaces supported by the subsystems and the corresponding interpretation in terms of the Hawking radiation are identical to the first phase of the previous sub case.\\

\noindent\textbf{Phase 2:} We proceed to the second phase where the dominant contributions to the entanglement entropy of the subsystems $A$, $B$ and $A\cup B$ arise from dome, $b$-bulk and $ab$-type RT surfaces respectively. Hence, this phase includes an entanglement negativity island $I_{\varepsilon}(B)$ corresponding to the subsystem $B$ as depicted in \cref{Adjcase4}. Here the negativity island $I_{\varepsilon}(B)$ contains the entire interior region of both the black holes. Note that the entanglement negativity between the subsystems $A$ and $B$ in this phase is governed by the degrees of freedom of the region $B\cup I_{\varepsilon}(B)$. These degrees of freedom are determined by the interior Hawking modes captured by the island $I_{\varepsilon}(B)$ whose exterior partners are located in $(B\cup I_{\varepsilon}(B))^c$. Now the number of outgoing Hawking modes is larger than the number of ingoing Hawking modes in the region $B\cup I_{\varepsilon}(B)$. Hence the degrees of freedom of the region $B\cup I_{\varepsilon}(B)$ increase with time at a smaller rate than the previous phase which corresponds to the increasing behaviour of the Page curve as depicted in \cref{fig:Adjcase3}.
\noindent\\\\
\textbf{Phase 3:}  Finally in the last phase, the RT surfaces supported by the subsystem $B$ are $ab$-type whereas the subsystems $A$ and $A\cup B$ still admit dome and $ab$-type RT surfaces respectively (\cref{Adjcase4}). Once again this phase includes an entanglement negativity island $I_{\varepsilon}(B)$ for the subsystem $B$ with a description similar to the previous phase. In this scenario, the entanglement negativity between the subsystems is again governed by the degrees of freedom of the region $B\cup I_{\varepsilon}(B)$ which are determined similarly as in the previous phase. At late times, the number of ingoing and outgoing Hawking modes in the region $B\cup I_{\varepsilon}(B)$ are equal such that its degrees of freedom remain constant throughout this phase. Consequently, the Page curve for the entanglement negativity exhibits a constant behaviour as depicted in \cref{fig:Adjcase3}.

\subsubsection{Disjoint subsystems}\label{disj}
Next we consider a mixed state configuration of two disjoint subsystems $A$ and $B$ with finite lengths $l_1$ and $l_2$ respectively where a subsystem $C$ with length $l_c$ is sandwiched between them. In this context, we utilize the \cref{NegDis0,HEE-Shaghoulian} to obtain the holographic entanglement negativity between the subsystems $A$ and $B$ for three distinct scenarios involving the subsystem sizes and the time. Furthermore, we describe the qualitative nature of the corresponding entanglement negativity profiles in these scenarios.

\subsubsection*{$\bm{(i)}$ Subsystem $\bm{A}$ fixed, $\bm{C}$ varied}\label{discase1}
\begin{figure}[h!]
	\centering
	\begin{subfigure}[t]{.45\textwidth}
		\includegraphics[width=1\linewidth]{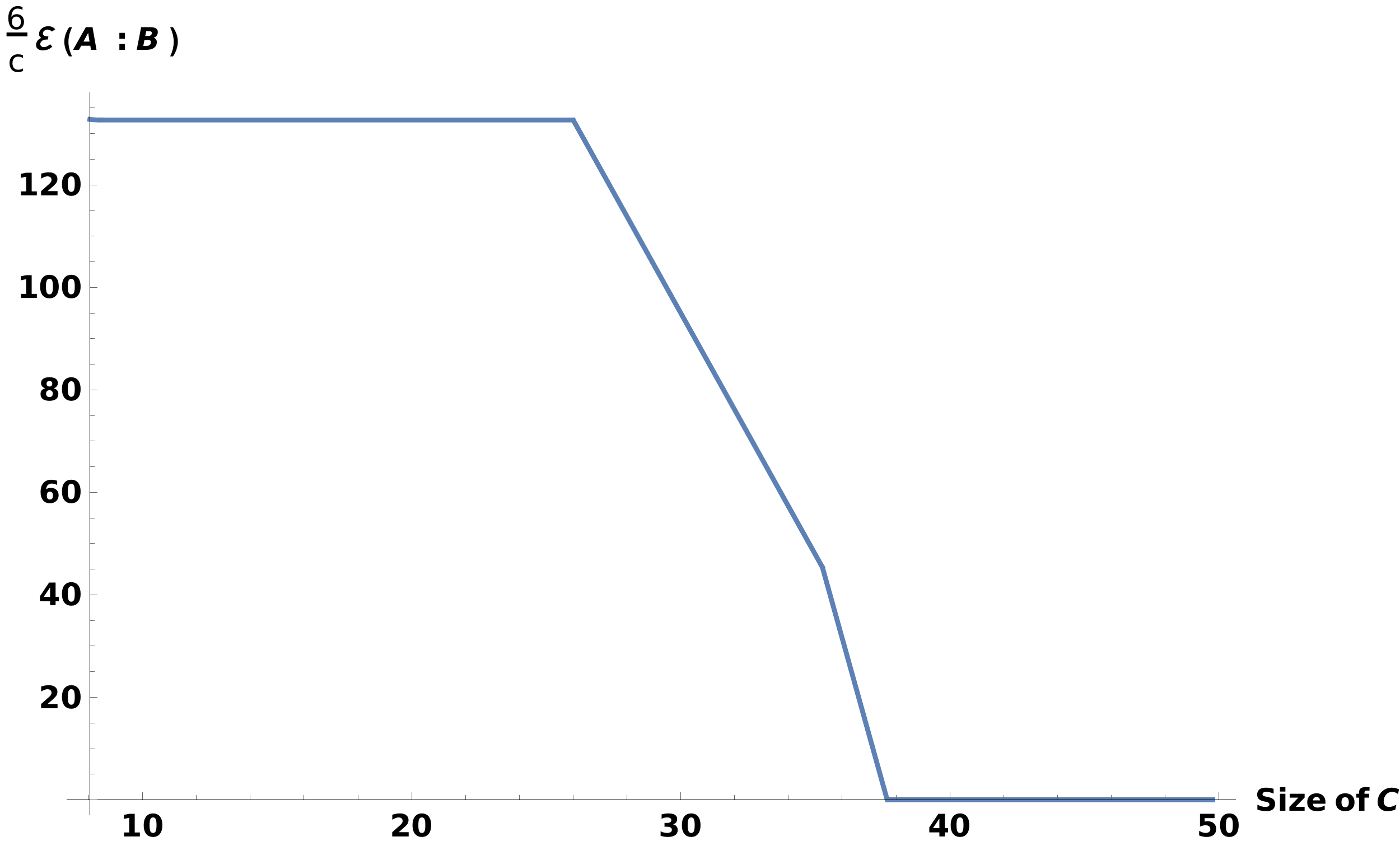}
		\caption{Entanglement negativity with respect to the size of $C$ where we have considered $B\cup C=[.15L,.99L]$ and $A\cup B\cup C=[.01L,.99L]$.}
		\label{fig:Discase1}
	\end{subfigure}
	\hspace{.1cm}
	\begin{subfigure}[t]{.45\textwidth}
		\includegraphics[width=1\linewidth]{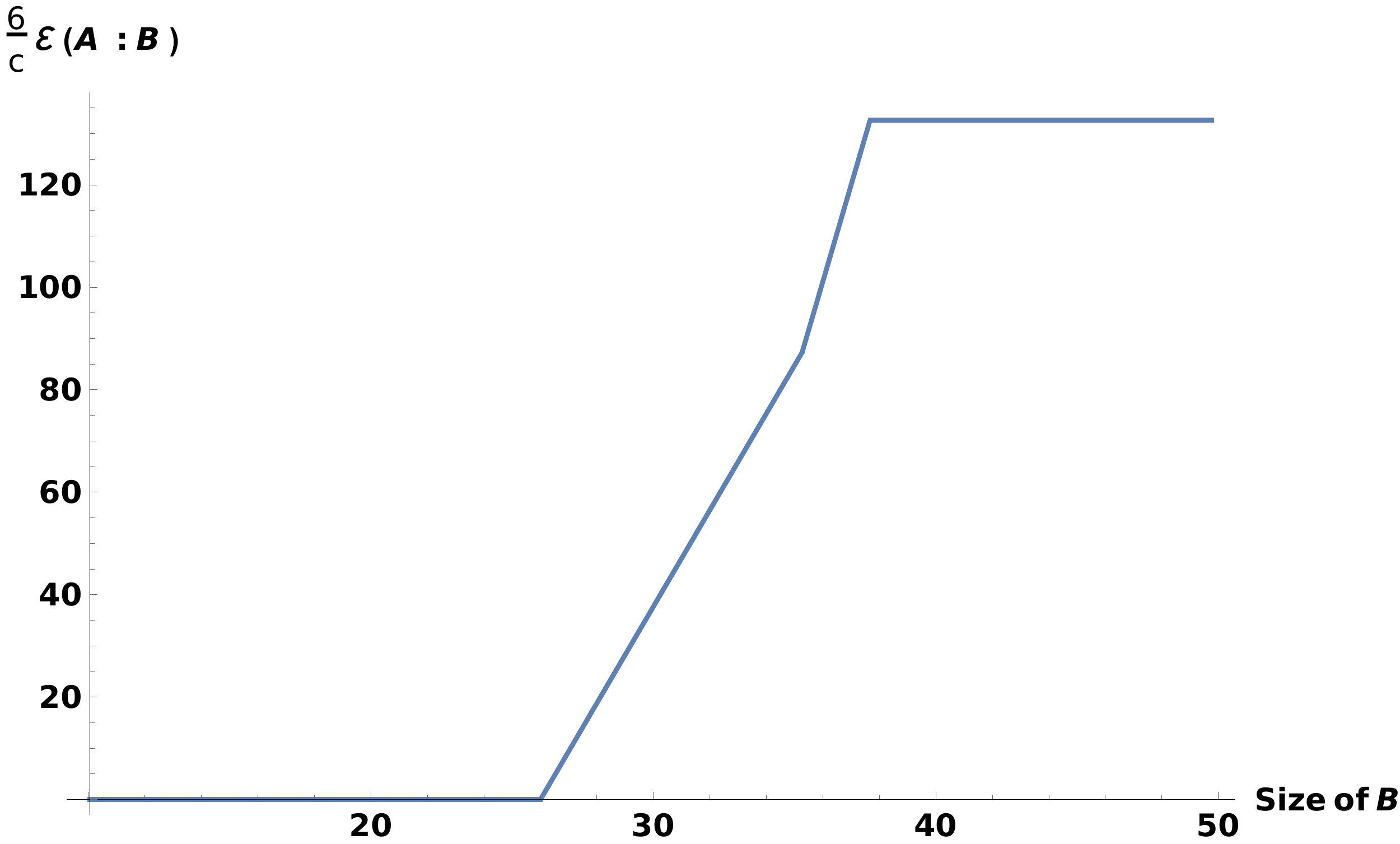}
		\caption{Entanglement negativity with respect to the size of $B$ where we have considered $C=[.15L,.2L]$.}
		\label{fig:Discase2}
	\end{subfigure}
	
	\hspace{.1cm}
	\begin{subfigure}[t]{.53\textwidth}
		\includegraphics[width=1\linewidth]{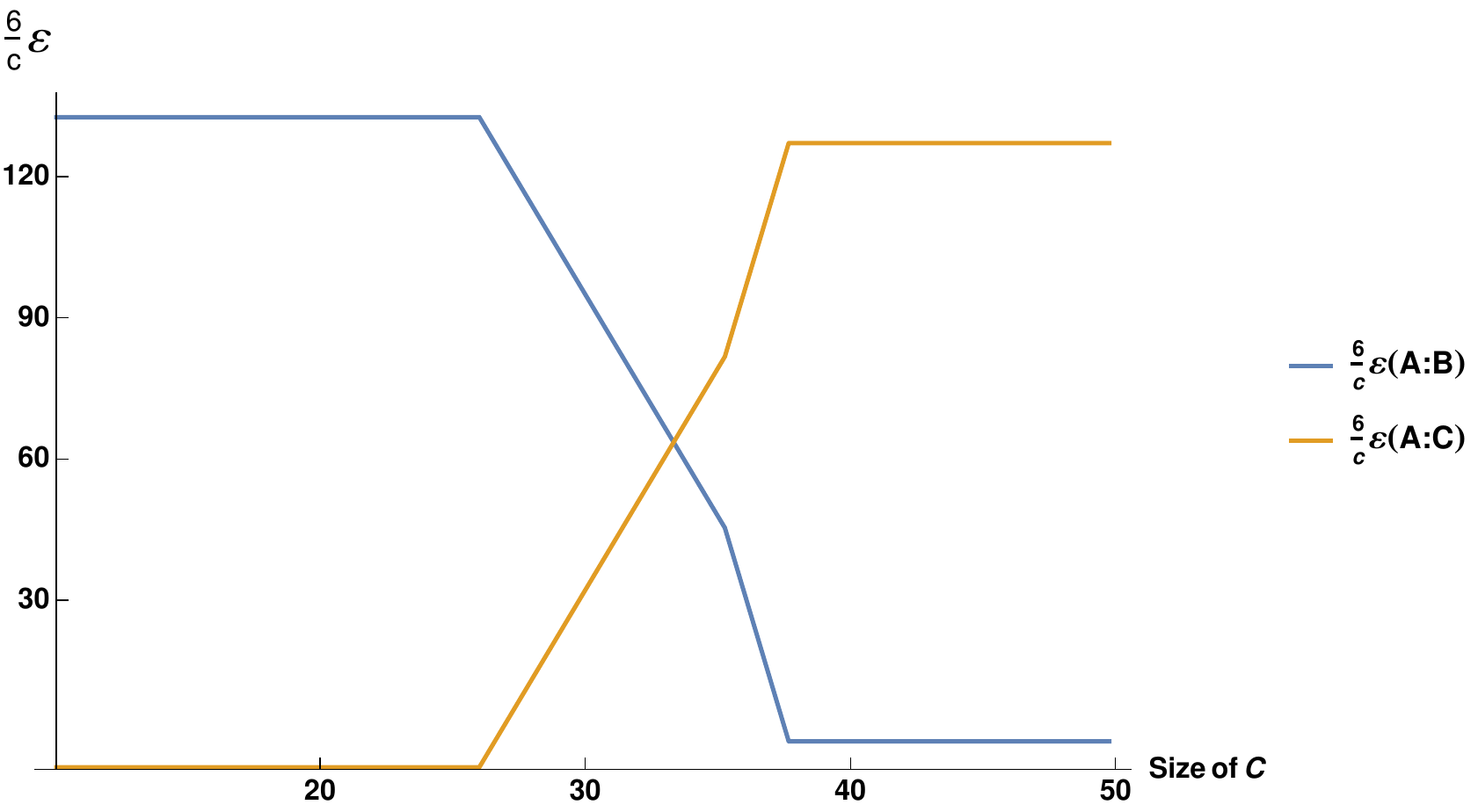}
		\caption{Entanglement negativity between different subsystems where we have considered $A=[.01L,.15L]$, $B\cup C=[.15L,.99L]$ and $A\cup B\cup C=[.01L,.99L]$.}
		\label{fig:Disjcase1}
	\end{subfigure}
	\caption{Here $\beta=1$, $c=500$, $t=20$, $\phi_0= \frac{30c}{6}$, $\phi_r= \frac{30}{\pi}$, $L=\frac{16\pi}{\beta}$, $\epsilon=.001$, $A=[.01L,.15L]$.}
	\label{EN_Shag_disj}
\end{figure}

We begin with the case where the length $l_1$ of the subsystem $A$ is fixed at a constant time slice and compute the holographic entanglement negativity between the subsystems $A$ and $B$ with an increasing length ${l_c}$ of the subsystem $C$. We observe four consecutive phases of the entanglement negativity profile as depicted in \cref{fig:Discase1} due to the various structures of the RT surfaces supported by the subsystems in question. Utilizing the \cref{NegDis0,HEE-Shaghoulian}, we obtain the expressions for the corresponding entanglement negativity in these phases which are listed in the appendix \ref{Appendix4}. In what follows, we describe these phases of the entanglement negativity profile in detail.
\noindent\\\\
\textbf{Phase 1:} In the first phase, the RT surfaces for the subsystems $A\cup C$ and $C$ are dome-type each whereas the subsystems $B\cup C$ and $A\cup B\cup C$ support $ab$-type RT surfaces. Here the size of the subsystem $A\cup C$ is very small compared to the size of $B$ such that this phase does not correspond to any island region for the subsystem $A\cup C$. In this phase, the entanglement negativity between the subsystems $A$ and $B$ is governed by the degrees of freedom of the subsystem $A$. Note that the corresponding degrees of freedom of the subsystem $A$ remain constant since the number of Hawking modes arriving from both the black holes together with the $CFT$ degrees of freedom do not change with the increasing length $l_c$. Consequently the entanglement negativity profile in this phase remains constant as depicted in \cref{fig:Discase1}.\\

\noindent\textbf{Phase 2:} Next we proceed to second phase where the subsystems $A\cup C$ and $C$ admit $a$-bulk and dome-type RT surfaces respectively whereas the subsystems $B\cup C$ and $A\cup B\cup C$ still support $ab$-type RT surfaces. Consequently, this phase involves an entanglement negativity island $I_{\varepsilon}(B)$ on the $b$-brane corresponding to the subsystem $B$. This island $I_{\varepsilon}(B)$ involves the entire interior region of the $b$-black hole. The entanglement negativity between the subsystems $A$ and $B$ in this phase is governed by the degrees of freedom of the region $B\cup I_{\varepsilon}(B)$. As we increase $l_c$, the number of Hawking modes arriving from the $a$-black hole leave the region $B\cup I_{\varepsilon}(B)$. However an equal number of interior Hawking modes captured by the negativity island $I_{\varepsilon}(B)$ whose partners are located in $(B\cup I_{\varepsilon}(B))^c$ increase simultaneously. Consequently, the degrees of freedom of the region $B\cup I_{\varepsilon}(B)$ is determined only by the $CFT_2$-degrees of freedom which decrease with the increasing length $l_c$ as the size of the subsystem $B$ decreases. Accordingly, the corresponding entanglement negativity profile decreases linearly as depicted in \cref{fig:Discase1}.
\begin{figure}[h!]
	\centering
	\begin{subfigure}[b]{0.45\textwidth}
		\centering
		\includegraphics[width=\textwidth]{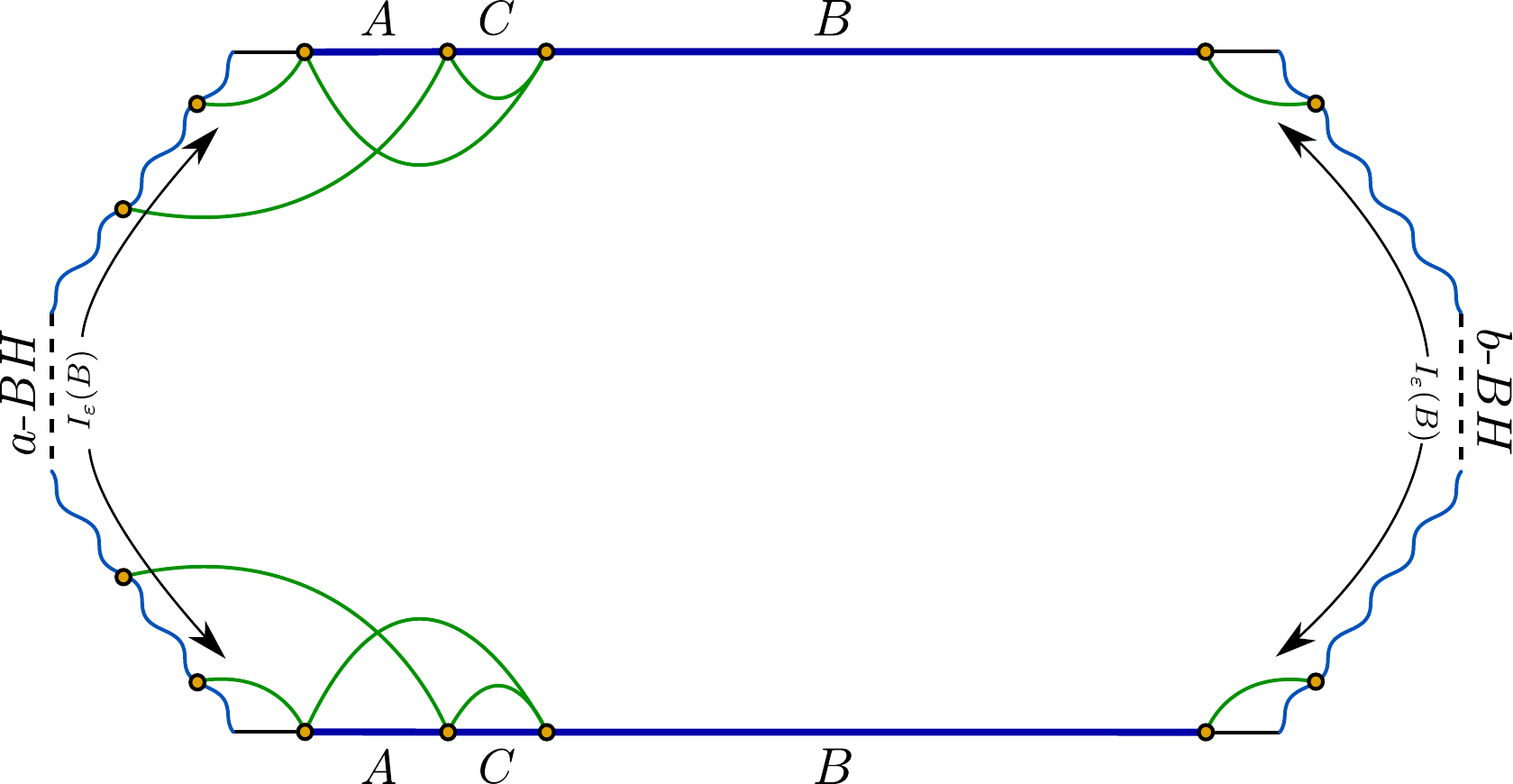}
		\caption{Phase-1}
		\label{}
	\end{subfigure}
	\vspace{.4cm}
	\hspace{.12cm}
	\begin{subfigure}[b]{0.45\textwidth}
		\centering
		\includegraphics[width=\textwidth]{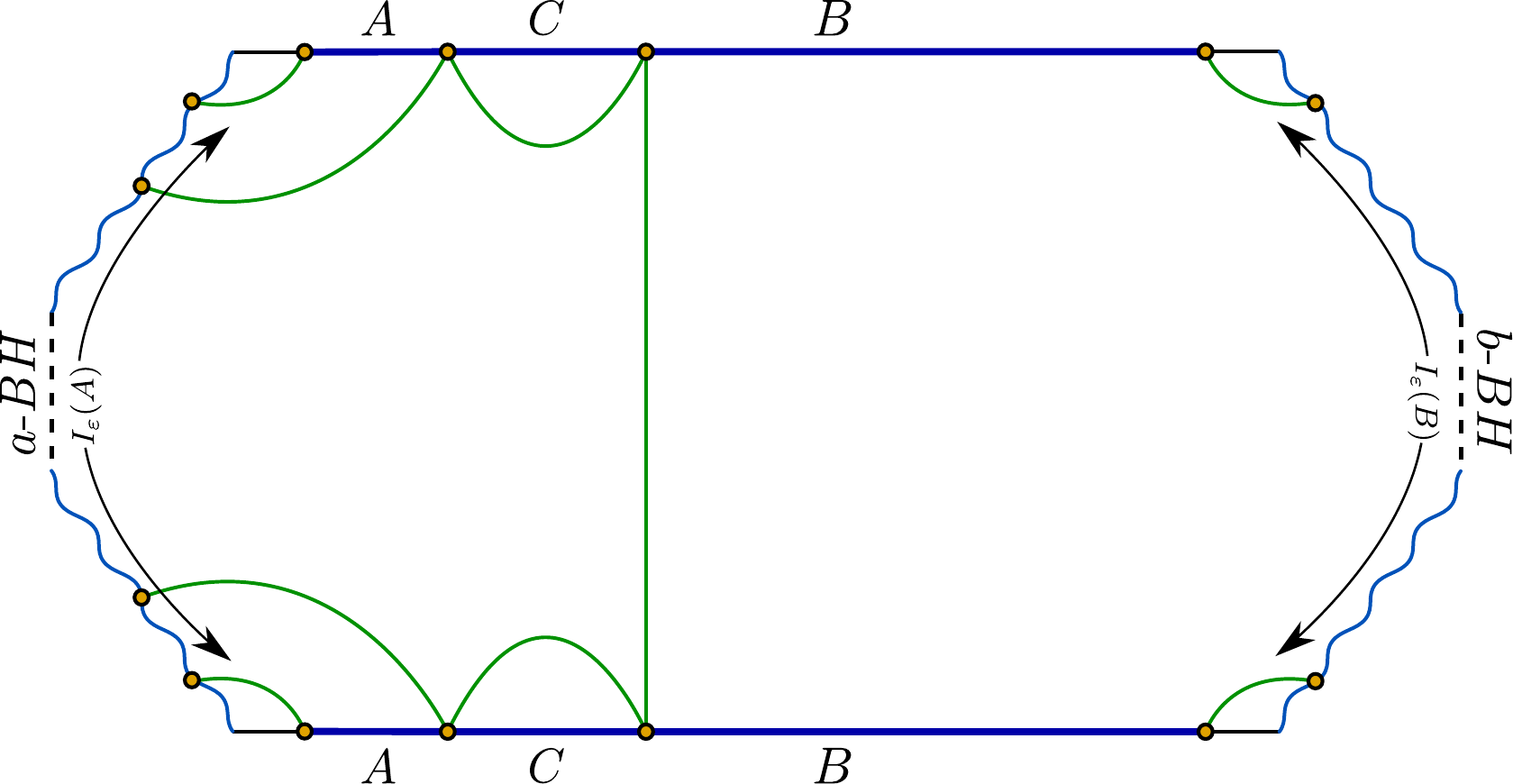}
		\caption{Phase-2}
		\label{}
	\end{subfigure}
	\begin{subfigure}[b]{0.45\textwidth}
		\centering
		\includegraphics[width=\textwidth]{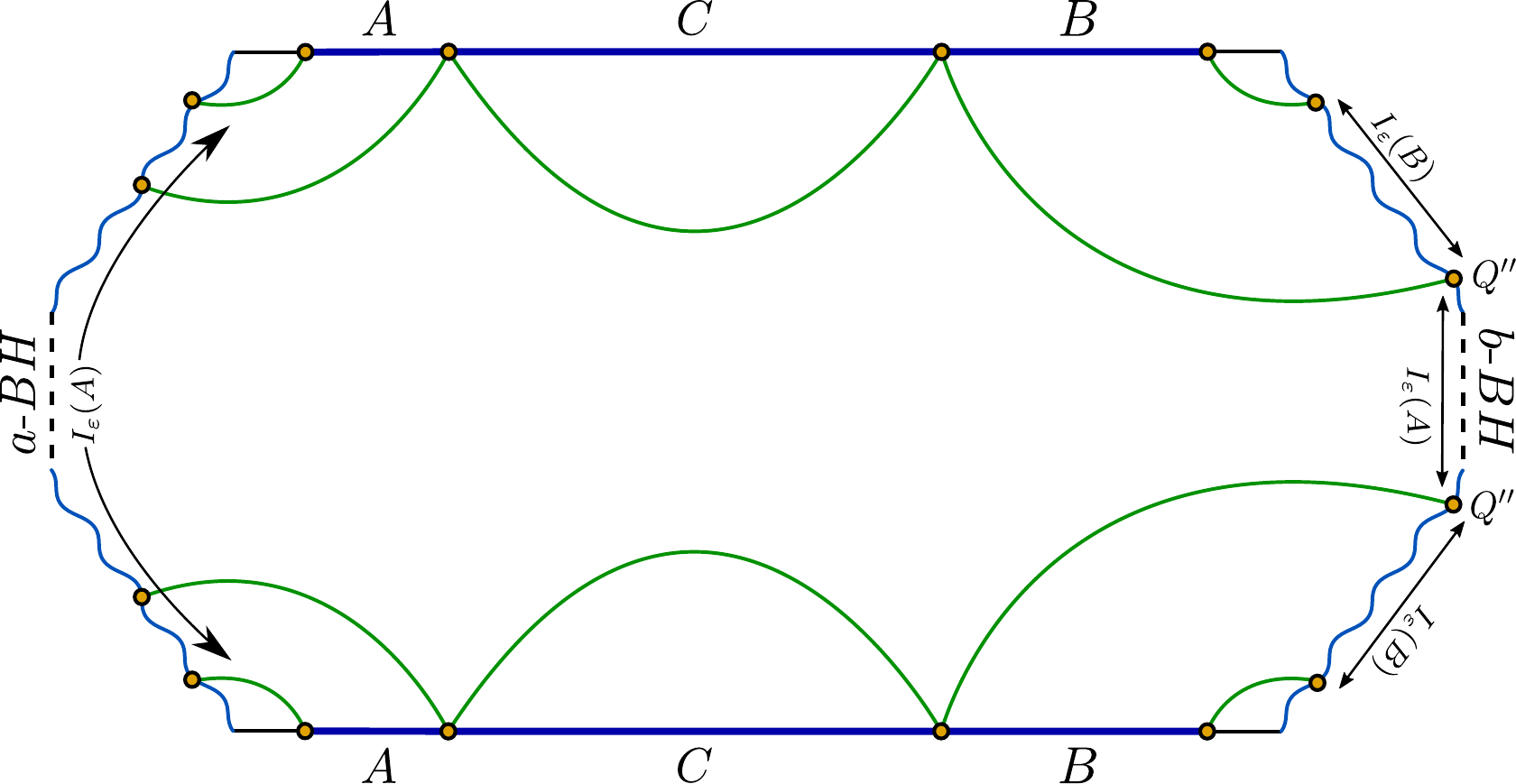}
		\caption{Phase-3}
		\label{}
	\end{subfigure}
	\hspace{.12cm}
	\begin{subfigure}[b]{0.45\textwidth}
		\centering
		\includegraphics[width=\textwidth]{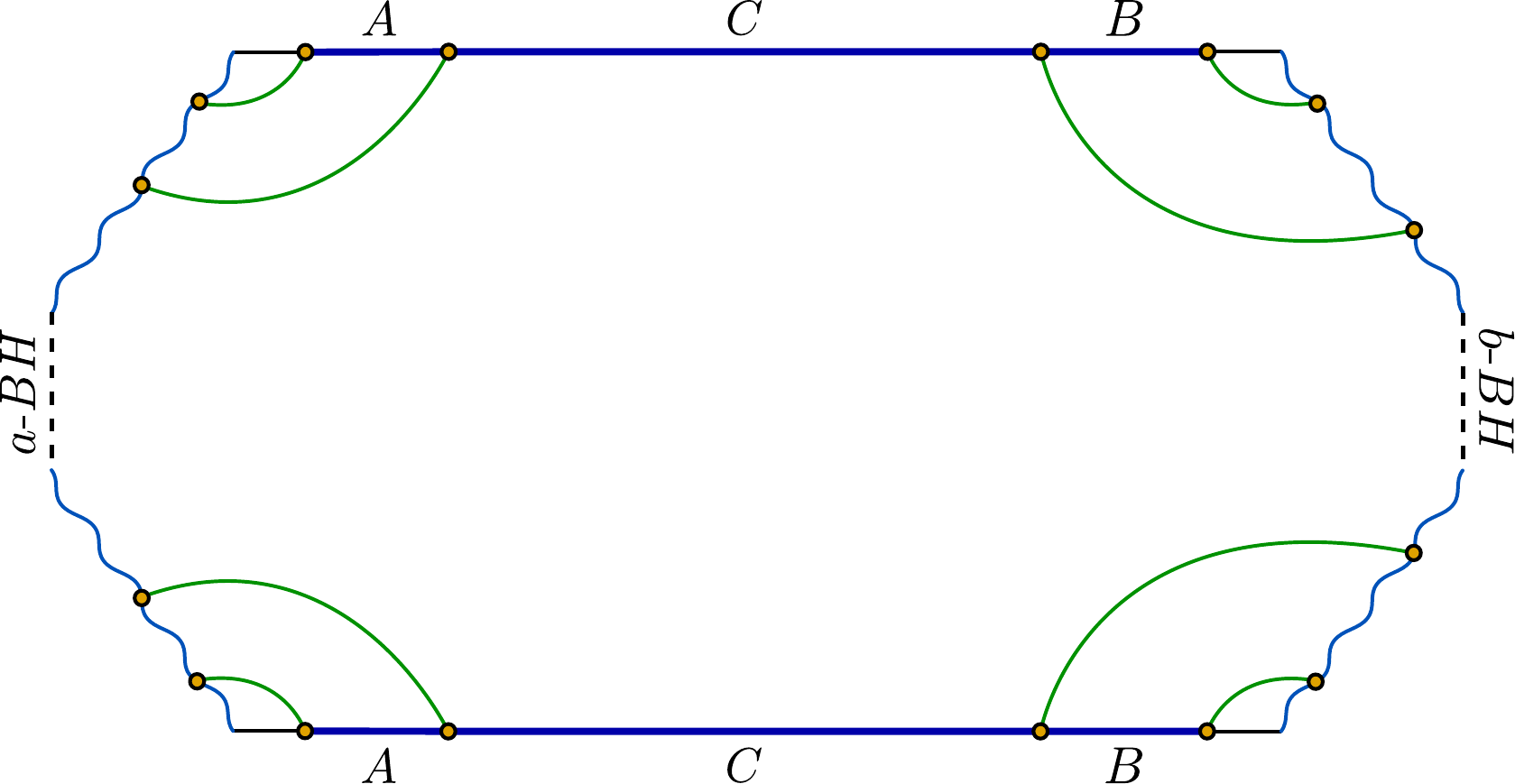}
		\caption{Phase-4}
		\label{}
	\end{subfigure}
	\caption{The possible phases of the entanglement negativity between two disjoint subsystems $A$ and $B$ while increasing the size of the subsystem $C$ sandwiched between them.}
	\label{Disjcase1}
\end{figure}
\noindent\\\\
\textbf{Phase 3:} In this phase the dominant contributions to the entanglement entropies of the subsystems $A\cup C$, $B\cup C$ and $A\cup B \cup C$ arise from $ab$-type RT surfaces each whereas the subsystem $C$ still supports dome-type RT surfaces. This phase includes an entanglement negativity island $I_{\varepsilon}(B)$ for the subsystem $B$ located on the exterior region of the $b$-black hole. Once again the entanglement negativity between the subsystems in this phase is governed by the degrees of freedom of the region $B\cup I_{\varepsilon}(B)$. Hence the number of Hawking modes arriving from both the black holes and the $CFT_2$-degrees of freedom present in the region $B\cup I_{\varepsilon}(B)$ decrease with increasing length $l_c$ of the subsystem $C$  as the size of the subsystem $B$ decreases. Consequently, we observe a linear decreasing profile of the corresponding entanglement negativity with a rate higher than the previous phase as depicted in \cref{fig:Discase1}.\\

\noindent\textbf{Phase 4:} In the last phase, the RT surfaces for the subsystems $A\cup C$, $B\cup C$, $A\cup B \cup C$ and $C$ are identified as $ab$-type each. Consequently, the corresponding entanglement wedges of the subsystems $A$ and $B$ are disconnected from each other in this phase which indicates a zero entanglement negativity between them. On a separate note, here the subsystems $A$ and $B$ are very small and located far away from each other such that they do not develop any entanglement between themselves.\\

We now discuss an interesting issue by comparing the above results with those discussed in sub\cref{adjcase2} where the entanglement negativity between two adjacent subsystems $A$ and $C$ was analyzed. This comparison is depicted in figure-\ref{fig:Disjcase1}.

We find that the entanglement negativity between the subsystems $A$ and $C$ follows identical behaviour as discussed in sub\cref{adjcase2}. Note that the negativity between the subsystems $A$ and $B\cup C$ is constant throughout our analysis since we have fixed their sizes $l_1$, $l_c+l_2$ respectively and the time $t$. Now, with the increasing length $l_c$, the size of the subsystem $B$ decreases accordingly. Hence the degrees of freedom of $B$ are eventually transferred to the subsystem $C$ as $l_c$ increases. As a result the entanglement negativity between the subsystems $A$ and $B$ decreases while the entanglement negativity between the subsystems $A$ and $C$ is increasing as depicted in \cref{fig:Disjcase1}.

\subsubsection*{$\bm{(ii)}$ Subsystems $\bm{A}$ and $\bm{C}$ fixed, $\bm{B}$ varied}\label{discase2}
In this scenario, the lengths $l_1$ and $l_c$ of the subsystems $A$ and $C$ are fixed at a constant time slice. We then compute the holographic entanglement negativity between the two disjoint subsystems $A$ and $B$ with an increase in the length $l_2$ of the subsystem $B$ utilizing the \cref{NegDis0,HEE-Shaghoulian}. In this case, we observe four consecutive phases of the entanglement negativity profile as exhibited in \cref{fig:Discase2}. Once again the expressions for the corresponding entanglement negativity in these distinct phases are listed in the appendix \ref{Appendix5}. It is interesting to note that in this scenario, the profile of the entanglement negativity follows the behaviour similar to the adjacent case discussed in the sub\cref{adjcase2} where we fixed the size of the subsystem $A$ and varied the size of $B$. Consequently in the present case, the corresponding phases may be explained analogously in terms of the Hawking radiations.

\subsubsection*{$\bm{(iii)}$ Subsystems $\bm{A}$, $\bm{B}$ and $\bm{C}$ fixed, time varied}\label{discase3}

We conclude our analysis with the following case where the sizes of all the subsystems are fixed. For this scenario we compute the holographic entanglement negativity between the disjoint subsystems $A$ and $B$ with increasing time utilizing the \cref{NegDis0,HEE-Shaghoulian}. In particular, we obtain two Page curves for the corresponding entanglement negativity for two sub cases with equal and unequal lengths of the subsystems $A$ and $B$ as depicted in \cref{fig:Discase3}. In these scenarios, the $CFT_2$-degrees of freedom are irrelevant for the description of the entanglement negativity profiles since all the subsystem sizes are fixed. Hence the only effect on these profiles arise from the Hawking modes arriving from both the black holes.

\subsubsection*{(a) For $\bm{l_1=l_2}$}\label{dis_p=1}
For the case of two equal lengths subsystems, the Page curve for the entanglement negativity between the subsystems $A$ and $B$ consists of three consecutive phases and the expressions for the same in these phases are listed in the appendix \ref{Appendix6a}. In the following, we analyze these phases in detail.\\

\begin{figure}[h!]
	\centering
	\begin{subfigure}[b]{0.45\textwidth}
		\centering
		\includegraphics[width=\textwidth]{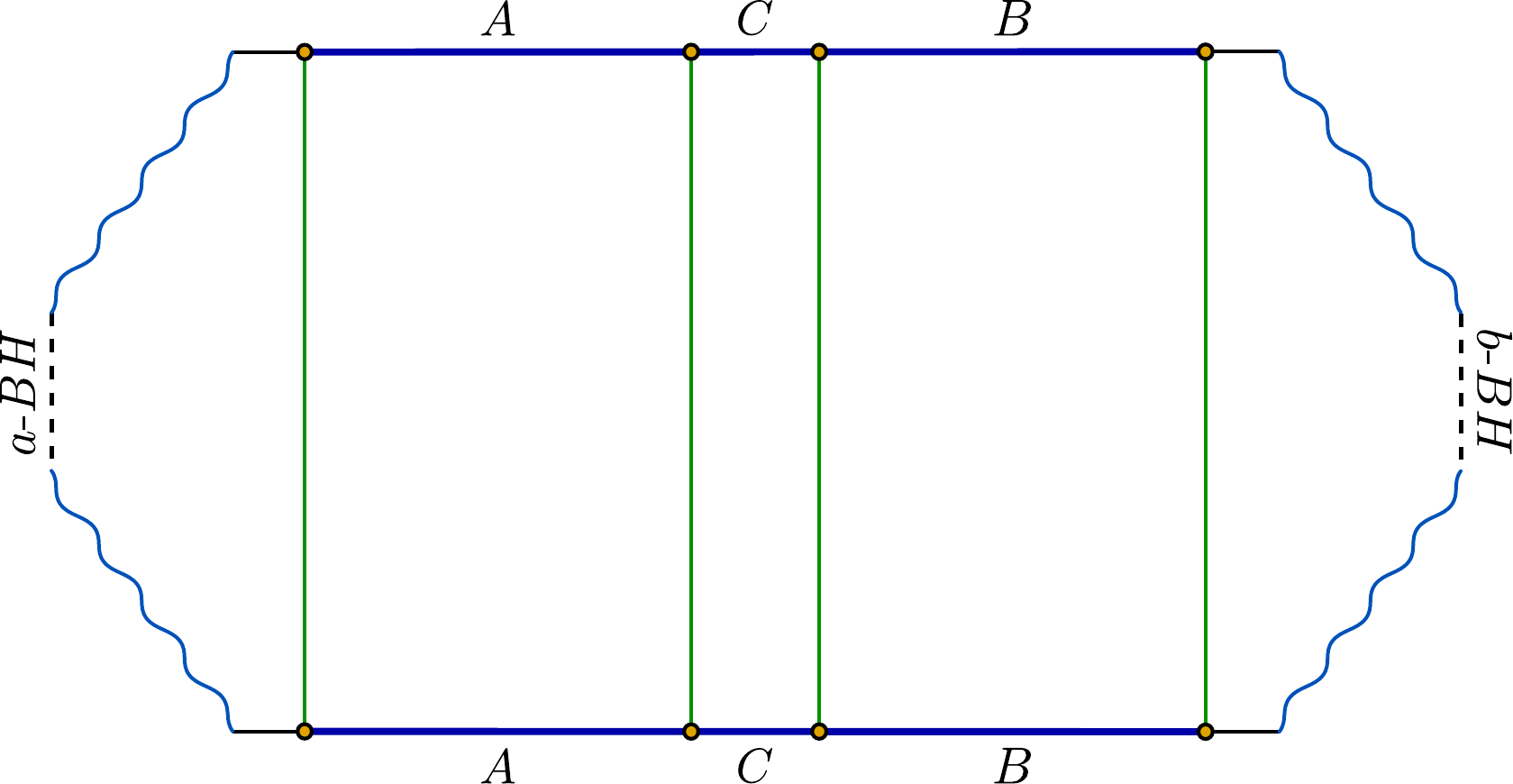}
		\caption{Phase-1}
		\label{}
	\end{subfigure}
	\vspace{.4cm}
	\hspace{.12cm}
	\begin{subfigure}[b]{0.45\textwidth}
		\centering
		\includegraphics[width=\textwidth]{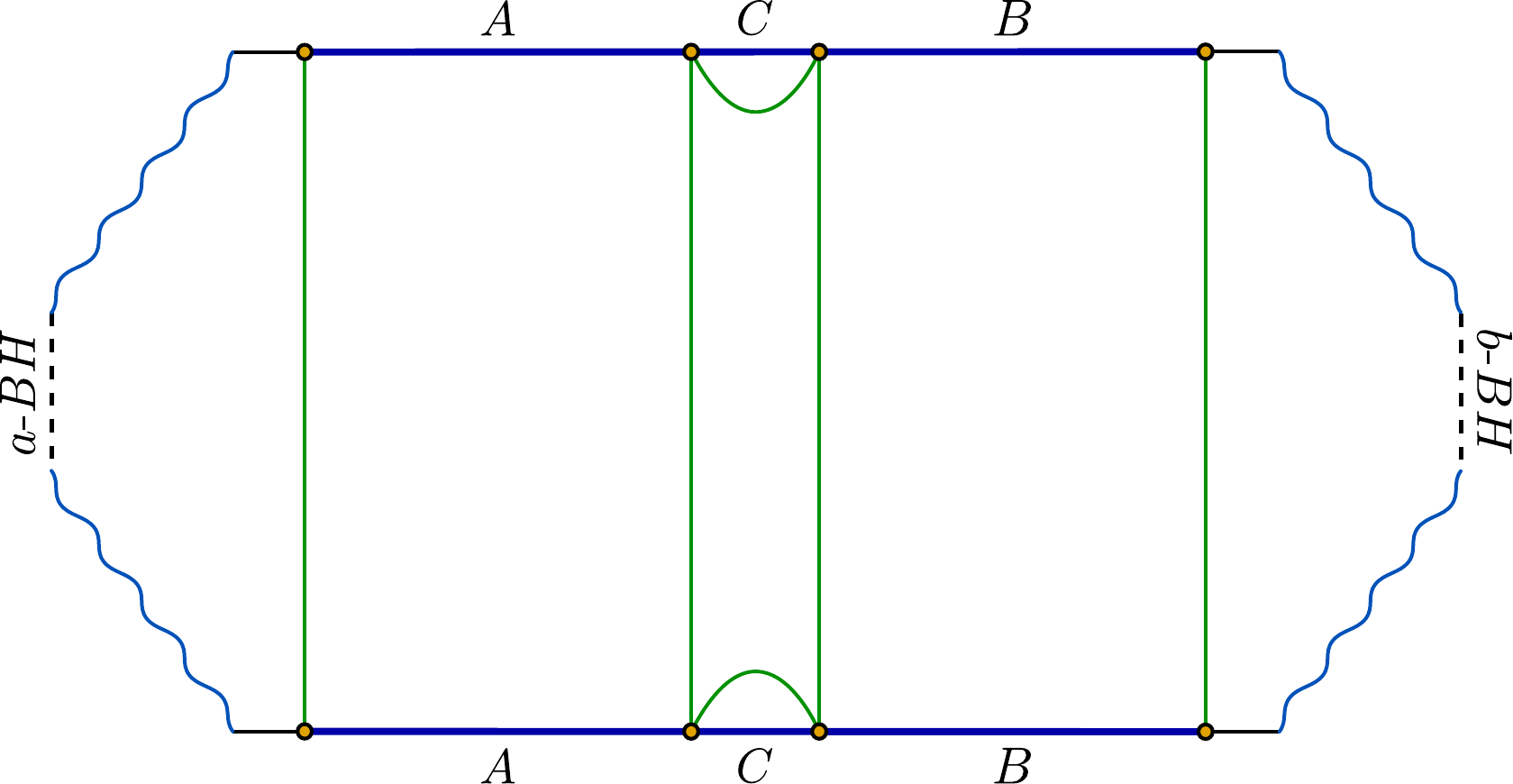}
		\caption{Phase-2a}
		\label{}
	\end{subfigure}
	\begin{subfigure}[b]{0.45\textwidth}
		\centering
		\includegraphics[width=\textwidth]{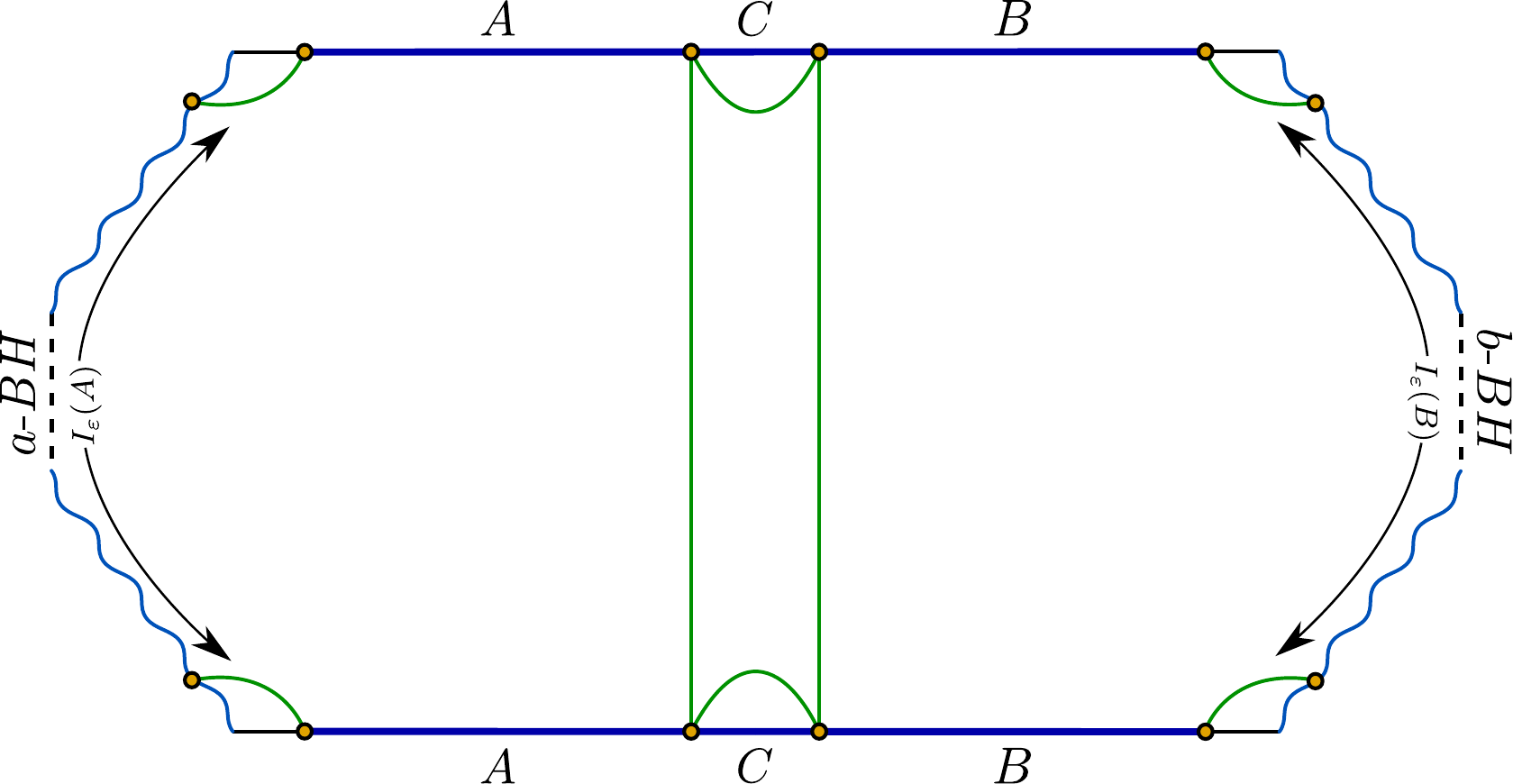}
		\caption{Phase-2b}
		\label{}
	\end{subfigure}
	\hspace{.12cm}
	\begin{subfigure}[b]{0.45\textwidth}
		\centering
		\includegraphics[width=\textwidth]{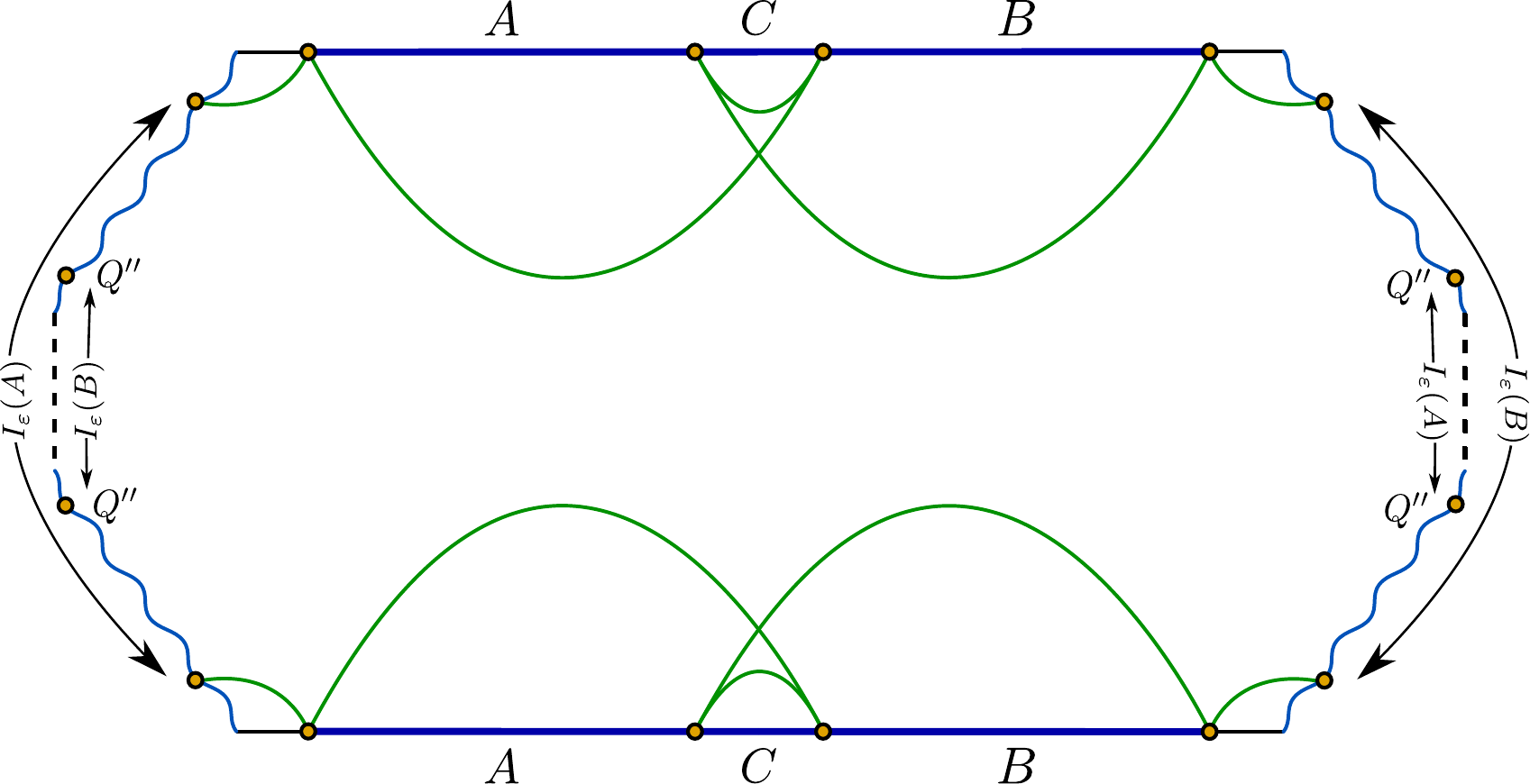}
		\caption{Phase-3}
		\label{}
	\end{subfigure}
	\caption{The diagram shows the possible phases of the entanglement negativity between two disjoint subsystems with equal sizes as time increases.}
	\label{Disjcase3(Peq1)}
\end{figure}

\noindent\textbf{Phase 1:} In the first phase (\cref{Disjcase3(Peq1)}), the dominant contributions to the entanglement entropies of all the subsystems arise from bulk-type RT surfaces each. Hence the entanglement wedges of the subsystems $A$ and $B$ are disconnected in this phase. Consequently the entanglement negativity between the subsystems in this case is zero as depicted in \cref{fig:Discase3}. On a separate note, at initial times the number of Hawking modes present in the subsystems $A$ and $B$ are very
small such that they do not lead to a significant entanglement between the subsystems.\\

\noindent\textbf{Phase 2:} This phase consists of two sub phases due to the various structures of the RT surfaces supported by the subsystems as depicted in \cref{Disjcase3(Peq1)}. In the first sub phase, the RT surfaces for the subsystems $A\cup C$, $B\cup C$ and $A\cup B\cup C$ are still identified as bulk-type whereas the subsystem $C$ now supports dome-type RT surfaces. In the second sub phase the dominant contributions to the entanglement entropies of the subsystems $A\cup C$, $B\cup C$, $A\cup B\cup C$ and $C$ arise from $a$-bulk, $b$-bulk, $ab$ and dome-type RT surfaces respectively. The increasing behaviour of the corresponding entanglement negativity profile in these two sub phases follows the interpretations which are analogous to the first phase of the adjacent case described in sub\cref{p=1}.\\

\noindent\textbf{Phase 3:} In the last phase, the Page curve for the entanglement negativity between the two disjoint subsystems $A$ and $B$ depicts a constant behaviour as exhibited in \cref{fig:Discase3}. Here the RT surfaces supported by the subsystems $A\cup C$, $B\cup C$ and $C$ are dome type each whereas the subsystem $A\cup B\cup C$ admits $ab$ type RT surfaces (\cref{Disjcase3(Peq1)}). Once again, we refer to the second phase of the adjacent case described in sub\cref{p=1} to explain the constant behaviour of the corresponding entanglement negativity profile.

\subsubsection*{(b) For $\bm{l_1 \neq l_2}$}\label{dis_p!=1}
For the case of two unequal lengths subsystems $A$ and $B$, we observe that the Page curve for the entanglement negativity between them consists of four different phases as depicted in \cref{fig:Discase3}. Once again the corresponding entanglement negativity expressions are listed in the appendix \ref{Appendix6b}. In what follows, we describe these phases in detail.\\

\begin{figure}[h!]
	\centering
	\begin{subfigure}[b]{0.45\textwidth}
		\centering
		\includegraphics[width=\textwidth]{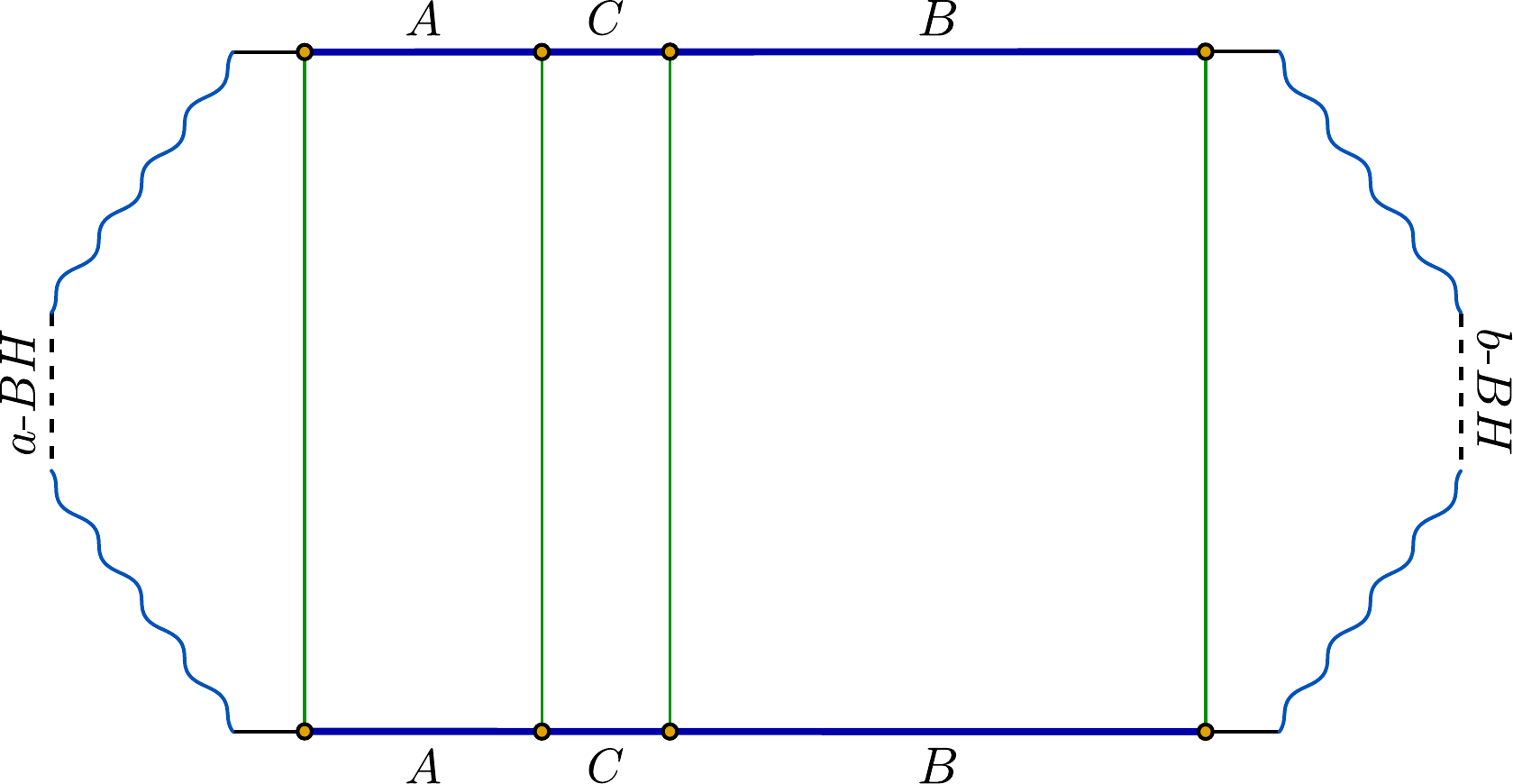}
		\caption{Phase-1}
		\label{}
	\end{subfigure}
	\vspace{.4cm}
	\hspace{.12cm}
	\begin{subfigure}[b]{0.45\textwidth}
		\centering
		\includegraphics[width=\textwidth]{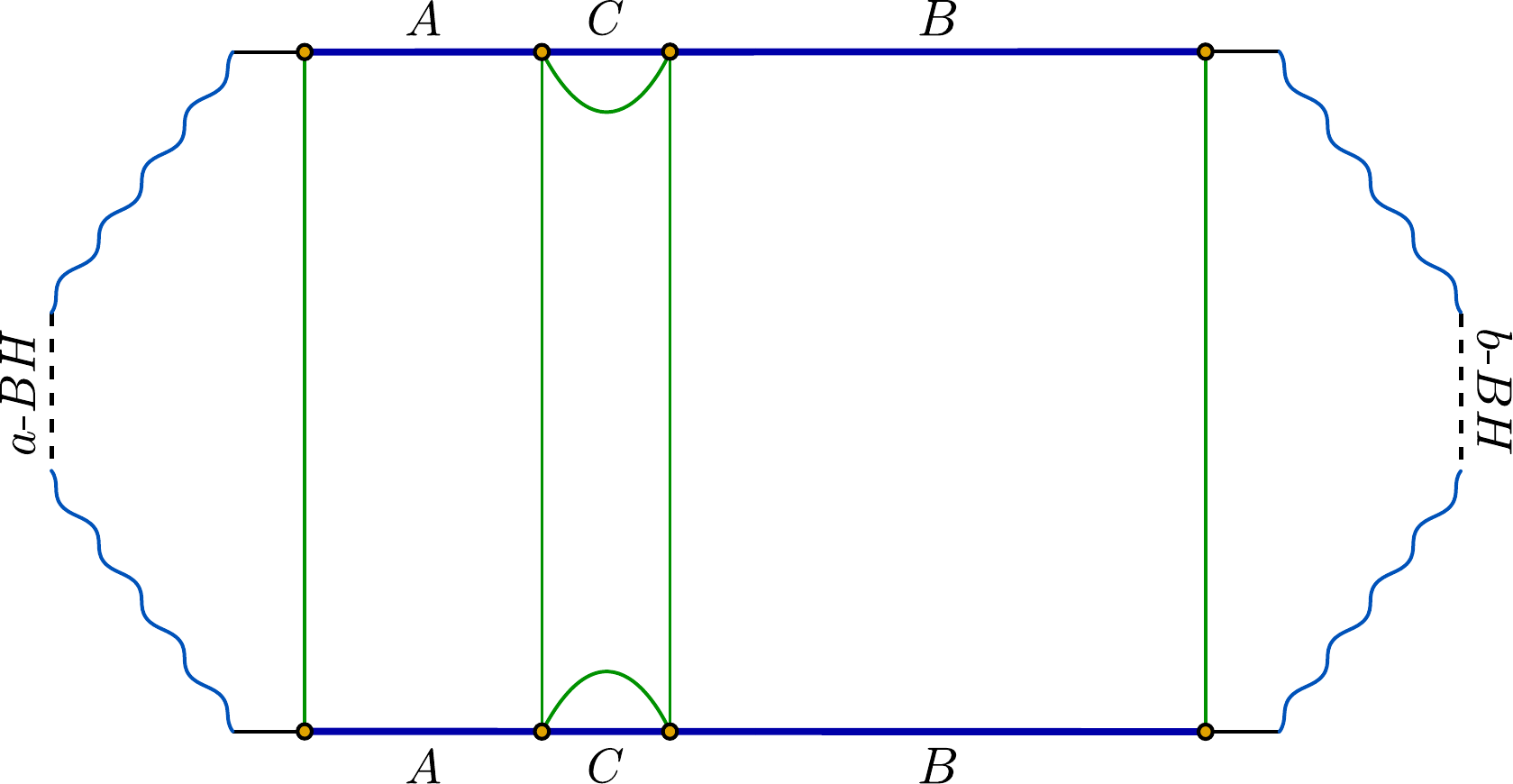}
		\caption{Phase-2a}
		\label{}
	\end{subfigure}
	\hspace{.12cm}
	\begin{subfigure}[b]{0.45\textwidth}
		\centering
		\includegraphics[width=\textwidth]{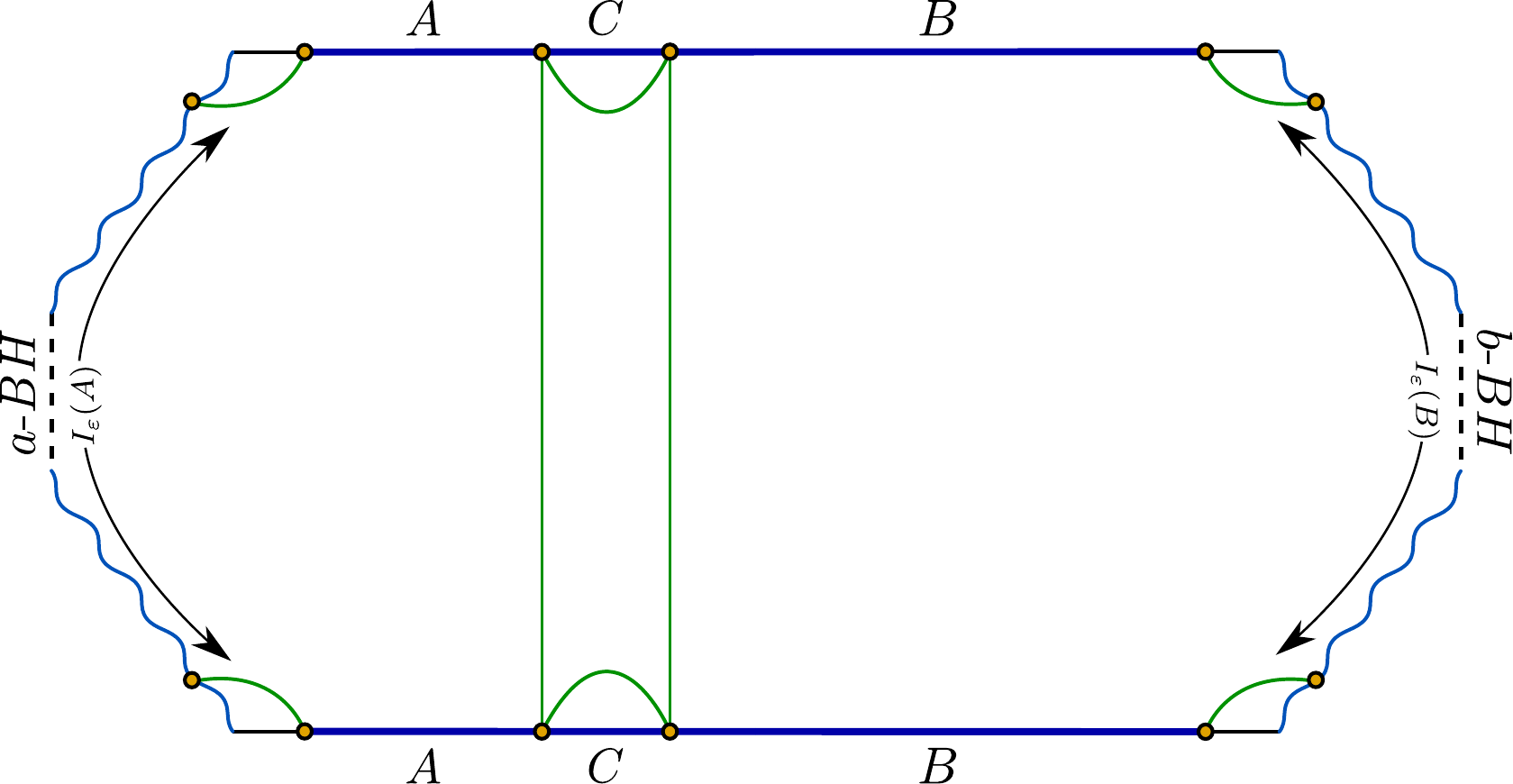}
		\caption{Phase-2b}
		\label{}
	\end{subfigure}
	\hspace{.1cm}
	\vspace{.4cm}
	\begin{subfigure}[b]{0.45\textwidth}
		\centering
		\includegraphics[width=\textwidth]{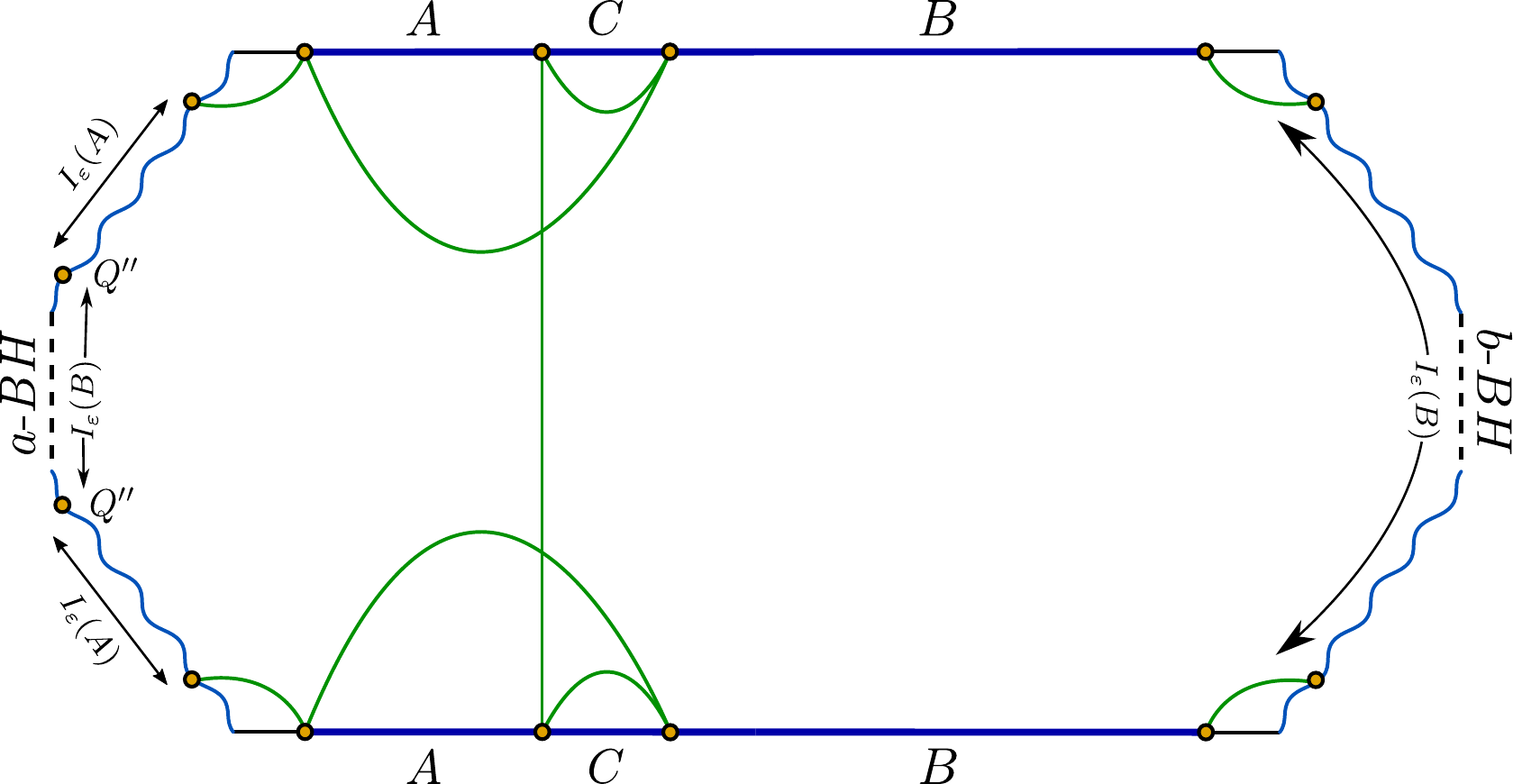}
		\caption{Phase-3}
		\label{}
	\end{subfigure}
	\begin{subfigure}[b]{0.45\textwidth}
		\centering
		\includegraphics[width=\textwidth]{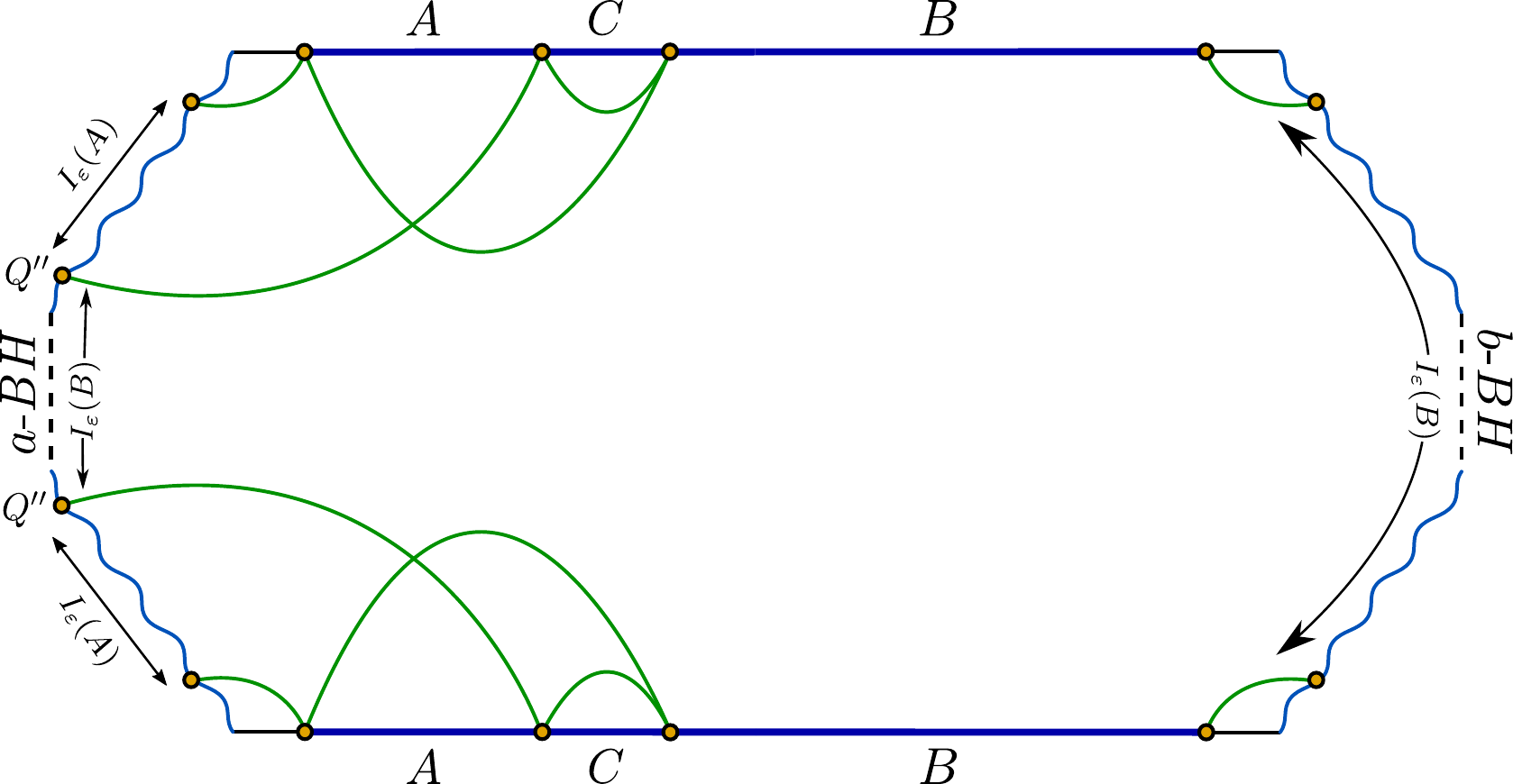}
		\caption{Phase-4}
		\label{}
	\end{subfigure}
	\caption{The possible phases of the entanglement negativity between two disjoint subsystems with unequal sizes as time increases.}
	\label{Disjcase3(PNeq1)}
\end{figure}

\noindent
\textbf{Phase 1 and Phase 2:} In these phases (\cref{Disjcase3(PNeq1)}), the RT surfaces supported by the subsystems and the description for the corresponding entanglement negativity profile are identical to the first two phases of the previous sub case.\\

\noindent\textbf{Phase 3:} In the third phase, the dominant contributions to the entanglement entropies of the subsystems $A \cup C$ and $C$ arise from dome-type RT surfaces each whereas the subsystems $B\cup C$ and $A\cup B\cup C$ admit $b$-bulk and $ab$ type RT surfaces respectively as shown in \cref{Disjcase3(PNeq1)}. This phase includes entanglement negativity island $I_{\varepsilon}(A)$ corresponding to the subsystem $A$ located in the exterior regions of the $a$-black hole. However, the entanglement negativity island $I_{\varepsilon}(B)$ corresponding to the subsystem $B$ involve the entire interior regions of both the black holes. The explanation for the increasing behaviour of the corresponding entanglement negativity profile in this phase is similar to the second phase of the adjacent case discussed in sub\cref{p!=1}.\\

\noindent\textbf{Phase 4:} Finally in the last phase (\cref{Disjcase3(PNeq1)}), the entanglement negativity between the subsystems $A$ and $B$ exhibits a constant behaviour which again may be explained similarly to that of  the third phase of the adjacent case as discussed in sub\cref{p!=1}.

\begin{figure}[H]
	\centering
	\includegraphics[width=10cm]{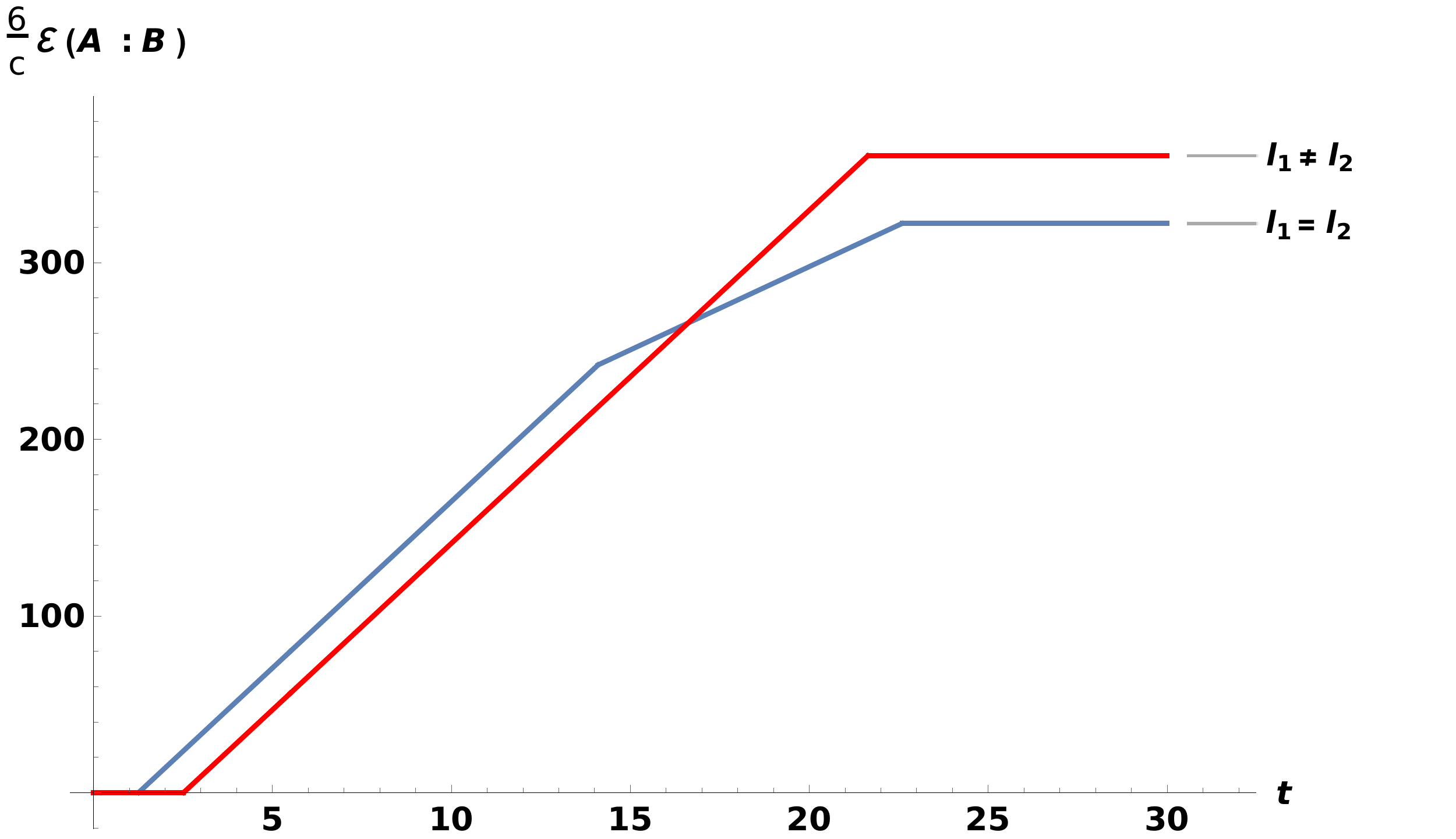}
	\caption{Page curves for entanglement negativity with respect to time $t$. Here $\beta=1$, $c=500$, $\phi_0= \frac{30c}{6}$, $\phi_r= \frac{30}{\pi}$, $L=\frac{16\pi}{\beta}$, $\epsilon=.001$, $A=[.01L,.45L]$ and $B=[.55L,.99L]$ (for $l_1=l_2$), $A=[.01L,.35L]$ and $B=[.4L,.99L]$ (for $l_1\neq l_2$).}
	\label{fig:Discase3}
\end{figure}

\section{Summary and Discussion}\label{discussion}

To summarize, we have investigated the holographic entanglement negativity for various finite temperature bipartite mixed states in a braneworld model of two communicating black holes. This construction involves two finite sized non-gravitating reservoirs coupled to two quantum dots at their boundaries at a finite temperature. These quantum dots constituted two copies of thermofield double states which interacted through the common reservoirs. The holographic dual of these quantum dots were described by JT gravity on two Plank branes with $AdS_2$ geometries. Interestingly, each non-gravitating reservoir together with a Planck brane appeared to be gravitating from the perspective of the other brane. These Planck branes involved two eternal JT black holes which were in communication  through the shared reservoirs. In this configuration, the black hole and the reservoir regions supported identical matter $CFT_2$s with transparent boundary conditions. In this context, we obtained the holographic entanglement negativity for various bipartite mixed states of two adjacent and disjoint subsystems in the reservoirs for the above configuration. In this connection, we have analyzed the profiles of the generalized entanglement negativity for different scenarios involving the subsystem sizes and the time. The behaviour of the corresponding entanglement negativity profiles observed for the above scenarios were similar to that described in \cite{KumarBasak:2021rrx} where the authors considered evaporating black holes in JT gravity through a geometrized island construction.

In appendix \ref{model1}, another model was described by a brane world geometry involving a bulk eternal $AdS_3$ BTZ black hole truncated by two \textit{Karch-Randall} (KR) branes with two dimensional black holes induced from the higher dimension. These induced  black holes were in communication through shared baths described by thermal $BCFT_2$s on a strip with different boundary conditions at either end constituting a thermofield double state \cite{Geng:2021iyq}. We have computed the entanglement entropy for a subsystem with both of its end points located in the bulk of the $BCFT_2$s for the above configuration. Furthermore we also computed the holographic entanglement negativity for two adjacent and disjoint subsystems in the bath $BCFT_2$s using the holographic proposals described in \cite{Jain:2017aqk} for different scenarios involving the subsystem sizes and the time. For these cases, we observed the behaviour of the entanglement negativity profiles similar to those described in \cite{KumarBasak:2021rrx}.

We would like to emphasize that the presence of two black holes in these braneworld models lead to the appearance of extra phases in the Page curves for both the entanglement entropy and the entanglement negativity compared to configurations which support a single black hole. Note that the braneworld models considered by us are structurally distinct although both supported black holes communicating through shared radiation reservoirs. We considered a braneworld model \cite{Balasubramanian:2021xcm} incorporating JT black holes on the Planck branes which were dual to quantum dots on either ends of finite sized $CFT_2$ radiation reservoirs. In this context it was possible to express an explicit island formula for the entanglement measures. However, in appendix \ref{model1}, we considered another braneworld model \cite{Geng:2021iyq} involving two dimensional black holes on the KR branes induced from a higher dimensional BTZ black hole where an explicit island prescription for the entanglement measures can not be utilized although the final results for the Page curves are consistent with the expected island scenario. Note that the overall behaviour of the entanglement profiles were similar in these braneworld geometries. However, there were some interesting distinctions such as for the variation of the subsystem size in one of the cases, the corresponding Page curve for the entanglement negativity involved a plateau region for the braneworld model \cite{Balasubramanian:2021xcm} due to the position independence of the entanglement entropy in contrast to the other model \cite{Geng:2021iyq}. Additionally an interesting feature of tripartite entanglement was also observed for subsystems with comparable sizes for the case of Planck braneworld geometries \cite{Balasubramanian:2021xcm}.

There are several interesting future directions to explore in connection with our results. A significant open issue involves the generalization of these braneworld construction to higher dimensions. Such an analysis although non trivial may reveal deeper insights into the structure of the mixed state entanglement in such communicating black hole/bath systems. Our analysis should also extend to models with defect $CFT$s on the EOW branes. Furthermore certain overlapping configurations observed for the entanglement negativity islands in some phases indicate some subtle characteristics of the structure of mixed state entanglement which needs more careful investigation using various toy models of black hole evaporation. We hope to return to these exciting issues in the near future.\\\\

\section{Acknowledgement}
We are grateful to Vinay Malvimat and Vinayak Raj for useful discussions. The work of GS is partially supported by the Dr. Jagmohan Garg Chair Professor position at the Indian Institute of Technology, Kanpur.

\begin{appendices}

	\section{Entanglement measures in braneworld model-II}\label{model1}
	In this section we first compute the entanglement entropy for a generic subsystem in the $BCFT_2$s \cite{Geng:2021iyq}. Note that 
	here we consider both the end points of the subsystem to be situated deep into the annular region of the $BCFT_2$s\footnote{This is in contrast to the article \cite{Geng:2021iyq} where the authors have examined the entanglement entropy for a subsystem with one end point located on the boundary.}. Subsequently, we analyze the holographic entanglement negativity for various bipartite mixed states in the $BCFT_2$s at a finite temperature. The dual bulk space time for this construction is very complicated but in the high temperature limit this reduces to an eternal $AdS_3$ BTZ black hole \cite{Maldacena:2001kr,Banados:1992wn} as demonstrated in \cite{Geng:2021iyq}. The bulk configuration involves two dimensional black holes on both the KR branes induced from the higher dimensional BTZ black hole \cite{Geng:2021iyq} but with different temperatures due to the distinct boundary conditions at the two boundaries of the two dual $BCFT_2$s. Interestingly, these entanglement measures characterize the communication between the two induced black holes on the KR branes \cite{Geng:2021iyq}. 
	
	\subsection{Entanglement entropy and Page curve}
	We begin with the computations of the different contributions to the entanglement entropy for a generic subsystem at a finite temperature as mentioned above. In this connection, we first describe the field theory analysis for the entanglement entropy in the dual $BCFT_2$s\footnote{These two $BCFT_2$s are termed as the $BCFT_L$ and the $BCFT_R$ referring to \cref{singlefinite}.} following the method discussed in \cite{Geng:2021iyq}. Furthermore we substantiate these field theory results from 
	holographic computations of the entanglement entropy in the dual bulk geometry utilizing wedge holography.
	
	\subsubsection{Field theory computations}\label{finiteTgenericA}
\begin{figure}[h!]
		\centering
		\includegraphics[scale=.85]{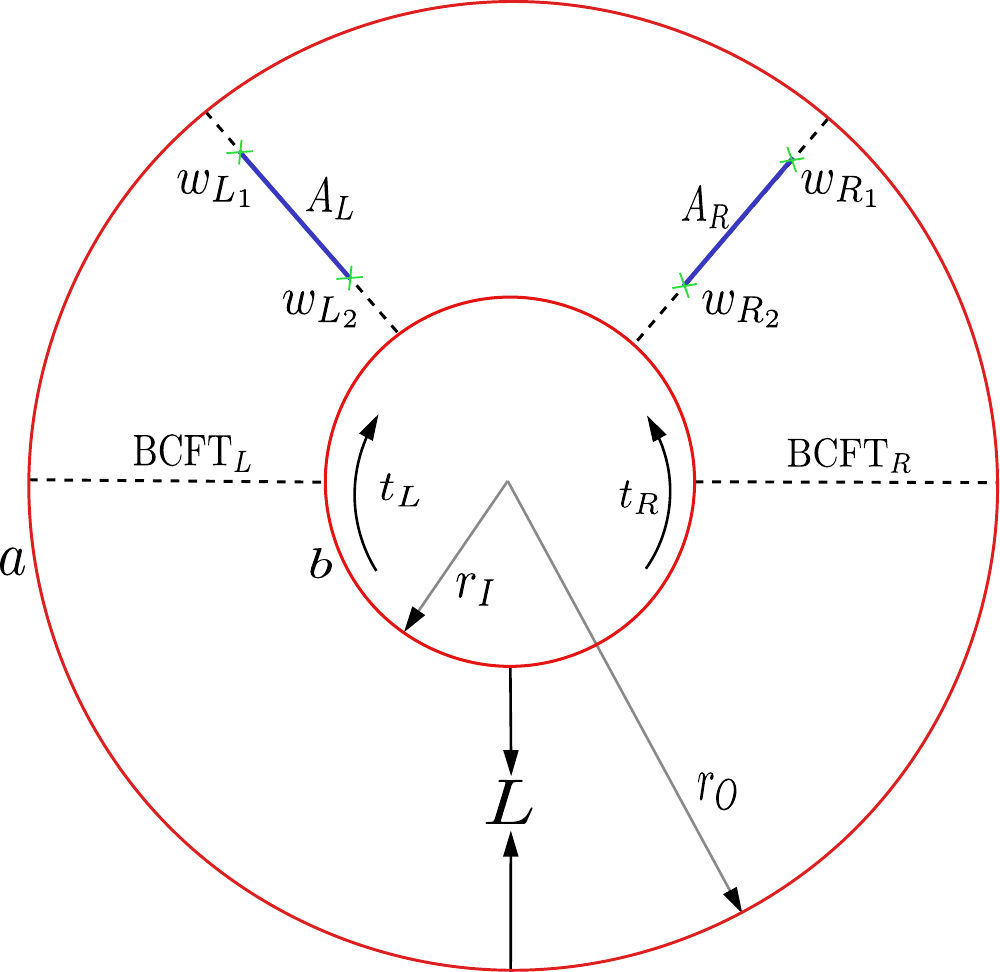}
		\caption{This schematic depicts a generic subsystem ($A=A_L\cup A_R$) with both the end points in the bulk of $BCFT_L$ and $BCFT_R$ at a constant time slice. (Figure modified from \cite{Geng:2021iyq})}\label{singlefinite}
	\end{figure}

	In this subsection, we consider a subsystem $A=A_L\cup A_R$ in the $BCFT_L$ and $BCFT_R$ with its end points located in the bulk of the annular region as shown in figure \ref{singlefinite}. These two $BCFT_2$s at a finite temperature constitute a thermofield double state as discussed in \cite{Geng:2021iyq}. The twist operators $\bar{\Phi}_n\left(w_{L_1},\bar{w}_{L_1}\right)$, $\Phi_n\left(w_{L_2},\bar{w}_{L_2}\right)$, $\Phi_n\left(w_{R_1},\bar{w}_{R_1}\right)$, and $\bar{\Phi}_n\left(w_{R_2},\bar{w}_{R_2}\right)$ in this context are situated at the end points of the subsystem $A$. Note that, the motivation behind the choice of a generic subsystem mentioned above is to explore all the possible channels for a four point twist field correlator in the entanglement entropy expression discussed in \cref{FieldTheoryEE} in contrast to the article \cite{Geng:2021iyq}. In the subsequent appendix, this would significantly lead to a rich phase structure for the holographic entanglement negativity of bipartite mixed states described by two adjacent and disjoint subsystems with all the end points located into the bulk of the bath $BCFT_2$s.
	
	The above configuration of a generic subsystem in the bath $BCFT_2$s is considered in the complex $w$-plane where the boundary conditions on the two boundaries are denoted as $a$ and $b$ with a separation $L$ between the inner ($r_I$) and the outer radii ($r_O$) as depicted in \cref{singlefinite}. At this moment, it would be beneficial to have a conformal map from annulus to a strip such that the techniques described in \cite{Geng:2021iyq} can be utilized to obtain the entanglement entropy. However, there can never be any such transformation since the annulus generally has non zero modular parameter which should be preserved under such conformal mapping \cite{Geng:2021iyq,Polchinski}. Interestingly, if we consider the high temperature limit, the above mentioned mapping can still be accomplished. This corresponds to the limit of fixed $L=r_O-r_I$ and considering the inner radius of the annulus close to zero ($r_I\slash L\rightarrow0$). The high temperature limit can also be achieved by considering the limit $r_O\rightarrow \infty$ and consequently $r_I\slash L \rightarrow0$\footnote{Specifically, this limit ($r_O\rightarrow \infty$) would be more favourable than considering the inner radius close to zero since in \cref{mapping}, $w (r_I, z)$ does not exist as $r_I\rightarrow 0$. Note that since we are considering the high temperature limit ($\beta\rightarrow0$), our results for the entanglement entropy and the entanglement negativity in the subsequent sections is correct only up to corrections $x\slash\beta$ where $x$ denotes at least some of the spatial scales. Higher order corrections to our results will correspond to the cases of finite temperature scenarios for which the bulk dual will be very complicated as mentioned in \cite{Geng:2021iyq}.}. The entanglement entropy of the subsystem $A=A_L\cup A_R$ can now be computed using a mapping of the $w$-plane to the UHP following the transformation 
	\begin{equation}\label{mapping}
		w=r_I\left(	\frac{1}{z-\frac{i}{2}}-i\right)\,,
	\end{equation} 
	with $r_I$ being the inner radius of the annulus. Note that the corresponding transformation maps the inner and the outer radii to the real axis and to a point $z=\frac{i}{2}$ respectively on the UHP. This reduces the computation of the four point twist correlator on the annulus to that on the UHP. Note that here we restrict to a subset of the available channels with the assumption that the pair of intervals in the TFD copy always mapped onto each other by a rotation of the annulus \footnote{We would like to thank the referee for this crucial comment.}. Consequently, utilizing the transformation in \cref{mapping}, we obtain the contributions to the entanglement entropy from the four point function as follows
	\begin{equation}\label{FieldTheoryEE}
		S_A=\lim_{n\rightarrow1}\frac{1}{1-n}\log(\left<\Phi_n\left(z_{R_1},\bar{z}_{R_1}\right)\bar{\Phi}_n\left(z_{R_2},\bar{z}_{R_2}\right)\Phi_n\left(z_{L_2},\bar{z}_{L_2}\right)\bar{\Phi}_n\left(z_{L_1},\bar{z}_{L_1}\right) \right>^b _{\text{UHP},a})\,,
	\end{equation} 
	where the lower and the upper indices $a,b$ represent the boundary conditions corresponding to the two boundaries of the $BCFT_2$s. It is possible to identify seven distinct contributions to the corresponding entanglement entropy from the connected and disconnected channels for the four point function.
	
	$\bm{(a)}$ We first discuss one of the connected channels which may be obtained by considering the OPEs of the twist operators located on the $BCFT_L$ and $BCFT_R$. Therefore, the entanglement entropy in this channel is expressed as 
	\begin{align}\label{bulk-type}
		S^\text{bulk}_{A}=&\lim_{n\rightarrow1}\frac{1}{1-n}\log(\left<\Phi_n\left(z_{R_1},\bar{z}_{R_1}\right)\bar{\Phi}_n\left(z_{R_2},\bar{z}_{R_2}\right)\Phi_n\left(z_{L_2},\bar{z}_{L_2}\right)\bar{\Phi}_n\left(z_{L_1},\bar{z}_{L_1}\right) \right>^b _{\text{UHP},a})\notag\\
		=&\lim_{n\rightarrow1}\frac{1}{1-n}\log\Big(\left<\bar{\Phi}_n\left(z_{L_1},\bar{z}_{L_1}\right)\Phi_n\left(z_{R_1},\bar{z}_{R_1}\right)\right>_{\text{UHP}}\left<\bar{\Phi}_n\left(z_{R_2},\bar{z}_{R_2}\right)\Phi_n\left(z_{L_2},\bar{z}_{L_2}\right) \right> _{\text{UHP}}\Big)\notag\\
		=&\frac{c}{3}\ln\left(\frac{2 r_1}{\epsilon}\cosh\frac{2 \pi t}{\beta}\right)+\frac{c}{3}\ln\left(\frac{2 r_2}{\epsilon}\cosh\frac{2 \pi t}{\beta}\right)\,,
	\end{align}
	where $r_1$ and $r_2$ define the endpoints of the subsystem on the $w$-plane. In the second line, we have utilized a large $c$ factorization of the four point function into two 2-point functions on the UHP as discussed in the articles \cite{Coser:2014gsa,Malvimat:2018cfe,Malvimat}.  The UHP twist correlators in the second line of the above expression is then computed in the whole complex plane \footnote{Similarly, in rest of the cases, we compute the UHP twist correlators in the full complex plane.}. Subsequently, in the last line is obtained through the transformations $w_L=r \exp(i\theta)\,,\,w_R=r \exp(i\pi-i\theta)$ followed by an analytic continuation of the Euclidean time $t=\frac{i \theta \beta}{2 \pi}$ on the $w$-plane \cite{Geng:2021iyq}. We utilize this same process in all the following cases.
	
	$\bm{(b)}$ Next we compute the contribution to the entanglement entropy from a disconnected channel which may be obtained through the BOEs of all the twist operators in \cref{FieldTheoryEE} with respect to one of the two boundaries on the UHP \footnote{In this case, we consider that both the endpoints of the subsystem are much closer to the boundary rather than being closer to each other. However, when the endpoints comes close to each other compared to the boundary, dome type contribution to the entanglement entropy dominates (described in configuration (g)) which we do not consider here. Similar arguments can be implemented in the other scenarios.},
	\begin{equation}\label{boundarychannel}
		\begin{aligned}
			S^{bb}_{A}=&\lim_{n\rightarrow1}\frac{1}{1-n}\log(\left<\bar{\Phi}_n\left(z_{L_1},\bar{z}_{L_1}\right)\Phi_n\left(z_{R_1},\bar{z}_{R_1}\right)\bar{\Phi}_n\left(z_{R_2},\bar{z}_{R_2}\right)\Phi_n\left(z_{L_2},\bar{z}_{L_2}\right)\right>^b_\text{UHP} )\,\\
			=&\frac{c}{3}\ln\left(\frac{r_1 ^2-r_I ^2}{r_I \epsilon}\right)+\frac{c}{3}\ln\left(\frac{r_2 ^2-r_I ^2}{r_I \epsilon}\right)+4\ln\left(g_b\right)\,.
		\end{aligned}	
	\end{equation}
	In the above equation, the superscript refers to the BOEs of the twist operators corresponding to the boundary $b$ and $\ln\left(g_b\right)$ corresponds to the boundary degrees of freedom called the \textit{boundary entropy ($S_{bdy}$)} which strictly depends on the boundary condition $b$ \footnote{As described in \cite{Geng:2021iyq}, the doubling trick corresponds to non vanishing of the one point function in eq. (2.9) of \cite{Geng:2021iyq} on the UHP with conformal weights $h=\bar{h}$ and it is constrained to have a similar form as a chiral two point function on the full complex plane  \cite{Geng:2021iyq,Sully:2020pza}. The coefficient described in eq. (2.10) of \cite{Geng:2021iyq} of this two point function implies the dependence on the boundary condition. Finally, one may compute the entanglement entropy for the subsystem $A$ of length $L_A$ as
		\begin{equation}\label{entropy2}
			S_A=\frac{c}{6}\ln\frac{2 L_A}{\epsilon}+\ln\left(g_b\right)\,,
		\end{equation}    
		where $\epsilon$ is a UV cutoff in the $BCFT_2$. In the above equation, the first term is the kinematic
		term which can be produced utilizing the standard doubling trick.}.
	
	$\bm{(c)}$ A similar computation of the entanglement entropy may be performed by considering the BOEs of all the twist operators with respect to the other boundary located at the operator insertion point $z=\frac{i}{2}$ on the UHP. Hence the corresponding entanglement entropy in this disconnected channel may be obtained on the $w$-plane as,
	\begin{align}
		S^{aa}_{A}=&\lim_{n\rightarrow1}\frac{1}{1-n}\log\Big(\left<\Phi_n\left(z_{R_1},\bar{z}_{R_1}\right)\bar{\Phi}_n\left(z_{L_1},\bar{z}_{L_1}\right)\Phi_n\left(z_{R_2},\bar{z}_{R_2}\right)\bar{\Phi}_n\left(z_{L_2},\bar{z}_{L_2}\right) \Phi_a\left(z_a,\bar{z}_a\right)\right>_\text{UHP}\Big)\notag\\
		=&\frac{c}{3}\ln\left(\frac{r_O ^2-r_1 ^2}{r_O \epsilon}\right)+\frac{c}{3}\ln\left(\frac{r_O ^2-r_2 ^2}{r_O \epsilon}\right)+4\ln\left(g_a\right)\,.
	\end{align} 
	
	$\bm{(d)}$ Now we discuss the case where both the OPE and the BOE of the twist operators may be considered simultaneously. The entanglement entropy in this connected channel may be written as
	\begin{align}\label{boundarybulk1}
		S^{b\text{-bulk}}_{A}=&\lim_{n\rightarrow1}\frac{1}{1-n}\log(\left<\bar{\Phi}_n\left(z_{L_1},\bar{z}_{L_1}\right)\Phi_n\left(z_{R_1},\bar{z}_{R_1}\right)\right>^b_\text{UHP}\left<\bar{\Phi}_n\left(z_{R_2},\bar{z}_{R_2}\right)\Phi_n\left(z_{L_2},\bar{z}_{L_2}\right)\right>_\text{UHP} )\notag\\
		=&\frac{c}{3}\ln\left(\frac{2 r_2}{\epsilon}\cosh\frac{2 \pi t}{\beta}\right)+\frac{c}{3}\ln\left(\frac{r_1 ^2-r_I ^2}{r_I \epsilon}\right)+2\ln\left(g_b\right)\,,
	\end{align}	
	where we have performed BOEs for the twist operators  $\Phi_n\left(z_{R_1},\bar{z}_{R_1}\right)\,,\,\bar{\Phi}_n\left(z_{L_1},\bar{z}_{L_1}\right)$ and considered the OPE of the rest $\bar{\Phi}_n\left(z_{R_2},\bar{z}_{R_2}\right)\,,\,\Phi_n\left(z_{L_2},\bar{z}_{L_2}\right)$ on the UHP. Once again, in the last line we have utilized the transformations $w_L=r \exp(i\theta)\,,\,w_R=r \exp(i\pi-i\theta)$ followed by an analytic continuation of the Euclidean time $t=\frac{i \theta \beta}{2 \pi}$ on the $w$-plane. 
	
	$\bm{(e)}$ Conversely we may employ the OPE of the twist operators $\Phi_n\left(z_{R_1},\bar{z}_{R_1}\right)\,,\,\bar{\Phi}_n\left(z_{L_1},\bar{z}_{L_1}\right)$ and BOEs to the remaining twist operators, resulting into an another possibility of the entanglement entropy in the connected channel as follows
	\begin{align}\label{boundarybulk2}
		S^{a\text{-bulk}}_{A}=&\lim_{n\rightarrow1}\frac{1}{1-n}\log\Big(\left<\bar{\Phi}_n\left(z_{L_1},\bar{z}_{L_1}\right)\Phi_n\left(z_{R_1},\bar{z}_{R_1}\right)\right>_\text{UHP}\notag\\
		&\qquad\qquad\qquad\left<\bar{\Phi}_n\left(z_{R_2},\bar{z}_{R_2}\right)\Phi_n\left(z_{L_2},\bar{z}_{L_2}\right)\Phi_a\left(z_a,\bar{z}_a\right)\right>_\text{UHP}\Big) \nonumber\\
		=&\frac{c}{3}\ln\left(\frac{2 r_1}{\epsilon}\cosh\frac{2 \pi t}{\beta}\right)+\frac{c}{3}\ln\left(\frac{r_O ^2-r_2 ^2}{r_I \epsilon}\right)+2\ln\left(g_a\right)\,.	
	\end{align}
	Note that, we have performed the corresponding BOEs in the above equation with respect to the boundary $a$. 
	
	$\bm{(f)}$ An another case may also be analyzed by considering the BOEs of the twist operators involving both the boundaries on the UHP simultaneously. Hence, the entanglement entropy corresponding to this connected channel may be reduced to 
	\begin{align}
		\begin{split}
			S^{ab}_{A}=&\lim_{n\rightarrow1}\frac{1}{1-n}\log\Big( \left<\bar{\Phi}_n\left(z_{L_1},\bar{z}_{L_1}\right)\Phi_n\left(z_{R_1},\bar{z}_{R_1}\right)\right>^b_\text{UHP}\Big. \\
			&\Big.~~~~~~~~~~~~~~~~~\left<\bar{\Phi}_n\left(z_{R_2},\bar{z}_{R_2}\right)\Phi_n\left(z_{L_2},\bar{z}_{L_2}\right)\Phi_a\left(z_a,\bar{z}_a\right)\right>_\text{UHP}\Big) \,\\
			=&\frac{c}{3}\ln\left(\frac{r_1 ^2-r_I ^2}{r_I \epsilon}\right)+\frac{c}{3}\ln\left(\frac{r_O ^2-r_2 ^2}{r_I \epsilon}\right)+2\ln\left(g_a\right)+2\ln\left(g_b\right)\,.
		\end{split}
	\end{align}
	
	$\bm{(g)}$ We now discuss a special case of the disconnected channels where the contribution to the entanglement entropy may be computed from the OPE of the twist operators $\bar{\Phi}_n\left(w_{L_1},\bar{w}_{L_1}\right)$, $\Phi_n\left(w_{L_2},\bar{w}_{L_2}\right)$ in the $BCFT_L$ and similarly for $\bar{\Phi}_n\left(w_{R_1},\bar{w}_{R_1}\right)$, $\Phi_n\left(w_{R_2},\bar{w}_{R_2}\right)$ in the $BCFT_R$ \footnote{The authors of \cite{Geng:2021iyq} did not encounter this type of contribution to the entanglement entropy due to the choice of the subsystem.}. The expression for the entanglement entropy in this channel is then given as
	\begin{align}
		S^{\text{dome}}_{A}=&\lim_{n\rightarrow1}\frac{1}{1-n}\log\Big(\left<\bar{\Phi}_n\left(z_{L_1},\bar{z}_{L_1}\right)\Phi_n\left(z_{L_2},\bar{z}_{L_2}\right)\right>_\text{UHP}\left<\Phi_n\left(z_{R_1},\bar{z}_{R_1}\right)\bar{\Phi}_n\left(z_{R_2},\bar{z}_{R_2}\right)\right>_\text{UHP}\Big)\notag\\
		=&\frac{c}{3}\ln \left(\frac{\beta }{\pi  \epsilon } \sinh \frac{\pi(r_2 - r_1) } {\beta }\right)\,.
	\end{align}
	
	Finally, the entanglement entropy of a generic subsystem $A$ may be obtained by considering the minimum of the above contributions.
	\begin{equation}
		\begin{aligned}
			S_A=\text{min}\left(S^\text{bulk}_{A},S^{bb}_{A},S^{aa}_{A},S^{b\text{-bulk}}_{A},S^{a\text{-bulk}}_{A},S^{ab}_{A},S^\text{dome}_{A}\right)\,.
		\end{aligned}
	\end{equation}
	In the following subsection, we will compute the corresponding entanglement entropy from the dual bulk geometry through wedge holography which will substantiate the above field theory results.

	\subsubsection{Holographic computations}\label{holographicEEs}
	We now compute the holographic entanglement entropy for a generic subsystem $A$ at a finite temperature using the areas of the RT surfaces described in \cite{Geng:2021iyq} for the bulk dual geometry involving an eternal $AdS_3$ BTZ black hole with two dimensional KR branes \cite{Geng:2021iyq}. Once more it is possible to identify seven contributions to the entanglement entropy of the subsystem $A$ arising from distinct bulk RT surfaces. In what follows we describe
	the areas of these various RT surfaces for the subsystem $A$ as mentioned above. In this context we have considered both the asymptotic regions of the bulk BTZ geometry for computing the areas of the corresponding RT surfaces in contrast to \cite{Geng:2021iyq} where only a single asymptotic region was considered and finally the authors doubled the results to obtain the areas of the RT surfaces.
	
\begin{figure}
		\centering
		\begin{subfigure}{0.38\textwidth}
			\centering
			\includegraphics[width=\textwidth]{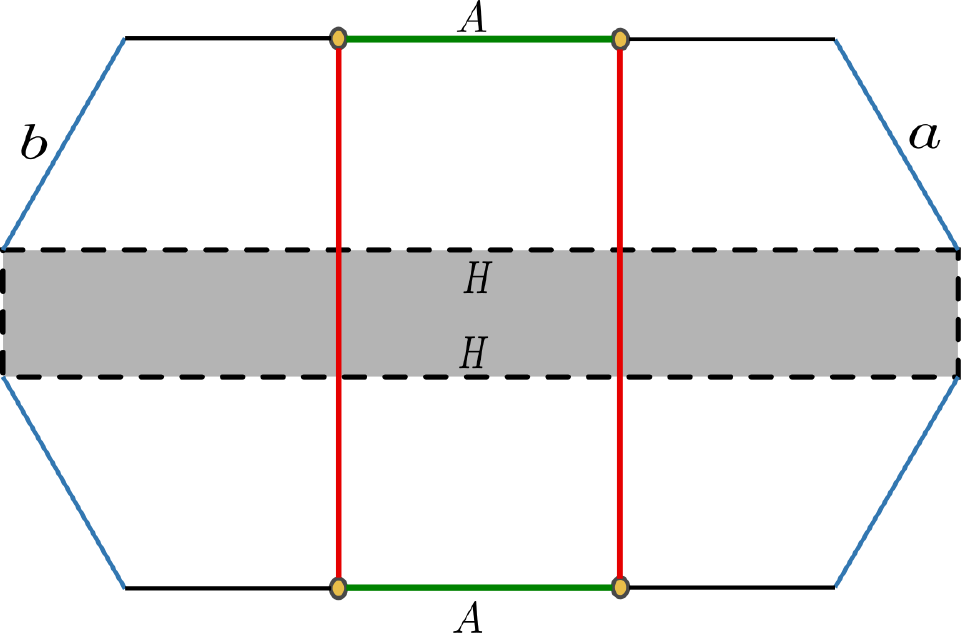}
			\caption{}
			\label{holographicsingle1}
		\end{subfigure}
		\hspace{1.5cm}
		\vspace{.2cm}
		\begin{subfigure}{0.38\textwidth}
			\centering
			\includegraphics[width=\textwidth]{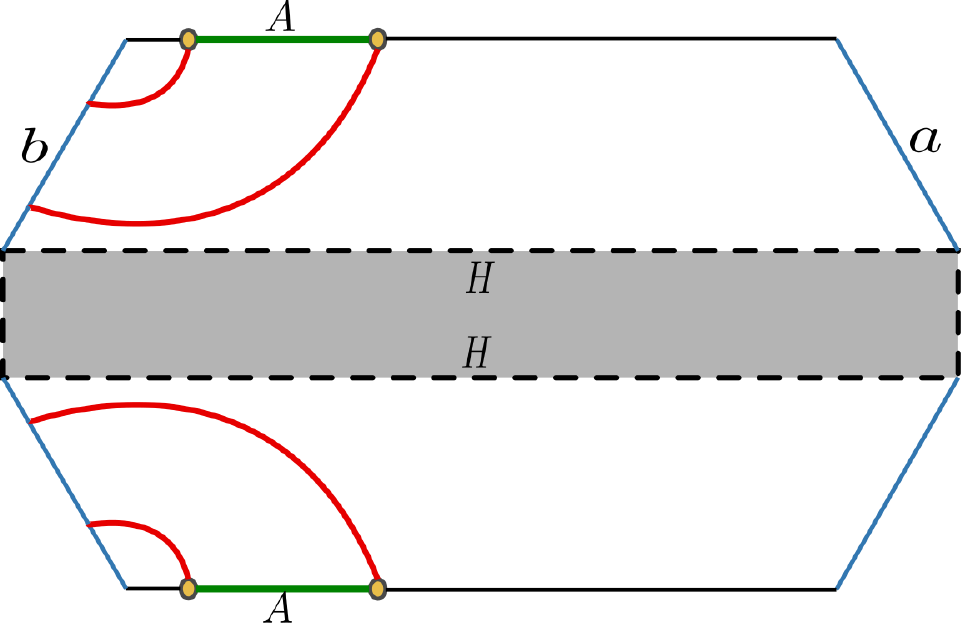}
			\caption{}
			\label{holographicsingle2}
		\end{subfigure}
		\hspace{1.5cm}
		\vspace{.2cm}
		\begin{subfigure}[b]{0.38\textwidth}
			\centering
			\includegraphics[width=\textwidth]{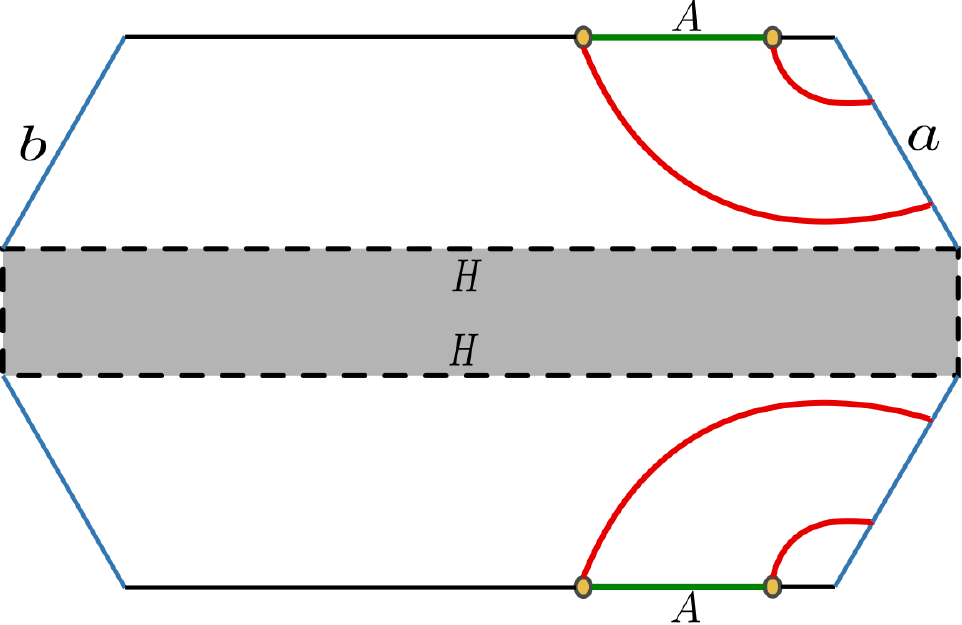}
			\caption{}
			\label{holographicsingle3}
		\end{subfigure}
		\hspace{1.5cm}
		\begin{subfigure}[b]{0.38\textwidth}
			\centering
			\includegraphics[width=\textwidth]{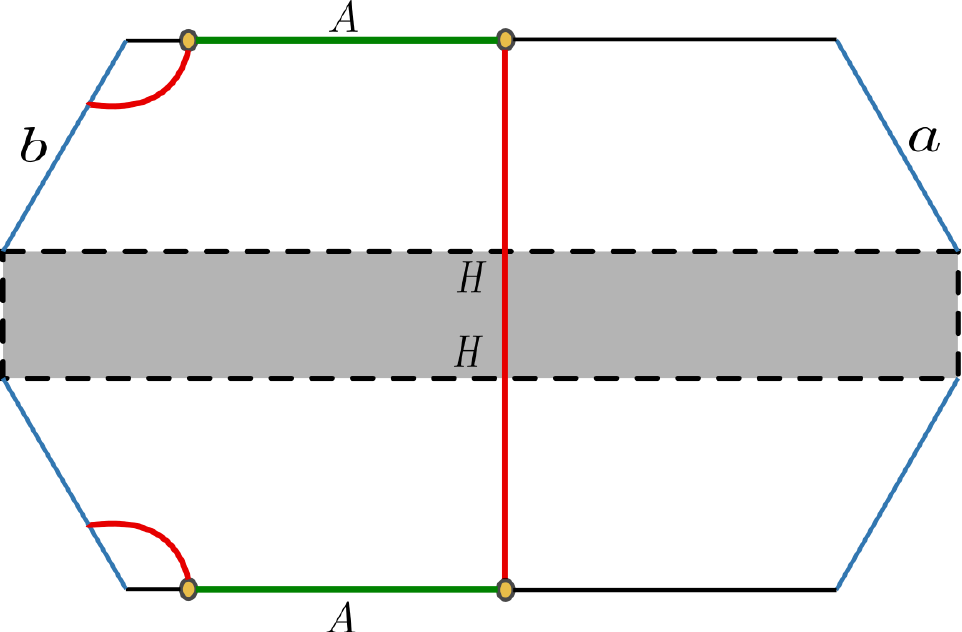}
			\caption{}
			\label{holographicsingle4}
		\end{subfigure}
		\hspace{1.5cm}
		\vspace{.2cm}
		\begin{subfigure}[b]{0.38\textwidth}
			\centering
			\includegraphics[width=\textwidth]{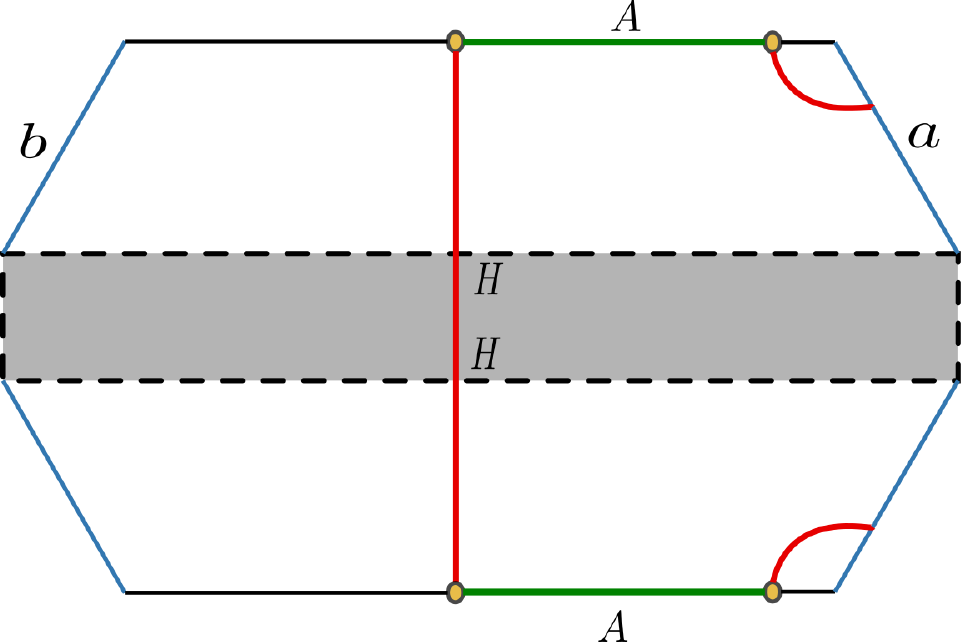}
			\caption{}
			\label{holographicsingle5}
		\end{subfigure}
		\hspace{1.5cm}
		\begin{subfigure}[b]{0.38\textwidth}
			\centering
			\includegraphics[width=\textwidth]{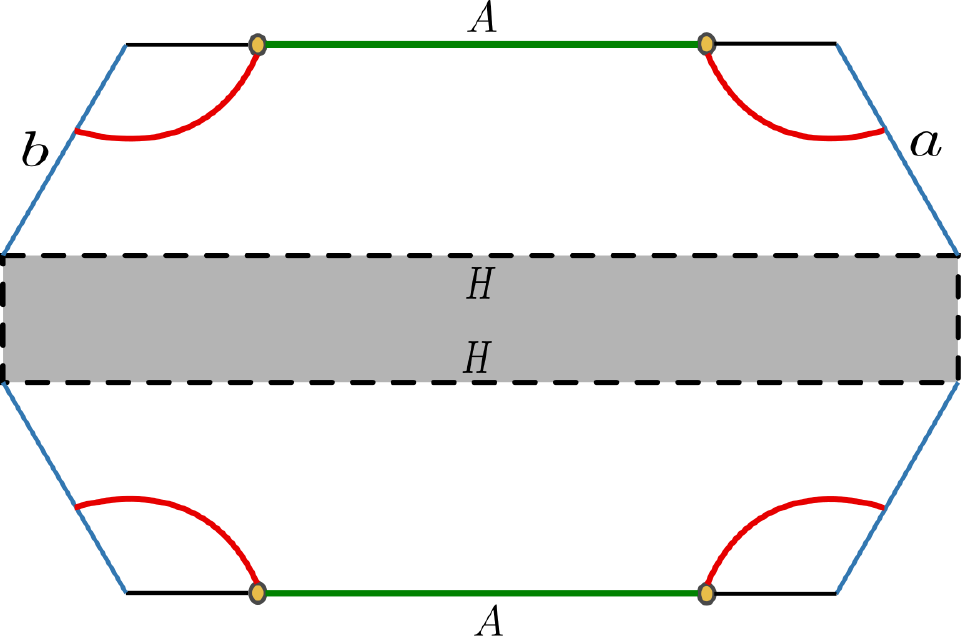}
			\caption{}
			\label{holographicsingle6}
		\end{subfigure}
		\hspace{1.5cm}
		\begin{subfigure}[b]{0.38\textwidth}
			\centering
			\includegraphics[width=\textwidth]{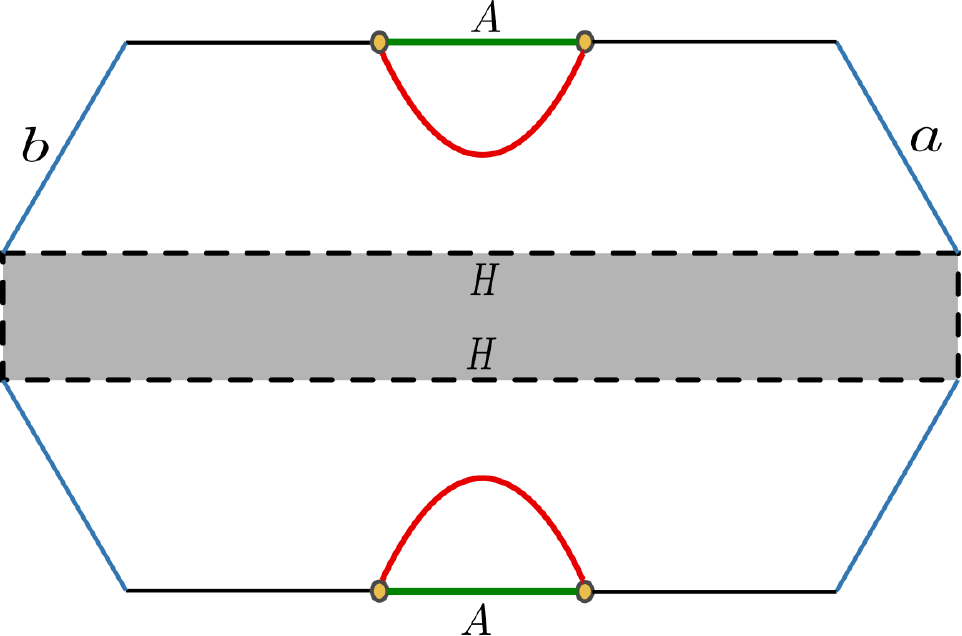}
			\caption{}
			\label{holographicsingle7}
		\end{subfigure}
		\caption{Diagram depicts all the possible RT surfaces corresponding to the subsystem $A$. Here the two asymptotic boundaries and the horizons of the bulk eternal BTZ black hole are denoted by the black solid lines and the grey shaded region respectively whereas the two KR branes are shown by the blue lines labeled as $a$ and $b$ corresponding to the different boundary conditions of the dual $BCFT_2$s. (Figure modified from \cite{Geng:2021hlu,Geng:2021iyq})}
		\label{holographic}
	\end{figure}

	$\bm{(a)}$ We start with the RT surfaces for the subsystem $A$ which consists of two HM surfaces connecting its end points in the two asymptotic boundaries as illustrated in the \cref{holographicsingle1}. We call this RT surface as bulk-type with an area contribution as follows
	\begin{equation}\label{HM}
		A_{\text{bulk}}=2\ln\left(\frac{2 r_1}{\epsilon}\cosh\frac{2 \pi t}{\beta}\right)+2\ln\left(\frac{2 r_2}{\epsilon}\cosh\frac{2 \pi t}{\beta}\right)\,.
	\end{equation} 
	
	$\bm{(b)}$ Next we consider the $bb$-type RT surfaces where both the geodesics start from $\partial A$ and end on the $b$-brane as shown in the \cref{holographicsingle2}. The corresponding area contribution may be obtained as
	\begin{equation}\label{bbtype}
		A_{bb}= 2\ln\left(\frac{r_1 ^2-r_I ^2}{r_I \epsilon}\right)+2\ln\left(\frac{r_2 ^2-r_I ^2}{r_I \epsilon}\right)+4\frac{6}{c}\ln\left(g_b\right)\,.
	\end{equation}
	
	$\bm{(c)}$ Similar to above case, if both the geodesics start from $\partial A$ and end on the $a$-brane, they are termed as the $aa$-type RT surfaces as described in \cref{holographicsingle3}. The area for these RT surfaces is given by
	\begin{equation}\label{aatype}
		A_{aa}= 2\ln\left(\frac{r_O ^2-r_1 ^2}{r_O \epsilon}\right)+2\ln\left(\frac{r_O ^2-r_2 ^2}{r_O \epsilon}\right)+4\frac{6}{c}\ln\left(g_a\right)\,.
	\end{equation} 
	
	$\bm{(d)}$ Next we describe the contribution to the EE of the subsystem $A$ from $b$-bulk type RT surfaces (\cref{holographicsingle4}) where both the geodesics start from $\partial A$, however two of them end on the $b$-brane while the other geodesic stretches between the two asymptotic boundaries as a HM surface. The area for this contribution is expressed as
	\begin{equation}\label{bbulk}
		A_{\text{$b$-bulk}}= 2\ln\left(\frac{2 r_2}{\epsilon}\cosh\frac{2 \pi t}{\beta}\right)+2\ln\left(\frac{r_1 ^2-r_I ^2}{r_I \epsilon}\right)+2\frac{6}{c}\ln\left(g_b\right)\,.
	\end{equation}
	
	$\bm{(e)}$ These RT surfaces are similar to the previous one but with the geodesics now ending on the $a$-brane instead of the $b$-brane (\cref{holographicsingle5}). We term these RT surfaces as $a$-bulk type with an area
	\begin{equation}\label{abulk}
		A_{\text{$a$-bulk}}= 2\ln\left(\frac{2 r_1}{\epsilon}\cosh\frac{2 \pi t}{\beta}\right)+2\ln\left(\frac{r_O ^2-r_2 ^2}{r_I \epsilon}\right)+2\frac{6}{c}\ln\left(g_a\right)\,.
	\end{equation}
	
	$\bm{(f)}$ We now consider the $ab$-type RT surfaces where the geodesics start from $\partial A$ and end on the two KR branes as depicted in the \cref{holographicsingle6} with an area contribution as follows
	\begin{equation}\label{abtype}
		A_{ab}= 2\ln\left(\frac{r_1 ^2-r_I ^2}{r_I \epsilon}\right)+2\ln\left(\frac{r_O ^2-r_2 ^2}{r_I \epsilon}\right)+2\frac{6}{c}\ln\left(g_a\right)+2\frac{6}{c}\ln\left(g_b\right)\,.
	\end{equation} 
	
	$\bm{(g)}$ Finally, we have the dome-type RT surfaces described by the geodesics homologous to the subsystem $A$ on both the asymptotic boundaries (\cref{holographicsingle7}). At finite temperature this configuration describes an area contribution to the EE given by
	\begin{equation}\label{dome}
		A_{\text{dome}}=4\ln \left(\frac{\beta }{\pi  \epsilon } \sinh \frac{\pi  (r_2 - r_1)}{\beta }\right)\,.
	\end{equation}
	
	Now we may compute the holographic entanglement entropy for the subsystem $A$ in the dual $BCFT_2$s by utilizing the wedge holography relation discussed in eq. (4.4) of	\cite{Geng:2021iyq} and \cref{aatype,bbtype,abtype,HM,abulk,bbulk,dome} \footnote{In the entropy expressions, we have identified the boundary entropies as $S_{bdya}=\ln (g_a)$ and $S_{bdyb}=\ln (g_b)$.}
	\begin{equation}
		\begin{aligned}\label{holographicentropy}
			S_A=&\text{min}\left(\frac{A_{\text{bulk}}}{4 G_N ^{(3)}},\frac{A_{bb}}{4 G_N ^{(3)}},\frac{A_{aa}}{4 G_N ^{(3)}},\frac{A_{b\text{-bulk}}}{4 G_N ^{(3)}},\frac{A_{a\text{-bulk}}}{4 G_N ^{(3)}},\frac{A_{ab}}{4 G_N ^{(3)}},\frac{A_{dome}}{4 G_N ^{(3)}}\right)\,\\ 
			=&\text{min}\left(S^\text{bulk}_{A},S^{bb}_{A},S^{aa}_{A},S^{b\text{-bulk}}_{A},S^{a\text{-bulk}}_{A},S^{ab}_{A},S^\text{dome}_{A}\right)\,.
		\end{aligned}
	\end{equation}

	In the following, we plot the entanglement entropies corresponding to the RT surfaces detailed above with respect to the size of the subsystem $A$ and time $t$.
	\begin{figure}[H]
		\centering
		\begin{subfigure}[t]{.45\textwidth}
			\includegraphics[width=1\linewidth]{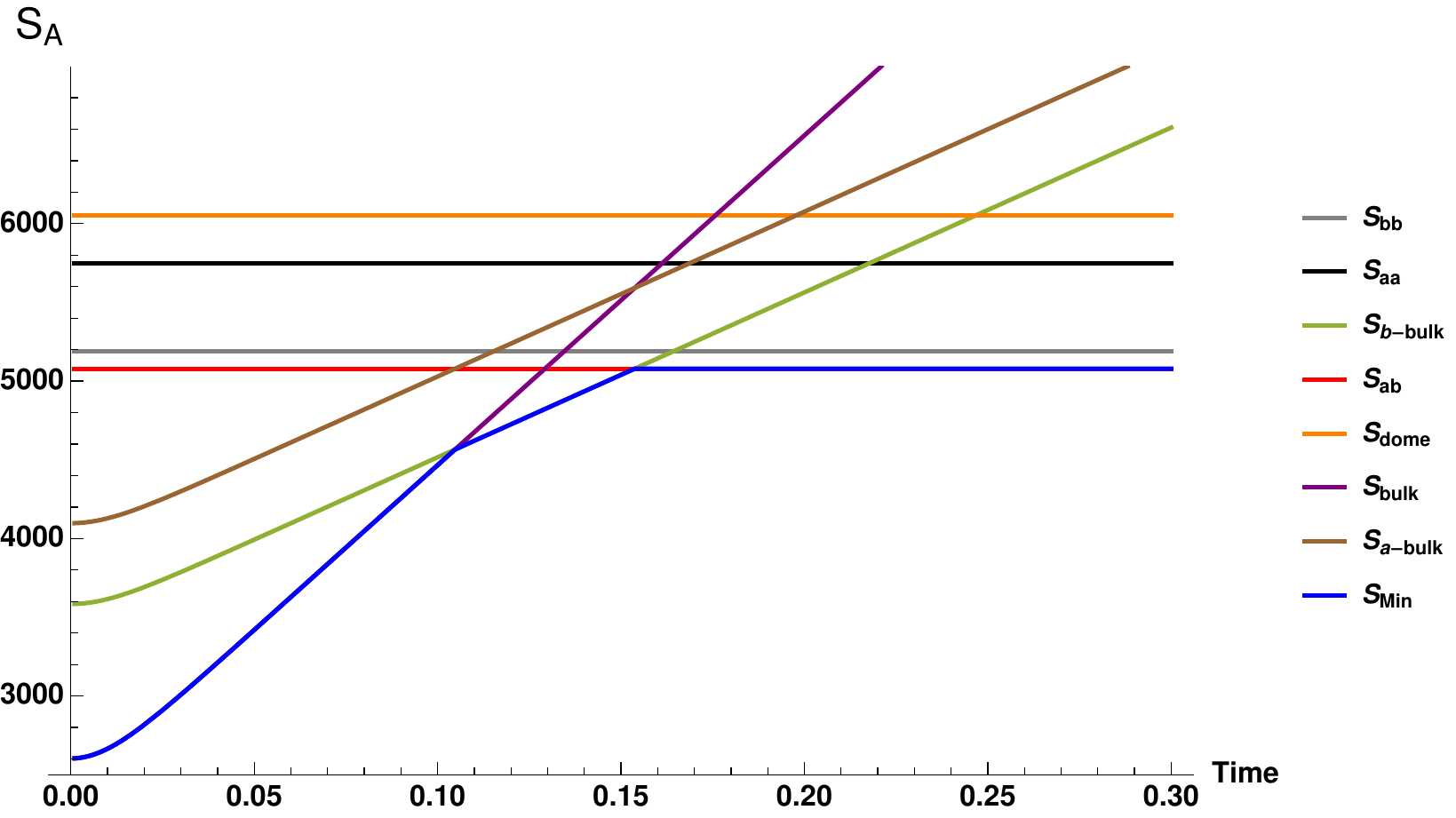}
			\caption{Entanglement entropies corresponding to the different RT surfaces w.r.t time. Here $A=[1.01,1.5]$ and time $t$ is varied from [0,\;.3]}
			\label{entropy t geng}
		\end{subfigure}
		\hspace{.1cm}
		\begin{subfigure}[t]{.45\textwidth}
			\includegraphics[width=1\linewidth]{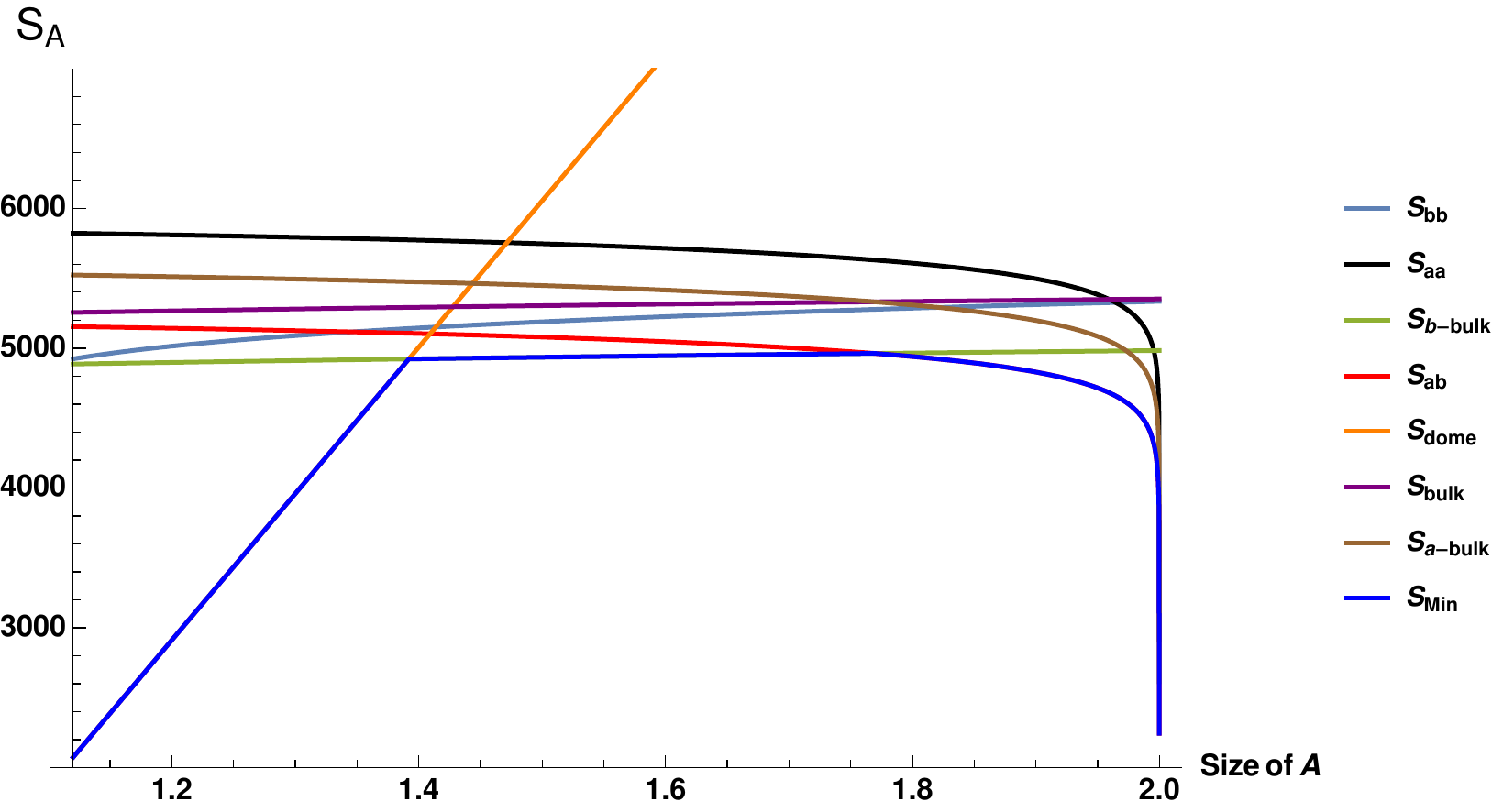}
			\caption{Entanglement entropies corresponding to the different RT surfaces w.r.t the size of the subsystem $A$. Here $t=.15$ and the size of $A$ is varied from $[1.01+\epsilon,r_O-\epsilon]$.}
			\label{entropy r geng}
		\end{subfigure}
		\caption{In the above figures, we have chosen $\beta=.1$, $c=500$, $\epsilon=.001$, $r_I=1$, $r_O=2$, $S_{bdyb}=875$, $S_{bdya}=850$.}
		\label{holographicEE}
	\end{figure}

	\subsection{Holographic entanglement negativity and Page curve}
	
	In this subsection we compute the holographic entanglement negativity for various bipartite mixed states in the context of the braneworld model-II from the proposals described in \cref{NegAdj1,NegDis0} and the expression for the holographic entanglement entropy in \cref{holographicentropy}. Subsequently we study the characteristics of the holographic entanglement negativity obtained for different scenarios involving the subsystem sizes and the time for this configuration. Some of these profiles describe the corresponding Page curves for the holographic entanglement negativity. It is important to note here that the first Page curve was obtained in the context of a bipartite quantum system in \cite{Page:1993df} using the Harr random average of the entanglement entropy for one of the subsystem as a function of the size of its Hilbert space. This Page curve was later interpreted in the context of black hole systems as the evolution of the EE for the Hawking radiation with respect to the time \cite{Page:1993wv,Page:2013dx}. Recently, the Page curves for the entanglement negativity of bipartite mixed states were obtained through the random matrix techniques in \cite{Shapourian:2020mkc}. Interestingly the corresponding Page curves for the entanglement negativity obtained through our holographic constructions for the configuration of two communicating black holes \cite{Geng:2021iyq} considered here are similar in nature.
	
	\subsubsection{Adjacent subsystems}
	In this subsection, we compute the holographic entanglement negativity for two generic adjacent subsystems $A$ and $B$ in the dual $BCFT_2$s by utilizing the \cref{NegAdj1,holographicentropy}. In particular, we investigate the qualitative nature of the entanglement negativity and discuss its profiles for three distinct scenarios involving the subsystem sizes and the time in the first model. In this context, we use the diagrams similar to those depicted in \cref{holographic} to study the various RT surfaces for the two adjacent subsystems under consideration.
	
	\subsubsection*{$\bm{(i)}$ Full system ($\bm{A\cup B}$) fixed, common point varied}
\begin{figure}[H]
		\centering
		\begin{subfigure}[t]{.45\textwidth}
			\includegraphics[width=1\linewidth]{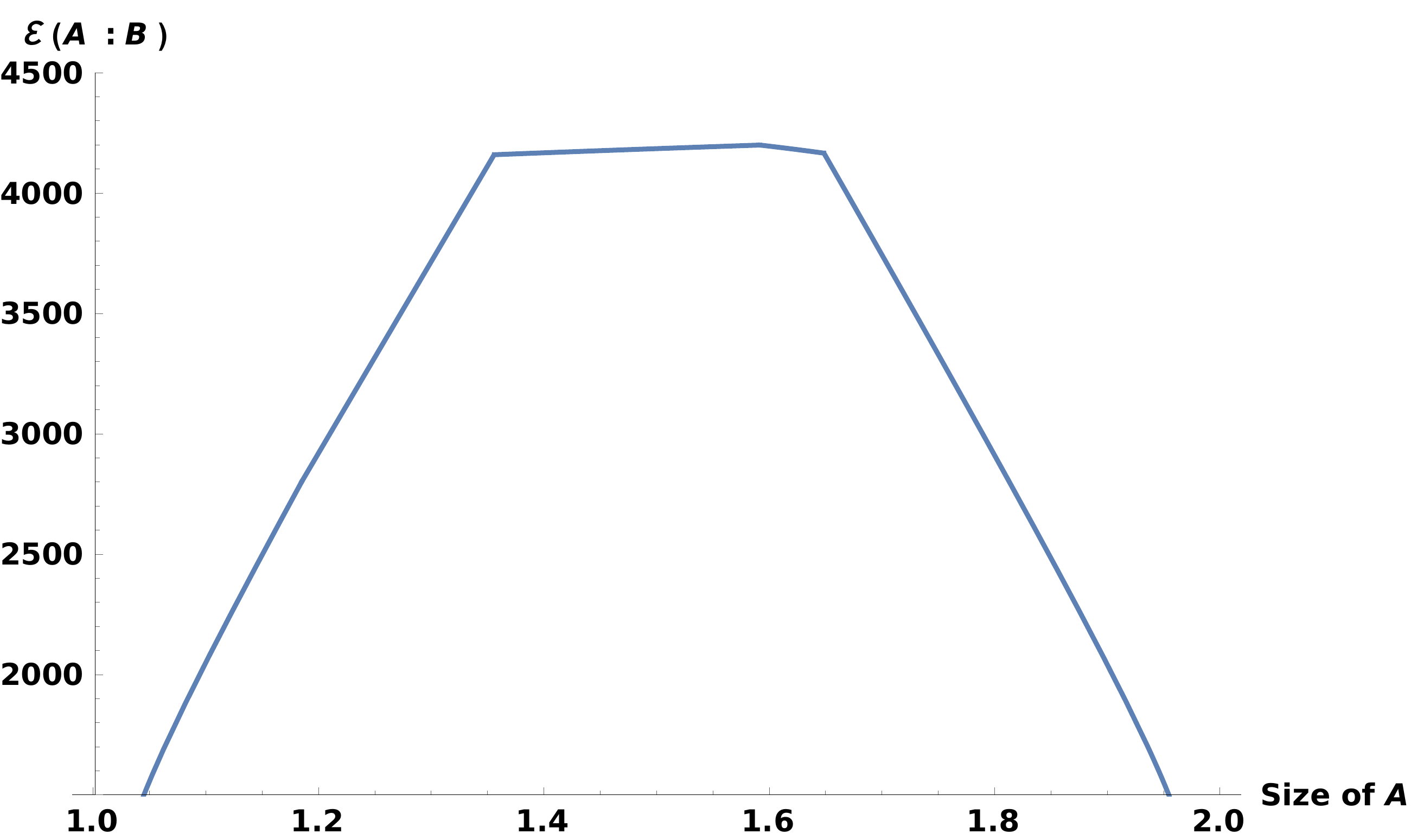}
			\caption{Page curve for the entanglement negativity with respect to the size of the subsystem $A$.}
			\label{fig1}
		\end{subfigure}
		\hspace{.1cm}
		\begin{subfigure}[t]{.45\textwidth}
			\includegraphics[width=1\linewidth]{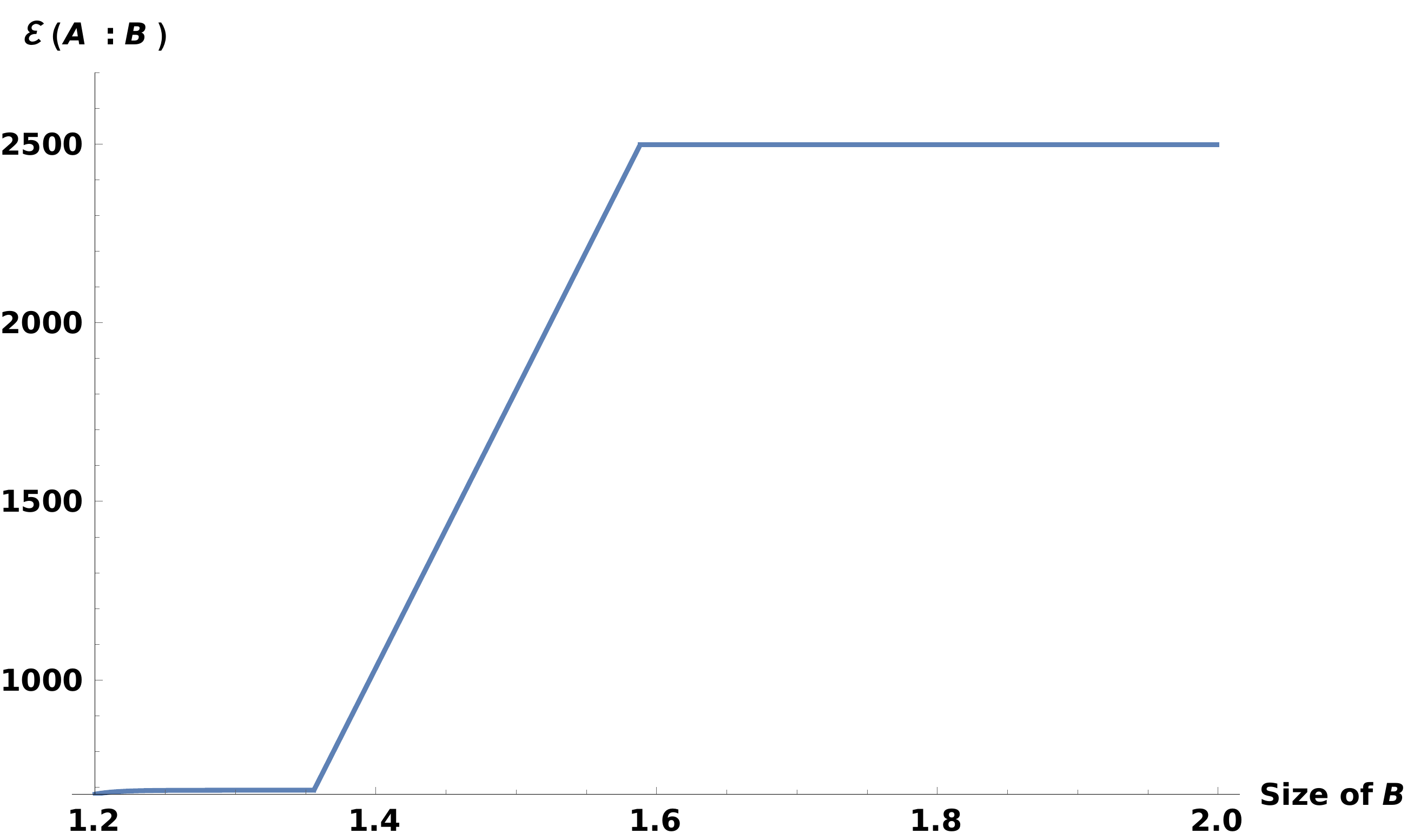}
			\caption{The entanglement negativity between the adjacent subsystems as a function of the size of $B$ where $r_1=1.15$.}
			\label{fig2}
		\end{subfigure}
		\caption{Here $r_I=1$, $r_O=2$, $\epsilon=.001$, $\beta=.1$, $c=500$, $t=.15$, $S_{bdyb}=875$ and $S_{bdya}=850$.}
		\label{holographicEN1}
	\end{figure}

	In the first case, we obtain the profile for the entanglement negativity between two adjacent subsystems $A=[r_I+\epsilon,r]$ and $B=[r,r_O-\epsilon]$ at a constant time slice while varying the common point $r$ as shown in \cref{fig1}. We compute the holographic entanglement negativity between $A$ and $B$ using the \cref{holographicentropy,NegAdj1} as follows\\
	\begin{equation}\label{adjacentcase1}
		\mathcal{E}(A:B) = 
		\begin{cases}
			\frac{c}{4} \log\left[\frac{\beta ^2 (r-r_I) (r+r_I) \sinh ^2\left(\frac{\pi  (r-r_I-\epsilon )}{\beta }\right)}{\pi ^2 \epsilon ^3 (2 r_I+\epsilon )}\right]\,,& \text{Phase-1} \vspace{2mm}\\ 
			\frac{c}{4}\log\left[\frac{2 \beta ^2 r r_I \cosh \left(\frac{2 \pi  t}{\beta }\right) \sinh ^2\left(\frac{\pi  (r-r_I-\epsilon )}{\beta }\right)}{\pi ^2 \epsilon ^3 (2 r_I+\epsilon )}\right]-2 S_{\text{bdyb}}\,, &  \text{Phase-2}\vspace{2mm} \\
			\frac{c}{4}\log\left[\frac{4 r^2 \cosh ^2\left(\frac{2 \pi  t}{\beta }\right)}{\epsilon ^2}\right]\,,&  \text{Phase-3} \vspace{2mm}\\
			\frac{c}{4}\log\left[\frac{(r_O-r)^2 (r+r_O)^2}{r_O^2 \epsilon ^2}\right]+ 4 S_{\text{bdya}}\,,&  \text{Phase-4}\vspace{2mm} \\
			\frac{c}{4}\log\left[\frac{\beta ^2 (r_O-r) (r+r_O) \sinh ^2\left(\frac{\pi  (r-r_O+\epsilon )}{\beta }\right)}{\pi ^2 \epsilon ^3 (2 r_O-\epsilon )}\right]\,,&  \text{Phase-5}
		\end{cases}       
	\end{equation}	
	where the boundary entropies corresponding to the two KR branes are denoted as $S_{\text{bdya}}$ and $S_{\text{bdyb}}$.  It is interesting to note that the behaviour of the Page curve for the entanglement negativity in this context is analogous to the one obtained in \cite{KumarBasak:2021rrx,Shapourian:2020mkc}. 
	
	In the present scenario, we identify all the possible contributions to the entanglement entropies of the subsystems $A$ and $B$ by using \cref{holographicentropy} to elucidate the phase transitions observed in \cref{fig1}. In this context it is possible to identify five distinct phases for the holographic entanglement negativity between the two adjacent subsystems which is described as follows.

	\subsubsection*{$\bm{(ii)}$ Subsystem $\bm{A}$ fixed, $\bm{B}$ varied}\label{adjcaseii}
	Next we analyze the behaviour of the entanglement negativity between the two adjacent subsystems at a constant time slice where we consider the subsystem $A=[r_I+\epsilon, r_1]$ with a fixed size and vary the size of the subsystem $B=[r_1+\epsilon, r]$ by shifting the point $r$. This is obtained by using the \cref{holographicentropy,NegAdj1} as follows
	\begin{equation} 
		\mathcal{E}(A:B) = 
		\begin{cases} 
			\frac{c}{2} \log\left[\frac{\beta  \sinh \left(\frac{\pi  \left(r_1-r\right)}{\beta }\right) \sinh \left(\frac{\pi  \left(-r_1+r_I+\epsilon \right)}{\beta }\right) \text{csch}\left(\frac{\pi  (r-r_I-\epsilon )}{\beta
				}\right)}{\pi  \epsilon }\right]\,,& \text{Phase-1} \vspace{2mm} \\
			\frac{c}{4}\log\left[\frac{\beta ^4 r_I \text{sech}\left(\frac{2 \pi  t}{\beta }\right) \sinh ^2\left(\frac{\pi  \left(r-r_1\right)}{\beta }\right) \sinh ^2\left(\frac{\pi  \left(-r_1+r_I+\epsilon
					\right)}{\beta }\right)}{2 \pi ^4 r \epsilon ^3 (2 r_I+\epsilon )}\right]+2 S_{\text{bdyb}} \,,&  \text{Phase-2}  \vspace{2mm}\\
			\frac{c}{4}\log\left[\frac{\beta ^2 \left(r_1^2-r_I ^2\right)  \sinh ^2\left(\frac{\pi  \left(-r_1+r_I+\epsilon \right)}{\beta }\right)}{\pi ^2 \epsilon ^3  (2 r_I+\epsilon )}\right]\,,&  \text{Phase-3} 
		\end{cases} 
	\end{equation}
	which corresponds to three possible phases for the entanglement negativity between $A$ and $B$ as depicted in \cref{fig2}. In what follows we analyze these phases in detail.

	\subsubsection*{$\bm{(iii)}$ Subsystems $\bm{A}$ and $\bm{B}$ fixed, time varied} \label{Adjtime}
\begin{figure}[H]
		\centering
		\includegraphics[width=10cm]{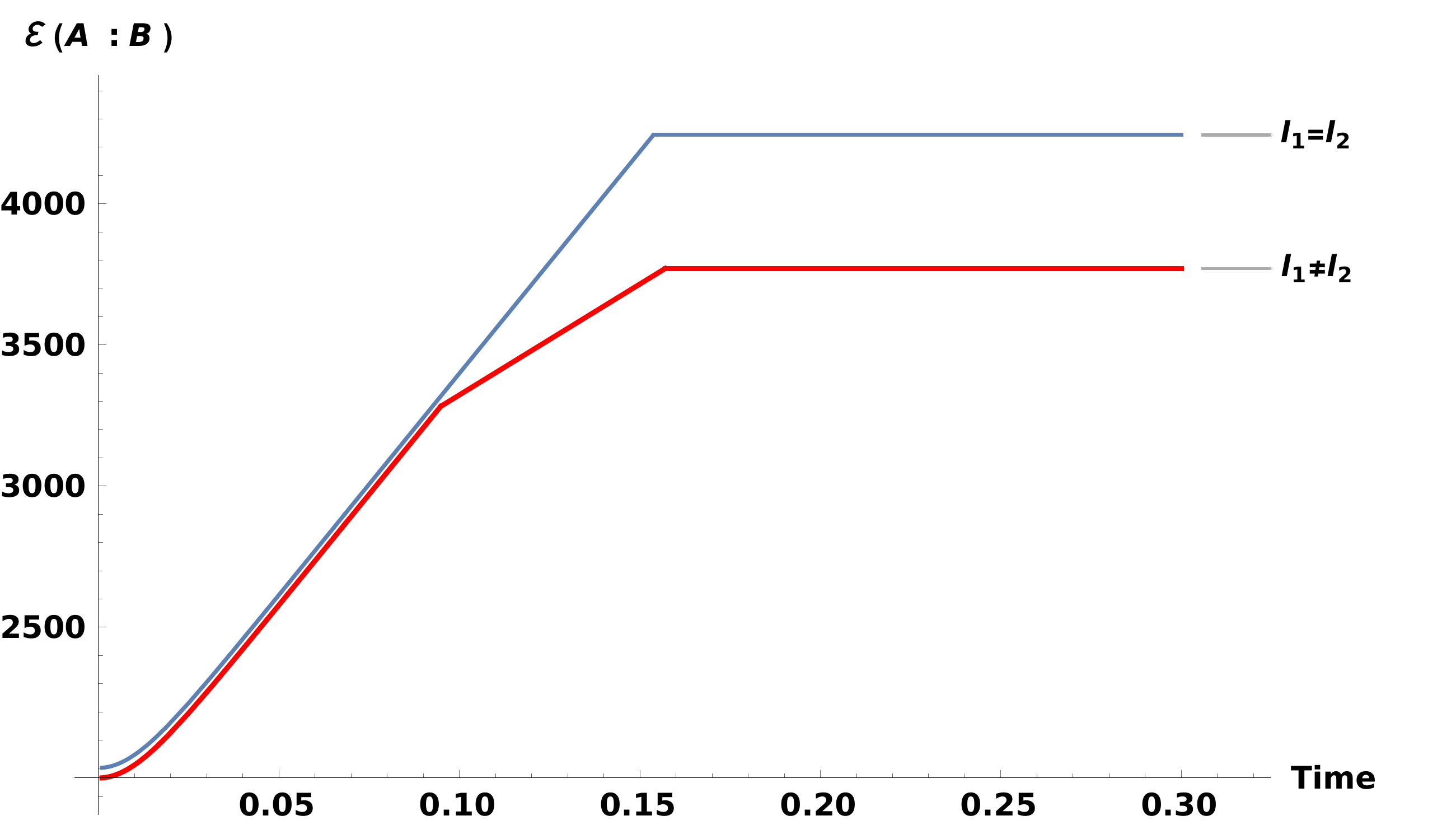}
		\caption{Page curves for the entanglement negativity between two adjacent subsystems $A$ and $B$ as a function of time. Here $r_I=1$, $r_O=2$, $\epsilon=.001$, $\beta=.1$, $c=500$, $S_{bdyb}=875$, $S_{bdya}=850$ and $A=[r_I+\epsilon ,r_1]$, $B=[r_1 ,r_O-\epsilon]$ with $r_1=1.5$ (for $l_1 = l_2$) and $A=[r_I+\epsilon ,r_1]$, $B=[r_1 ,r_O-\epsilon]$ with $r_1=1.3$ (for $l_1\neq l_2$).}\label{fig3}
	\end{figure}

	Finally we investigate the holographic entanglement negativity between the adjacent subsystems $A$ and $B$ with fixed lengths $l_1$ and $l_2$ respectively while varying the time. In particular, we will study two sub cases of equal and unequal lengths of the two subsystems in question and obtain the corresponding entanglement negativities as depicted in the \cref{fig3}. 
	\noindent
	\textbf{(a) For $\bm{l_1=l_2}$}\\  
	For the case of two equal length subsystems $A$ and $B$, we observe two phases in the entanglement negativity profile as depicted in \cref{fig3}. The expressions for these may be obtained by utilizing the \cref{holographicentropy,NegAdj1} as follows
	
	\begin{equation}\label{adjacentcase3} 
		\mathcal{E}(A:B) = 
		\begin{cases} 
			\frac{c}{4} \log\left[\frac{4 r_1^2 \cosh ^2\left(\frac{2 \pi  t}{\beta }\right)}{\epsilon ^2}\right]\,,&  \text{Phase-1} \vspace{3mm}\\
			\frac{c}{4} \log\left[\frac{ \left(r_O^2-r_1^2\right)^2 }{r_O^2 \epsilon ^2 }\right]+4 S_{\text{bdya}}\,.& \text{Phase-2}  
		\end{cases} 
	\end{equation}
	\noindent
	\textbf{(b) For $\bm{l_1\neq l_2}$}\\
	Here we consider two unequal lengths of the subsystems $A$ and $B$ and compute the holographic entanglement negativity between them utilizing the \cref{holographicentropy,NegAdj1}. We observe three consecutive phases in the entanglement negativity profile as depicted in \cref{fig3}. The expressions for the corresponding entanglement negativity in these phases are given as follows
	\begin{equation}\label{adjacentcase4}
		\mathcal{E}(A:B) = 
		\begin{cases} 
			\frac{c}{4} \log\left[\frac{4 r_1^2 \cosh ^2\left(\frac{2 \pi  t}{\beta }\right)}{\epsilon ^2}\right]\,, &  \text{Phase-1} \vspace{2mm}\\
			\frac{c}{4} \log\left[\frac{2 \beta ^2 r_1 r_I \cosh \left(\frac{2 \pi  t}{\beta }\right) \sinh ^2\left(\frac{\pi  \left(-r_1+r_I+\epsilon \right)}{\beta }\right)}{\pi ^2 \epsilon ^3 (2 r_I+\epsilon )}\right]-2 S_{\text{bdyb}}\,\,, &  \text{Phase-2} \vspace{2mm}\\
			\frac{c}{4} \log\left[\frac{\beta ^2 \left(r_1^2-r_I^2\right) \sinh ^2\left(\frac{\pi  \left(-r_1+r_I+\epsilon \right)}{\beta }\right)}{\pi ^2 \epsilon ^3 (2 r_I+\epsilon )}\right]\,. &  \text{Phase-3} 
		\end{cases} 
	\end{equation}

	\subsubsection{Disjoint subsystems}
	Now we consider two generic disjoint subsystems $A$ and $B$ with a subsystem $C$ enclosed between them in the dual $BCFT_2$s and compute the holographic entanglement negativity between $A$ and $B$ using the \cref{NegDis0,holographicentropy}. In particular, we will investigate the qualitative feature of the entanglement negativity profile for three different scenarios of the disjoint subsystems involving the subsystem sizes and the time in the context of the braneworld model-II. Once again we utilize diagrams similar to those depicted in \cref{holographic} to study the various RT surfaces for the subsystems in question.

	\subsubsection*{$\bm{(i)}$ Subsystem $\bm{A}$ fixed, $\bm{C}$ varied}
\begin{figure}[H]
		\centering
		\begin{subfigure}[t]{.45\textwidth}
			\includegraphics[width=1\linewidth]{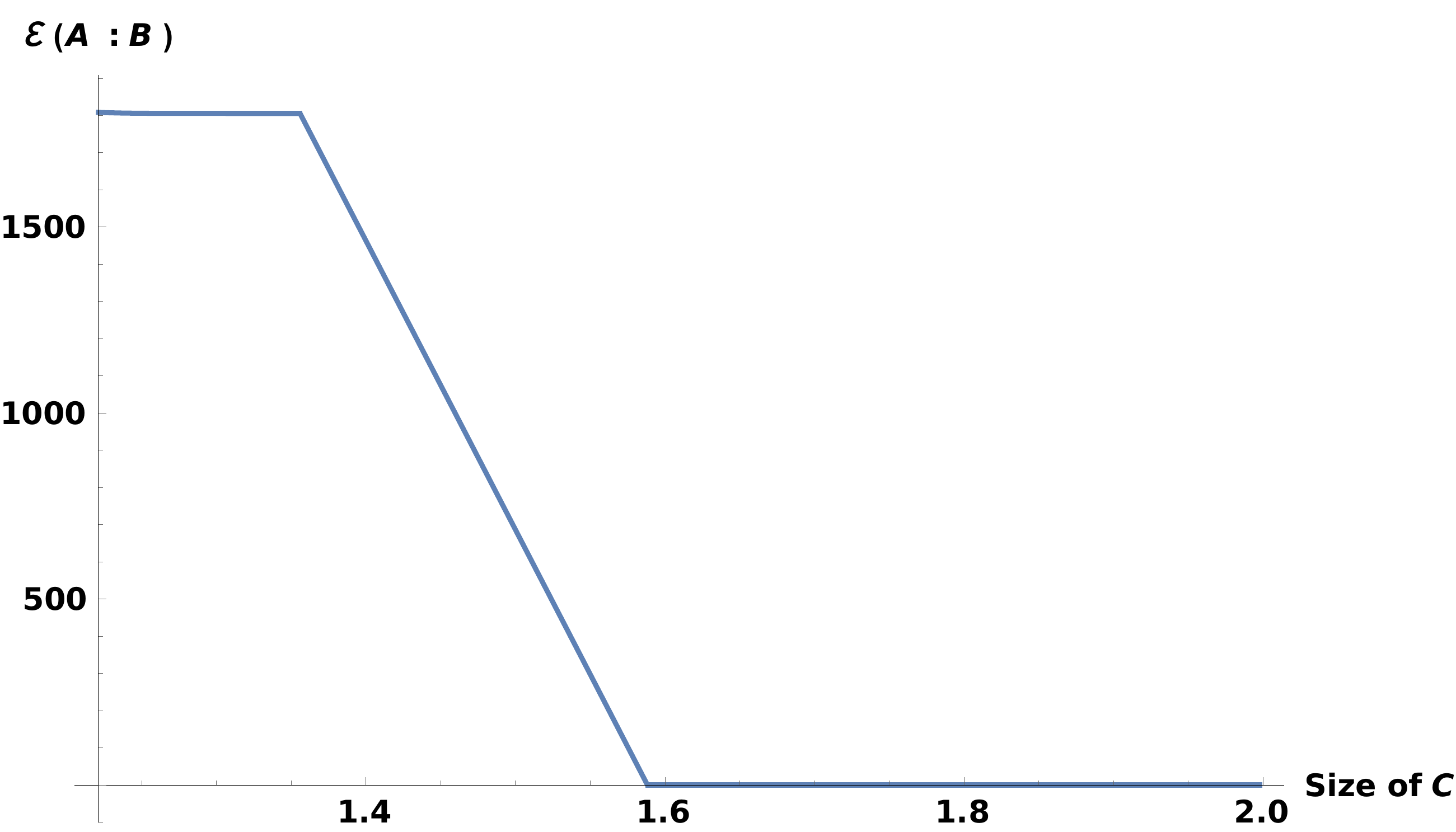}
			\caption{Entanglement negativity between two disjoint subsystems with the variation of size $C$ where $r_1=1.15$.}
			\label{fig6}
		\end{subfigure}
		\hspace{.1cm}
		\begin{subfigure}[t]{.45\textwidth}
			\includegraphics[width=1\linewidth]{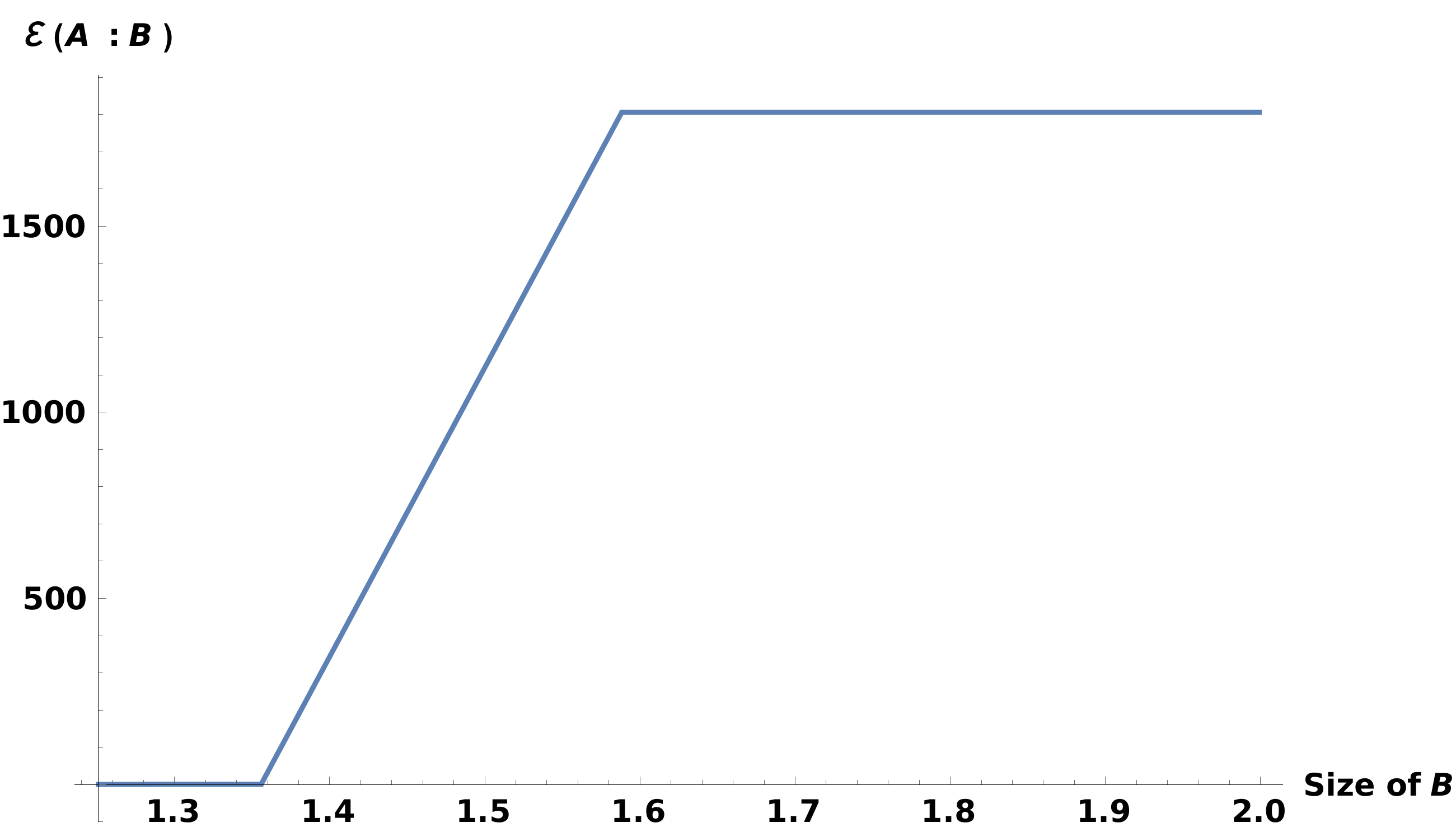}
			\caption{The entanglement negativity between two disjoint subsystems $A\,\text{and}\, B$ with a variation in the size of $B$ where $r_1=1.15$ and $r_2=1.25$.}
			\label{fig5}
		\end{subfigure}
		\caption{Here, $r_I=1$, $r_O=2$, $\epsilon=.001$, $\beta=.1$, $c=500$, $t=.15$, $S_{bdyb}=875$, $S_{bdya}=850$.}
		\label{holographicEN2}
	\end{figure}

	In the first case, we fix the size of the subsystem $A=[r_I+\epsilon, r_1]$ at a constant time slice and vary the size of $C=[r_1,r]$ by shifting the point $r$ from $r_1+\epsilon$ to $r_O-\epsilon$ to examine the holographic entanglement negativity between the subsystems $A$ and $B$. To this end, we may compute the entanglement negativity by utilizing the \cref{NegDis0,holographicentropy} as follows
	
	\begin{equation}
		\mathcal{E}(A:B) = 
		\begin{cases} 
			\frac{c}{4} \log\left[\frac{\left(-r_I^2+r_1^2\right) \sinh ^2\left(\frac{\pi  \left(r-r_I-\epsilon \right)}{\beta}\right) }{\epsilon  (2 r_I+\epsilon ) \sinh^2\left(\frac{\pi  \left(r-r_1\right)}{\beta }\right) }\right] \;,&\text{Phase-1} \vspace{3mm}\\
			\frac{c}{4} \log\left[\frac{2 \pi ^2 \, r \, \left(-r_I^2+r_1^2\right) \cosh \left(\frac{2 \pi  t}{\beta }\right)}{\beta ^2 r_I \epsilon   \sinh^2\left(\frac{\pi \left(r-r_1\right)}{\beta }\right)}\right] +2 S_{\text{bdyb}}\;\;,& \text{Phase-2} \vspace{3mm}\\
			0\;.&\text{Phase-3}
		\end{cases} 
	\end{equation} 
	Here we observe three phases from the corresponding entanglement negativity profile as depicted in \cref{fig6} which we now analyze in details.\\\\

	\subsubsection*{$\bm{(ii)}$ Subsystems $\bm{A}$ and $\bm{C}$ fixed, $\bm{B}$ varied}\label{discasei}
	In this case, we consider the subsystem sizes of $A=[r_I+\epsilon ,r_1]$ and $C=[r_1,r_2]$ to be fixed and vary the size of $B=[r_2+\epsilon,r]$ by shifting the point $r$. The entanglement negativity between the two disjoint subsystems $A$ and $B$ corresponds to three consecutive phases as shown in \cref{fig5}. In this context, the size of $C$ is considered to be very small such that the dominant contribution to its entanglement entropy arises from dome-type RT surfaces throughout this case. Finally, the expressions for the entanglement negativity in this scenario may be given as \\
	\begin{equation}\label{disjointcase1} 
		\mathcal{E}(A:B) = 
		\begin{cases} 
			\frac{c}{2} \log \left[\frac{\sinh \left(\frac{\pi  \left(-r_1+r\right)}{\beta }\right) \sinh \left(\frac{\pi  \left(-r_2+r_I+\epsilon \right)}{\beta }\right)}{\sinh\left(\frac{\pi  \left(r_1-r_2\right)}{\beta }\right)\sinh\left(\frac{\pi  \left(r-r_I-\epsilon \right)}{\beta }\right)}\right]\;, &  \text{Phase-1} \vspace{2mm} \\
			\frac{c}{4} \log\left[\frac{\beta ^2 r_I  \text{sech}\left(\frac{2 \pi  t}{\beta }\right) \sinh ^2\left(\frac{\pi  \left(-r_1+r\right)}{\beta }\right) \sinh
				^2\left(\frac{\pi  \left(-r_2+r_I+\epsilon \right)}{\beta }\right)}{2 \pi ^2 \epsilon \, r \, (2 r_I+\epsilon )\sinh^2\left(\frac{\pi  \left(r_1-r_2\right)}{\beta }\right)}\right]-2 S_{\text{bdyb}}\;\;, &  \text{Phase-2} \vspace{2mm}\\
			\frac{c}{4} \log\left[\frac{\left(r_1^2-r_I^2\right)  \sinh ^2\left(\frac{\pi  \left(-r_2+r_I+\epsilon
					\right)}{\beta }\right)}{\epsilon  (2 r_I+\epsilon ) \sinh^2\left(\frac{\pi  \left(r_1-r_2\right)}{\beta }\right)}\right]\;. & \text{Phase-3}
		\end{cases} 
	\end{equation}

	\noindent

	Note that in the present scenario, we increase only the size of the subsystem $B$ at a constant time slice while fixing the size of $A$ and $C$. This is similar to the adjacent case discussed in sub\cref{adjcaseii} where we have fixed the size of the subsystem $A$ and increased the size of $B$. Hence all the above phases may be described in terms of the Hawking modes using the explanations analogous to the adjacent case.

	\subsubsection*{$\bm{(iii)}$ Subsystems $\bm{A}$, $\bm{B}$ and $\bm{C}$ fixed, time varied}
	We end our analysis with the final case where we investigate the nature of the entanglement negativity between the two disjoint subsystems $A$ and $B$ with lengths $l_1$ and $l_2$ respectively while varying the time. Here we consider the size of the subsystem $C$ to be very small such that the dominant contribution to the entanglement entropy arises from dome-type RT surfaces. Note that, the entanglement wedges of the subsystems $A$ and $B$ are connected in this scenario which corresponds to a non-zero entanglement negativity between them. In  what follows, we explore two sub cases of equal and unequal lengths of the subsystems $A$ and $B$ while varying the time.
	\begin{figure}[H]\label{disjointtime1}
		\centering
		\includegraphics[width=10cm]{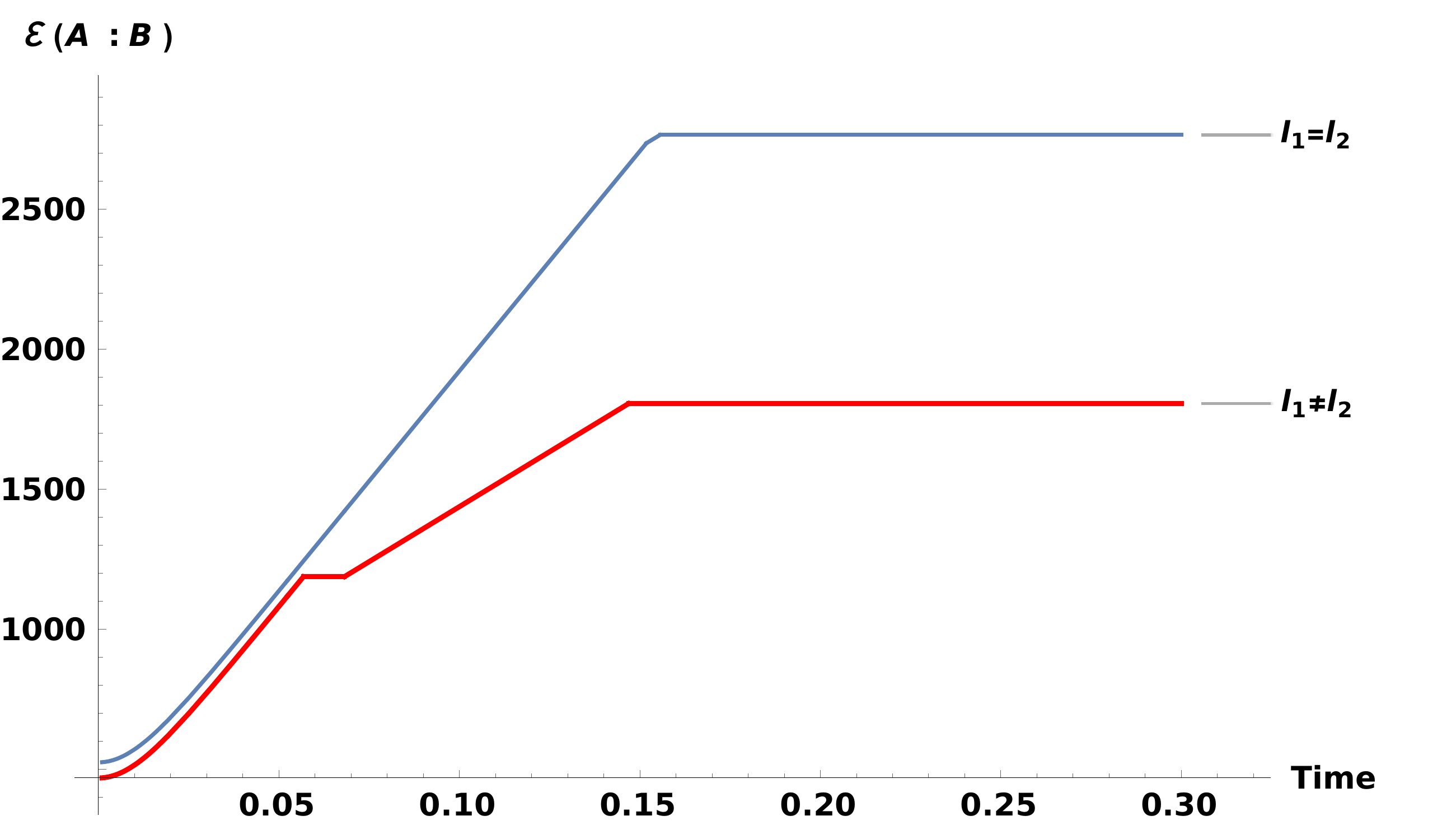}
		\caption{The Page curves for the entanglement negativity for two disjoint subsystems as a function of time. In this case we have chosen $r_I=1$, $r_O=2$, $\epsilon=.001$, $\beta=.1$, $c=500$, $S_{bdyb}=875$, $S_{bdya}=850$ and $A=[r_I+\epsilon ,r_1]$, $B=[r_2 ,r_O-\epsilon]$ with $r_1=1.45$, $r_2=1.55$ (for $l_1 = l_2$) and
			$A=[r_I+\epsilon ,r_1]$, $B=[r_2 ,r_O-\epsilon]$ with $r_1=1.15$, $r_2=1.25$ (for $l_1\neq l_2$).}\label{fig7}
	\end{figure}

	\subsubsection*{(a) For $\bm{l_1=l_2}$}\label{peq1} 
	For the case of two equal length subsystems $A$ and $B$, we examine the qualitative profile of the entanglement negativity between them which corresponds to three different phases as depicted in \cref{fig7}. In this context, we may compute the entanglement negativity by utilizing \cref{NegDis0,holographicentropy} as follows
	\begin{equation}\label{disjointtime3}
		\mathcal{E}(A:B) = 
		\begin{cases} 
			\frac{c}{4} \log\left[\frac{4 \pi ^2 r_1 \, r_2 \,  \cosh ^2\left(\frac{2 \pi  t}{\beta }\right)}{\beta ^2 \sinh^2\left(\frac{\pi  \left(r_1-r_2\right)}{\beta }\right)}\right]\,, & \text{Phase-1}\vspace{2mm}\\
			\frac{c}{4} \log\left[\frac{2 \pi ^2 r_2 \left(r_1^2-r_I^2\right)\cosh \left(\frac{2 \pi  t}{\beta }\right)}{\beta ^2 r_I \sinh^2\left(\frac{\pi  \left(r_1-r_2\right)}{\beta }\right) }\right]+2 S_{\text{bdyb}} \,\,,& \text{Phase-2}\vspace{2mm} \\
			\frac{c}{4} \log\left[\frac{\pi ^2 \left(r_1^2-r_I^2\right) \left(r_O^2-r_2^2\right) }{\beta ^2 r_I\,\,r_O\sinh^2\left(\frac{\pi  \left(r_1-r_2\right)}{\beta }\right)}\right]+2 S_{\text{bdyb}}+2 S_{\text{bdya}}\,\,.&\text{Phase-3} \\
		\end{cases} 
	\end{equation}

	\subsubsection*{(b) For $\bm{l_1\neq l_2}$}\label{pneq1} 
	Finally we consider two disjoint subsystems $A$ and $B$ with unequal lengths while increasing the time and compute the entanglement negativity between them utilizing \cref{NegDis0,holographicentropy}. In this context, we obtain four consecutive phases of the corresponding entanglement negativity profile which may be expressed as follows
	
	\begin{equation}\label{disjointtime}
		\mathcal{E}(A:B) = 
		\begin{cases} 
			\frac{c}{4} \log\left[\frac{4 \pi ^2 r_1 \, r_2 \,  \cosh ^2\left(\frac{2 \pi  t}{\beta }\right)}{\beta ^2 \sinh^2\left(\frac{\pi  \left(r_1-r_2\right)}{\beta }\right)}\right]\,, & \text{Phase-1}\vspace{2mm}\\
			\frac{c}{4}\log\left[\frac{r_O (r_O-\epsilon)\left(-r_I^2+r_1^2\right) \sinh ^2\left(\frac{\pi  \left(-r_2+r_I+\epsilon \right)}{\beta }\right)}{r_I \epsilon  (r_I+\epsilon ) (2 r_O-\epsilon) \sinh^2\left(\frac{\pi  \left(r_1-r_2\right)}{\beta
				}\right)}\right]-2 S_{\text{bdya}}+2 S_{\text{bdyb}}\,,&\text{Phase-2} \vspace{2mm}\\
			\frac{c}{4}\log\left[\frac{2 r_O (r_O-\epsilon)\left(-r_I^2+r_1^2\right) \sinh ^2\left(\frac{\pi  \left(-r_2+r_I+\epsilon \right)}{\beta }\right)\cosh \left(\frac{2 \pi  t}{\beta }\right)}{r_I \epsilon^2  (2 r_I+\epsilon ) (2 r_O-\epsilon) \sinh^2\left(\frac{\pi  \left(r_1-r_2\right)}{\beta
				}\right)}\right]-2 S_{\text{bdya}}\,\,,& \text{Phase-3} \vspace{2mm}\\
			\frac{c}{4} \log\left[\frac{\left(r_1^2-r_I^2\right)  \sinh ^2\left(\frac{\pi  \left(-r_2+r_I+\epsilon
					\right)}{\beta }\right)}{\epsilon  (2 r_I+\epsilon )\sinh^2 \left(\frac{\pi  \left(r_1-r_2\right)}{\beta }\right)}\right]\,\,.&\text{Phase-4} 
		\end{cases} 
	\end{equation}

	\section{Expressions for holographic entanglement negativity for braneworld model}\label{formula_model2}
	In this appendix, we show the expressions for the holographic entanglement negativity between two adjacent and disjoint subsystems obtained through the \cref{NegAdj1,NegDis0,HEE-Shaghoulian} in the braneworld model.
	\subsection*{Adjacent subsystems}	
	Here we list the expressions of the holographic entanglement negativity between two adjacent subsystems utilizing the \cref{NegAdj1,HEE-Shaghoulian} in three distinct scenarios involving the subsystem sizes and the time as discussed in \cref{Adj}.
	\subsubsection*{$\bm{(i)}$ Full system ($\bm{A\cup B}$) fixed, common point varied}\label{Appendix1}
	In the first scenario, the size of the subsystem $A\cup B$ is fixed which covers the whole radiation reservoirs and the common point between them is varied at a constant time slice. The corresponding expressions for the entanglement negativity between $A$ and $B$ in the different phases is given as follows\\
	
	\textbf{Phase 1}
	\begin{eqnarray}
		\mathcal{E}(A:B) &=& \frac{c}{2} \log \left[\frac{\beta}{\pi } \sinh \frac{\pi}{\beta } |a-r| \right]+\frac{3\pi\phi_r}{\beta } \coth \left(\frac{2\pi}{\beta} r+  \log \frac{24 \pi \phi_r }{c \beta }\right)\nonumber\\
		&+&\frac{c}{4}\log \left[\frac{\beta}{\pi}\;\frac{\cosh \left(\frac{4 \pi}{\beta} r+  \log \frac{24 \pi  \phi_r }{c \beta } \right)-1}{  \sinh \left(\frac{2 \pi}{\beta} r+ \log \frac{24 \pi  \phi_r }{c \beta } \right)} \right] - \frac{3\pi\phi_r}{\beta } \coth \left(\frac{2\pi}{\beta} a+  \log \frac{24 \pi \phi_r }{c \beta }\right)\nonumber\\
		&-&\frac{c}{4}\log \left[\frac{\beta}{\pi}\;\frac{\cosh \left(\frac{4 \pi}{\beta} a+  \log \frac{24 \pi  \phi_r }{c \beta } \right)-1}{  \sinh \left(\frac{2 \pi}{\beta} a+ \log \frac{24 \pi  \phi_r }{c \beta } \right)} \right]  ,
	\end{eqnarray}
	
	\textbf{Phase 2}
	\begin{eqnarray}
		\mathcal{E}(A:B) &=& \frac{c}{2} \log \left[\frac{\beta}{\pi } \sinh \frac{\pi}{\beta } |a-r| \right] - \frac{3}{2}\phi_0 - \frac{3\pi\phi_r}{\beta } \coth \left(\frac{2\pi}{\beta} a+  \log \frac{24 \pi \phi_r }{c \beta }\right)\nonumber\\
		&-&\frac{c}{4}\log \left[\frac{\beta}{\pi}\;\frac{\cosh \left(\frac{4 \pi}{\beta} a+  \log \frac{24 \pi  \phi_r }{c \beta } \right)-1}{  \sinh \left(\frac{2 \pi}{\beta} a+ \log \frac{24 \pi  \phi_r }{c \beta } \right)} \right]+\frac{c}{4} \log \left[\frac{\beta}{\pi } \cosh \frac{2\pi }{\beta }t\right] ,
	\end{eqnarray}
	
	\textbf{Phase 3}
	\begin{eqnarray}
		\mathcal{E}(A:B) &=& \frac{c}{2} \log \left[\frac{\beta}{\pi } \cosh \frac{2\pi }{\beta }t\right] ,
	\end{eqnarray}
	
	\textbf{Phase 4}
	\begin{eqnarray}
		\mathcal{E}(A:B) &=& \frac{c}{2} \log \left[\frac{\beta}{\pi } \sinh \frac{\pi}{\beta } |r-b| \right]- \frac{3}{2}\phi_0 - \frac{3\pi\phi_r}{\beta } \coth \left(\frac{2\pi}{\beta}(L-b)+  \log \frac{24 \pi \phi_r }{c \beta } \right)\nonumber\\
		&-& \frac{c}{4}\log \left[\frac{\beta}{\pi}\;\frac{\cosh \left(\frac{4 \pi}{\beta} (L-b)+  \log \frac{24 \pi  \phi_r }{c \beta } \right)-1}{\sinh \left(\frac{2 \pi}{\beta} (L-b)+  \log \frac{24 \pi  \phi_r }{c \beta } \right)} \right]+\frac{c}{4} \log \left[\frac{\beta}{\pi } \cosh \frac{2\pi }{\beta }t\right] ,
	\end{eqnarray}
	
	\textbf{Phase 5}
	\begin{eqnarray}
		\mathcal{E}(A:B) &=& \frac{c}{2} \log \left[\frac{\beta}{\pi } \sinh \frac{\pi}{\beta } |r-b| \right] + \frac{c}{4}\log \left[\frac{\beta}{\pi}\;\frac{\cosh \left(\frac{4 \pi}{\beta} (L-r)+  \log \frac{24 \pi  \phi_r }{c \beta } \right)-1}{\sinh \left(\frac{2 \pi}{\beta} (L-r)+  \log \frac{24 \pi  \phi_r }{c \beta } \right)} \right]\nonumber\\
		&-&\frac{c}{4}\log \left[\frac{\beta}{\pi}\;\frac{\cosh \left(\frac{4 \pi}{\beta} (L-b)+  \log \frac{24 \pi  \phi_r }{c \beta } \right)-1}{\sinh \left(\frac{2 \pi}{\beta} (L-b)+  \log \frac{24 \pi  \phi_r }{c \beta } \right)} \right].
	\end{eqnarray}
	
	\subsubsection*{$\bm{(ii)}$ Subsystem $\bm{A}$ fixed, $\bm{B}$ varied}\label{Appendix2}
	In the second scenario, the size of the subsystem $A$ is fixed at a constant time slice while the size of the subsystem $B$ is varied. In what follows, the corresponding expressions for the entanglement negativity between the subsystems $A$ and $B$ in the different phases are given by\\
	
	\textbf{Phase 1}
	\begin{eqnarray}
		\mathcal{E}(A:B) &=& \frac{c}{2} \log \left[\frac{\beta}{\pi } \sinh \frac{\pi}{\beta } |a-r| \right]+\frac{c}{2} \log \left[\frac{\beta}{\pi } \sinh \frac{\pi}{\beta } |r-b| \right]\nonumber\\
		&-&\frac{c}{2} \log \left[\frac{\beta}{\pi } \sinh \frac{\pi}{\beta } |a-b| \right],
	\end{eqnarray}
	
	\textbf{Phase 2}
	\begin{eqnarray}
		\mathcal{E}(A:B) &=& \frac{c}{2} \log \left[\frac{\beta}{\pi } \sinh \frac{\pi}{\beta } |a-r| \right]+\frac{c}{2} \log \left[\frac{\beta}{\pi } \sinh \frac{\pi}{\beta } |r-b| \right]-\frac{c}{4} \log \left[\frac{\beta}{\pi } \cosh \frac{2\pi }{\beta }t\right]-\frac{3}{2}\phi_0\nonumber\\
		&-&\frac{3\pi\phi_r}{\beta } \coth \left(\frac{2\pi}{\beta} a+  \log \frac{24 \pi \phi_r }{c \beta }\right) - \frac{c}{4}\log \left[\frac{\beta}{\pi}\;\frac{\cosh \left(\frac{4 \pi}{\beta} a+  \log \frac{24 \pi  \phi_r }{c \beta } \right)-1}{  \sinh \left(\frac{2 \pi}{\beta} a+ \log \frac{24 \pi  \phi_r }{c \beta } \right)} \right],\nonumber\\
		&&
	\end{eqnarray}
	
	\textbf{Phase 3}
	\begin{eqnarray}
		\mathcal{E}(A:B) &=& \frac{c}{2} \log \left[\frac{\beta}{\pi } \sinh \frac{\pi}{\beta } |a-r| \right]+\frac{c}{2} \log \left[\frac{\beta}{\pi } \sinh \frac{\pi}{\beta } |r-b| \right]-3\phi_0\nonumber\\
		&-& \frac{3\pi\phi_r}{\beta } \coth \left(\frac{2\pi}{\beta} a+  \log \frac{24 \pi \phi_r }{c \beta }\right) - \frac{c}{4}\log \left[\frac{\beta}{\pi}\;\frac{\cosh \left(\frac{4 \pi}{\beta} a+  \log \frac{24 \pi  \phi_r }{c \beta } \right)-1}{  \sinh \left(\frac{2 \pi}{\beta} a+ \log \frac{24 \pi  \phi_r }{c \beta } \right)} \right]\nonumber\\
		&-&\frac{3\pi\phi_r}{\beta } \coth \left(\frac{2\pi}{\beta}(L-b)+  \log \frac{24 \pi \phi_r }{c \beta }\right)\nonumber\\
		&-& \frac{c}{4}\log \left[\frac{\beta}{\pi}\;\frac{\cosh \left(\frac{4 \pi}{\beta} (L-b)+  \log \frac{24 \pi  \phi_r }{c \beta } \right)-1}{\sinh \left(\frac{2 \pi}{\beta} (L-b)+  \log \frac{24 \pi  \phi_r }{c \beta } \right)} \right],
	\end{eqnarray}
	
	\textbf{Phase 4}
	\begin{eqnarray}
		\mathcal{E}(A:B) &=&\frac{3\pi\phi_r}{\beta } \coth \left(\frac{2\pi}{\beta} r+  \log \frac{24 \pi \phi_r }{c \beta }\right) + \frac{c}{4}\log \left[\frac{\beta}{\pi}\;\frac{\cosh \left(\frac{4 \pi}{\beta} r+  \log \frac{24 \pi  \phi_r }{c \beta } \right)-1}{\sinh \left(\frac{2 \pi}{\beta} r+ \log \frac{24 \pi  \phi_r }{c \beta } \right)} \right]\nonumber\\
		&-&\frac{3\pi\phi_r}{\beta } \coth \left(\frac{2\pi}{\beta} a+  \log \frac{24 \pi \phi_r }{c \beta }\right) - \frac{c}{4}\log \left[\frac{\beta}{\pi}\;\frac{\cosh \left(\frac{4 \pi}{\beta} a+  \log \frac{24 \pi  \phi_r }{c \beta } \right)-1}{  \sinh \left(\frac{2 \pi}{\beta} a+ \log \frac{24 \pi  \phi_r }{c \beta } \right)} \right]\nonumber\\
		&+& \frac{c}{2} \log \left[\frac{\beta}{\pi } \sinh \frac{\pi}{\beta } |a-r| \right].
	\end{eqnarray}
	
	\subsubsection*{$\bm{(iii)}$ Subsystems $\bm{A}$, $\bm{B}$ and $\bm{C}$ fixed, time varied}\label{Appendix3}
	The third scenario involves two sub cases of equal and unequal lengths of the subsystems $A$ and $B$ with increasing time where the lengths $l_1$ and $l_2$ of the two adjacent subsystems are fixed. The expression for the corresponding entanglement negativity between the subsystems $A$ and $B$ for these two sub cases in different phases are given as 
	\subsubsection*{(a) For $\bm{l_1=l_2}$}\label{Appendix3a}
	
	\textbf{\;\,\,\,\, Phase 1}
	\begin{eqnarray}
		\mathcal{E}(A:B) &=& \frac{c}{2} \log \left[\frac{\beta}{\pi } \cosh \frac{2\pi }{\beta }t\right],
	\end{eqnarray}
	
	\textbf{Phase 2}
	\begin{eqnarray}
		\mathcal{E}(A:B) &=& -3\phi_0 - \frac{3\pi\phi_r}{\beta } \coth \left(\frac{2\pi}{\beta} a+  \log \frac{24 \pi \phi_r }{c \beta }\right) - \frac{c}{4}\log \left[\frac{\beta}{\pi}\;\frac{\cosh \left(\frac{4 \pi}{\beta} a+  \log \frac{24 \pi  \phi_r }{c \beta } \right)-1}{  \sinh \left(\frac{2 \pi}{\beta} a+ \log \frac{24 \pi  \phi_r }{c \beta } \right)} \right]\nonumber\\
		&-&\frac{3\pi\phi_r}{\beta } \coth \left(\frac{2\pi}{\beta}(L-b)+  \log \frac{24 \pi \phi_r }{c \beta }\right)\nonumber\\
		&-& \frac{c}{4}\log \left[\frac{\beta}{\pi}\;\frac{\cosh \left(\frac{4 \pi}{\beta} (L-b)+  \log \frac{24 \pi  \phi_r }{c \beta } \right)-1}{\sinh \left(\frac{2 \pi}{\beta} (L-b)+  \log \frac{24 \pi  \phi_r }{c \beta } \right)} \right]\nonumber\\
		&+&\frac{c}{2} \log \left[\frac{\beta}{\pi } \sinh \frac{\pi}{\beta } |a-r| \right]+\frac{c}{2} \log \left[\frac{\beta}{\pi } \sinh \frac{\pi}{\beta } |r-b| \right].
	\end{eqnarray}
	
	\subsubsection*{(b) For $\bm{l_1\neq l_2}$}\label{Appendix3b}
	
	\textbf{\;\,\,\,\, Phase 1}
	\begin{eqnarray}
		\mathcal{E}(A:B) &=& \frac{c}{2} \log \left[\frac{\beta}{\pi } \cosh \frac{2\pi }{\beta }t\right],
	\end{eqnarray}
	
	\textbf{Phase 2}
	\begin{eqnarray}
		\mathcal{E}(A:B) &=&  \frac{c}{4} \log \left[\frac{\beta}{\pi } \cosh \frac{2\pi }{\beta }t\right]+\frac{c}{2} \log \left[\frac{\beta}{\pi } \sinh \frac{\pi}{\beta } |r-b| \right] - \frac{3}{2}\phi_0\nonumber\\
		&-&\frac{3\pi\phi_r}{\beta } \coth \left(\frac{2\pi}{\beta}(L-b)+  \log \frac{24 \pi \phi_r }{c \beta }\right)\nonumber\\
		&-& \frac{c}{4}\log \left[\frac{\beta}{\pi}\;\frac{\cosh \left(\frac{4 \pi}{\beta} (L-b)+  \log \frac{24 \pi  \phi_r }{c \beta } \right)-1}{\sinh \left(\frac{2 \pi}{\beta} (L-b)+  \log \frac{24 \pi  \phi_r }{c \beta } \right)} \right],
	\end{eqnarray}
	
	\textbf{Phase 3}
	\begin{eqnarray}
		\mathcal{E}(A:B) &=& \frac{c}{4}\log \left[\frac{\beta}{\pi}\;\frac{\cosh \left(\frac{4 \pi}{\beta} (L-r)+  \log \frac{24 \pi  \phi_r }{c \beta } \right)-1}{\sinh \left(\frac{2 \pi}{\beta} (L-r)+  \log \frac{24 \pi  \phi_r }{c \beta } \right)} \right]+\frac{c}{2} \log \left[\frac{\beta}{\pi } \sinh \frac{\pi}{\beta } |r-b| \right]\nonumber\\
		&-&\frac{c}{4}\log \left[\frac{\beta}{\pi}\;\frac{\cosh \left(\frac{4 \pi}{\beta} (L-b)+  \log \frac{24 \pi  \phi_r }{c \beta } \right)-1}{\sinh \left(\frac{2 \pi}{\beta} (L-b)+  \log \frac{24 \pi  \phi_r }{c \beta } \right)} \right].
	\end{eqnarray}

	\subsection*{Disjoint subsystems}	
	We now list the expressions of the holographic entanglement negativity between the two disjoint subsystems $A$ and $B$ in three different scenarios involving subsystem sizes and the time by utilizing the \cref{NegDis0,HEE-Shaghoulian} as described in sub\cref{disj}.
	\subsubsection*{$\bm{(i)}$ Subsystem $\bm{A}$ fixed, $\bm{C}$ varied}\label{Appendix4}
	In the first scenario, the size of the subsystem $A$ is fixed at a constant time slice while the size of the subsystem $C$ is varied. The expressions of the holographic entanglement negativity between the subsystems $A$ and $B$ in the different phases are given as follows
	
	\textbf{\;\,\,\,\, Phase 1}
	\begin{eqnarray}
		\mathcal{E}(A:B) &=& \frac{c}{2} \log \left[\frac{\beta}{\pi } \sinh \frac{\pi}{\beta } |a-r_2| \right]-\frac{c}{2} \log \left[\frac{\beta}{\pi } \sinh \frac{\pi}{\beta } |r_1-r_2| \right]\nonumber\\
		&+&\frac{3\pi\phi_r}{\beta } \coth \left(\frac{2\pi}{\beta} r_1+  \log \frac{24 \pi \phi_r }{c \beta }\right)+\frac{c}{4}\log \left[\frac{\beta}{\pi}\;\frac{\cosh \left(\frac{4 \pi}{\beta} r_1+  \log \frac{24 \pi  \phi_r }{c \beta } \right)-1}{  \sinh \left(\frac{2 \pi}{\beta} r_1+ \log \frac{24 \pi  \phi_r }{c \beta } \right)} \right]\nonumber\\
		&-&\frac{3\pi\phi_r}{\beta } \coth \left(\frac{2\pi}{\beta} a+  \log \frac{24 \pi \phi_r }{c \beta }\right)-\frac{c}{4}\log \left[\frac{\beta}{\pi}\;\frac{\cosh \left(\frac{4 \pi}{\beta} a+  \log \frac{24 \pi  \phi_r }{c \beta } \right)-1}{  \sinh \left(\frac{2 \pi}{\beta} a+ \log \frac{24 \pi  \phi_r }{c \beta } \right)} \right],
	\end{eqnarray}
	
	\textbf{Phase 2}
	\begin{eqnarray}
		\mathcal{E}(A:B) &=& \frac{c}{4} \log \left[\frac{\beta}{\pi } \cosh \frac{2\pi }{\beta }t\right]-\frac{c}{2} \log \left[\frac{\beta}{\pi } \sinh \frac{\pi}{\beta } |r_1-r_2| \right]+\frac{3}{2}\phi_0\nonumber\\
		&+&\frac{3\pi\phi_r}{\beta } \coth \left(\frac{2\pi}{\beta} r_1+  \log \frac{24 \pi \phi_r }{c \beta }\right)+\frac{c}{4}\log \left[\frac{\beta}{\pi}\;\frac{\cosh \left(\frac{4 \pi}{\beta} r_1+  \log \frac{24 \pi  \phi_r }{c \beta } \right)-1}{  \sinh \left(\frac{2 \pi}{\beta} r_1+ \log \frac{24 \pi  \phi_r }{c \beta } \right)} \right]\nonumber\\
		&-&\frac{3\pi\phi_r}{\beta } \coth \left(\frac{2\pi}{\beta}(L-b)+  \log \frac{24 \pi \phi_r }{c \beta }\right)\nonumber\\
		&-&\frac{c}{4}\log \left[\frac{\beta}{\pi}\;\frac{\cosh \left(\frac{4 \pi}{\beta} (L-b)+  \log \frac{24 \pi  \phi_r }{c \beta } \right)-1}{\sinh \left(\frac{2 \pi}{\beta} (L-b)+  \log \frac{24 \pi  \phi_r }{c \beta } \right)} \right]\nonumber\\
		&+&\frac{3\pi\phi_r}{\beta } \coth \left(\frac{2\pi}{\beta} a+  \log \frac{24 \pi \phi_r }{c \beta }\right)+\frac{c}{4}\log \left[\frac{\beta}{\pi}\;\frac{\cosh \left(\frac{4 \pi}{\beta} a+  \log \frac{24 \pi  \phi_r }{c \beta } \right)-1}{  \sinh \left(\frac{2 \pi}{\beta} a+ \log \frac{24 \pi  \phi_r }{c \beta } \right)} \right],
	\end{eqnarray}
	
	\textbf{Phase 3}
	\begin{eqnarray}
		\mathcal{E}(A:B) &=& \frac{3\pi\phi_r}{\beta } \coth \left(\frac{2\pi}{\beta}(L-r_2)+  \log \frac{24 \pi \phi_r }{c \beta }\right)\nonumber\\
		&+&\frac{c}{4}\log \left[\frac{\beta}{\pi}\;\frac{\cosh \left(\frac{2 \pi}{\beta} (L-r_2)+  \log \frac{24 \pi  \phi_r }{c \beta }\right)-1}{\sinh \left(\frac{2 \pi}{\beta} (L-r_2)+  \log \frac{24 \pi  \phi_r }{c \beta } \right)} \right]\nonumber\\
		&-&\frac{c}{2} \log \left[\frac{\beta}{\pi } \sinh \frac{\pi}{\beta } |r_1-r_2| \right]+3\phi_0+ \frac{3\pi\phi_r}{\beta } \coth \left(\frac{2\pi}{\beta} r_1+  \log \frac{24 \pi \phi_r }{c \beta }\right)\nonumber\\
		&+&\frac{c}{4}\log \left[\frac{\beta}{\pi}\;\frac{\cosh \left(\frac{4 \pi}{\beta} r_1+  \log \frac{24 \pi  \phi_r }{c \beta } \right)-1}{  \sinh \left(\frac{2 \pi}{\beta} r_1+ \log \frac{24 \pi  \phi_r }{c \beta } \right)} \right],
	\end{eqnarray}
	
	\textbf{Phase 4}
	\begin{eqnarray}
		\mathcal{E}(A:B) &=& 0.
	\end{eqnarray}
	
	\subsubsection*{$\bm{(ii)}$ Subsystems $\bm{A}$ and $\bm{C}$ fixed, $\bm{B}$ varied}\label{Appendix5}
	In the second scenario, the size of the subsystems $A$ and $C$ are fixed at a constant time slice while the size of the subsystems $B$ is being varied. The expressions of the entanglement negativity between the subsystems $A$ and $B$ in the different phases are indicated as
	
	\textbf{Phase 1}
	\begin{eqnarray}
		\mathcal{E}(A:B) &=& \frac{c}{2} \log \left[\frac{\beta}{\pi } \sinh \frac{\pi}{\beta } |a-r_2| \right]+\frac{c}{2} \log \left[\frac{\beta}{\pi } \sinh \frac{\pi}{\beta } |r_1 - b| \right]\nonumber\\
		&-&  \frac{c}{2} \log \left[\frac{\beta}{\pi } \sinh \frac{\pi}{\beta } |r_1-r_2| \right]- \frac{c}{2} \log \left[\frac{\beta}{\pi } \sinh \frac{\pi}{\beta } |a - b| \right],
	\end{eqnarray}
	
	\textbf{Phase 2}
	\begin{eqnarray}
		\mathcal{E}(A:B) &=& \frac{c}{2} \log \left[\frac{\beta}{\pi } \sinh \frac{\pi}{\beta } |a-r_2| \right]+\frac{c}{2} \log \left[\frac{\beta}{\pi } \sinh \frac{\pi}{\beta } |r_1 - b| \right] - \frac{c}{2} \log \left[\frac{\beta}{\pi } \sinh \frac{\pi}{\beta } |r_1-r_2| \right]\nonumber\\
		&-&\frac{c}{4} \log \left[\frac{\beta}{\pi } \cosh \frac{2\pi }{\beta }t\right] - \frac{3\pi\phi_r}{\beta } \coth \left(\frac{2\pi}{\beta} a+  \log \frac{24 \pi \phi_r }{c \beta }\right)\nonumber\\
		&-& \frac{c}{4}\log \left[\frac{\beta}{\pi}\;\frac{\cosh \left(\frac{4 \pi}{\beta} a+  \log \frac{24 \pi  \phi_r }{c \beta } \right)-1}{  \sinh \left(\frac{2 \pi}{\beta} a+ \log \frac{24 \pi  \phi_r }{c \beta } \right)} \right],
	\end{eqnarray}
	
	\textbf{Phase 3}
	\begin{eqnarray}
		\mathcal{E}(A:B) &=& \frac{c}{2} \log \left[\frac{\beta}{\pi } \sinh \frac{\pi}{\beta } |a-r_2| \right]+\frac{c}{2} \log \left[\frac{\beta}{\pi } \sinh \frac{\pi}{\beta } |r_1 - b| \right]-\frac{c}{2} \log \left[\frac{\beta}{\pi } \sinh \frac{\pi}{\beta } |r_1-r_2| \right]\nonumber\\
		&-&\frac{3\pi\phi_r}{\beta } \coth \left(\frac{2\pi}{\beta}(L-b)+  \log \frac{24 \pi \phi_r }{c \beta }\right)\nonumber\\
		&-&\frac{c}{4}\log \left[\frac{\beta}{\pi}\;\frac{\cosh \left(\frac{4 \pi}{\beta} (L-b)+  \log \frac{24 \pi  \phi_r }{c \beta } \right)-1}{\sinh \left(\frac{2 \pi}{\beta} (L-b)+  \log \frac{24 \pi  \phi_r }{c \beta } \right)} \right] - 3\phi_0\nonumber\\
		&-& \frac{3\pi\phi_r}{\beta } \coth \left(\frac{2\pi}{\beta} a+  \log \frac{24 \pi \phi_r }{c \beta }\right)-\frac{c}{4}\log \left[\frac{\beta}{\pi}\;\frac{\cosh \left(\frac{4 \pi}{\beta} a+  \log \frac{24 \pi  \phi_r }{c \beta } \right)-1}{  \sinh \left(\frac{2 \pi}{\beta} a+ \log \frac{24 \pi  \phi_r }{c \beta } \right)} \right],
	\end{eqnarray}
	
	\textbf{Phase 4}
	\begin{eqnarray}
		\mathcal{E}(A:B) &=& \frac{c}{2} \log \left[\frac{\beta}{\pi } \sinh \frac{\pi}{\beta } |a-r_2| \right]-\frac{c}{2} \log \left[\frac{\beta}{\pi } \sinh \frac{\pi}{\beta } |r_1-r_2| \right]\nonumber\\
		&+&\frac{3\pi\phi_r}{\beta } \coth \left(\frac{2\pi}{\beta} r_1+  \log \frac{24 \pi \phi_r }{c \beta }\right)+\frac{c}{4}\log \left[\frac{\beta}{\pi}\;\frac{\cosh \left(\frac{4 \pi}{\beta} r_1+  \log \frac{24 \pi  \phi_r }{c \beta } \right)-1}{  \sinh \left(\frac{2 \pi}{\beta} r_1+ \log \frac{24 \pi  \phi_r }{c \beta } \right)} \right]\nonumber\\
		&-&\frac{3\pi\phi_r}{\beta } \coth \left(\frac{2\pi}{\beta} a+  \log \frac{24 \pi \phi_r }{c \beta }\right)-\frac{c}{4}\log \left[\frac{\beta}{\pi}\;\frac{\cosh \left(\frac{4 \pi}{\beta} a+  \log \frac{24 \pi  \phi_r }{c \beta } \right)-1}{  \sinh \left(\frac{2 \pi}{\beta} a+ \log \frac{24 \pi  \phi_r }{c \beta } \right)} \right].
	\end{eqnarray}
	
	\subsubsection*{$\bm{(iii)}$ Subsystems $\bm{A}$, $\bm{B}$ and $\bm{C}$ fixed, time varied}\label{Appendix6}
	In the third scenario, the lengths $l_1$, $l_2$ and $l_c$ of the subsystems $A$, $B$ and $C$ are fixed respectively with increasing time. Here we consider two sub cases of equal and unequal lengths of the subsystems $A$ and $B$. The expressions of the corresponding entanglement negativity between the subsystems $A$ and $B$ in distinct phases are given as

	\subsubsection*{(a) For $\bm{l_1=l_2}$}\label{Appendix6a}
	
	\textbf{\;\,\,\,\, Phase 1}
	\begin{eqnarray}
		\mathcal{E}(A:B) &=& 0,
	\end{eqnarray}
	
	\textbf{Phase 2}
	\begin{eqnarray}
		\mathcal{E}(A:B) &=& \frac{c}{2} \log \left[\frac{\beta}{\pi } \cosh \frac{2\pi }{\beta }t\right]-\frac{c}{2} \log \left[\frac{\beta}{\pi } \sinh \frac{\pi}{\beta } |r_1-r_2| \right],
	\end{eqnarray}
	
	\textbf{Phase 3}
	\begin{eqnarray}
		\mathcal{E}(A:B) &=& \frac{c}{2} \log \left[\frac{\beta}{\pi } \sinh \frac{\pi}{\beta } |a-r_2| \right]+\frac{c}{2} \log \left[\frac{\beta}{\pi } \sinh \frac{\pi}{\beta } |r_1 - b| \right]-\frac{c}{2} \log \left[\frac{\beta}{\pi } \sinh \frac{\pi}{\beta } |r_1-r_2| \right]\nonumber\\
		&-&\frac{3\pi\phi_r}{\beta } \coth \left(\frac{2\pi}{\beta}(L-b)+  \log \frac{24 \pi \phi_r }{c \beta }\right)\nonumber\\
		&-&\frac{c}{4}\log \left[\frac{\beta}{\pi}\;\frac{\cosh \left(\frac{4 \pi}{\beta} (L-b)+  \log \frac{24 \pi  \phi_r }{c \beta } \right)-1}{\sinh \left(\frac{2 \pi}{\beta} (L-b)+  \log \frac{24 \pi  \phi_r }{c \beta } \right)} \right] - 3\phi_0\nonumber\\
		&-& \frac{3\pi\phi_r}{\beta } \coth \left(\frac{2\pi}{\beta} a+  \log \frac{24 \pi \phi_r }{c \beta }\right)-\frac{c}{4}\log \left[\frac{\beta}{\pi}\;\frac{\cosh \left(\frac{4 \pi}{\beta} a+  \log \frac{24 \pi  \phi_r }{c \beta } \right)-1}{  \sinh \left(\frac{2 \pi}{\beta} a+ \log \frac{24 \pi  \phi_r }{c \beta } \right)} \right].
	\end{eqnarray}
	
	\subsubsection*{(b) For $\bm{l_1\neq l_2}$}\label{Appendix6b}
	
	\textbf{\;\,\,\,\, Phase 1}
	\begin{eqnarray}
		\mathcal{E}(A:B) &=& 0,
	\end{eqnarray}
	
	\textbf{Phase 2}
	\begin{eqnarray}
		\mathcal{E}(A:B) &=& \frac{c}{2} \log \left[\frac{\beta}{\pi } \cosh \frac{2\pi }{\beta }t\right]-\frac{c}{2} \log \left[\frac{\beta}{\pi } \sinh \frac{\pi}{\beta } |r_1-r_2| \right],
	\end{eqnarray}
	
	\textbf{Phase 3}
	\begin{eqnarray}
		\mathcal{E}(A:B) &=& \frac{c}{2} \log \left[\frac{\beta}{\pi } \sinh \frac{\pi}{\beta } |a-r_2| \right]+\frac{c}{4} \log \left[\frac{\beta}{\pi } \cosh \frac{2\pi }{\beta }t\right]-\frac{c}{2} \log \left[\frac{\beta}{\pi } \sinh \frac{\pi}{\beta } |r_1-r_2| \right]-\frac{3}{2}\phi_0\nonumber\\
		&-&\frac{3\pi\phi_r}{\beta } \coth \left(\frac{2\pi}{\beta} a+  \log \frac{24 \pi \phi_r }{c \beta }\right)-\frac{c}{4}\log \left[\frac{\beta}{\pi}\;\frac{\cosh \left(\frac{4 \pi}{\beta} a+  \log \frac{24 \pi  \phi_r }{c \beta } \right)-1}{  \sinh \left(\frac{2 \pi}{\beta} a+ \log \frac{24 \pi  \phi_r }{c \beta } \right)} \right],
	\end{eqnarray}
	
	\textbf{Phase 4}
	\begin{eqnarray}
		\mathcal{E}(A:B) &=& \frac{c}{2} \log \left[\frac{\beta}{\pi } \sinh \frac{\pi}{\beta } |a-r_2| \right]-\frac{c}{2} \log \left[\frac{\beta}{\pi } \sinh \frac{\pi}{\beta } |r_1-r_2| \right]\nonumber\\
		&+&\frac{3\pi\phi_r}{\beta } \coth \left(\frac{2\pi}{\beta} r_1+  \log \frac{24 \pi \phi_r }{c \beta }\right)+\frac{c}{4}\log \left[\frac{\beta}{\pi}\;\frac{\cosh \left(\frac{4 \pi}{\beta} r_1+  \log \frac{24 \pi  \phi_r }{c \beta } \right)-1}{  \sinh \left(\frac{2 \pi}{\beta} r_1+ \log \frac{24 \pi  \phi_r }{c \beta } \right)} \right]\nonumber\\
		&-&\frac{3\pi\phi_r}{\beta } \coth \left(\frac{2\pi}{\beta} a+  \log \frac{24 \pi \phi_r }{c \beta }\right)\nonumber\\
		&-&\frac{c}{4}\log \left[\frac{\beta}{\pi}\;\frac{\cosh \left(\frac{4 \pi}{\beta} a+  \log \frac{24 \pi  \phi_r }{c \beta } \right)-1}{  \sinh \left(\frac{2 \pi}{\beta} a+ \log \frac{24 \pi  \phi_r }{c \beta } \right)} \right].
	\end{eqnarray}

\end{appendices}

\bibliographystyle{JHEP}

\bibliography{CommBH}

\end{document}